\documentclass{aa}  

\usepackage{graphicx}
\usepackage{txfonts}
\begin{document} 

   \title{How the cool-core population transitions \\ from galaxy groups to massive clusters}

   \subtitle{A comparison of the largest Magneticum simulation \\ with eROSITA, XMM-Newton,  Chandra and LOFAR observations}

   \author{Justo Antonio González Villalba
          \inst{1,} \inst{2} 
          \and
          Klaus Dolag
          \inst{1,} \inst{3} 
          \and
          Veronica Biffi
          \inst{4,} \inst{5}
          }

   \institute{
Universitäts-Sternwarte, Fakultät für Physik, Ludwig-Maximilians-Universität München, Scheinerstr. 1, 81679 München, Germany
   \and
   European Southern Observatory,
   Karl-Schwarzschildstr 2, D-85748 Garching bei München, Germany   
   \and
Max-Planck-Institut für Astrophysik, Karl-Schwarzschild-Straße 1, 85741 Garching, Germany
   \and
   INAF — Osservatorio Astronomico di Trieste, via Tiepolo 11, 34143 Trieste, Italy
   \and
   IFPU — Institute for Fundamental Physics of the Universe, via Beirut 2, 34014 Trieste, Italy
   }

   \date{Received 11 September 2024 / Accepted 12 December 2024}

\titlerunning{How the cool-core population transitions from galaxy groups to massive clusters}
\authorrunning{Justo Antonio González Villalba et. al}

 
  \abstract
   {}
   {Our aim is to understand how the interplay between AGN feedback and merge processes can effectively turn cool-core galaxy clusters into hot-core clusters in the modern universe. Additionally, we also aim to clarify which parameters of the AGN feedback model used in simulations can cause an excess of feedback at the scale of galaxy groups while not efficiently suppressing star formation at the scale of galaxy clusters.}
   {To obtain robust statistics of the cool-core population, we compare the modern Universe snapshot (z = 0.25) of the largest Magneticum simulation ({\it Box2b/hr}) with the eROSITA eFEDS survey and Planck SZ-selected clusters observed with XMM-Newton. Additionally, we compare the AGN feedback injected by the simulation in radio mode with Chandra observations of X-ray cavities, and LOFAR observations of radio emission.}
   {We confirm a decreasing trend in cool-core fractions towards the most massive galaxy clusters, which is well reproduced by the Magneticum simulations. This evolution is connected with an increased merge activity that injects high-energy particles into the core region, but it also requires thermalization and conductivity to enhance mixing through the ICM core, where both factors are increasingly efficient towards the high mass end. On the other hand, AGN feedback remains as the dominant factor at the scale of galaxy groups, while its relative impact decreases towards the most massive clusters.}
   {The problems suppressing star formation in simulations are not caused by low AGN feedback efficiencies. They root in the definition of the black hole sphere of influence used to distribute the feedback, which decreases as density and accretion rate increase. Actually, a decreasing AGN feedback efficiency towards low-mass galaxy groups is required to prevent overheating. These problems can be addressed in simulations by using relations between accretion rate, cavity power, and cavity reach derived from X-ray observations.}


   \maketitle
    
%

\section{Introduction} \label{sec:Introduction}

The cooling flow problem has been one of the major drivers pushing towards a better understanding of the key physical processes that the plasma hosted in the cores of galaxy clusters is undergoing. In the present work, we review the status of the cooling flow problem in galaxy clusters, the results obtained with the Magneticum simulations, and how the interplay between dynamical state and active galactic nuclei (AGN) feedback shapes the population of cool-core galaxy clusters in the modern universe.

As described by \cite{fabian2002cooling}, the intracluster medium gas (ICM) hosted in the central regions of galaxy clusters exhibits short adiabatic cooling times of less than 1 Gyr and mass deposition rates of $\sim [100-1000] M_\odot / yr$, which should theoretically lead to a runaway 'catastrophic cooling' situation with significant amounts of cooling gas and star formation.

However, spectra from the XMM-Newton Reflection Grating Spectrometer (RGS) detected increasingly less emission at lower temperatures than the cooling flow model would predict, with the coldest gas detected at around $\sim 6\cdot10^6 K / 0.5 \text{keV}$, corresponding to $\text{Fe}_{\text{XVII}}$ emission (\cite{peterson2003high}, \cite{sanders2010deep}). 

Similarly, star formation has been detected in bright cluster galaxies (BCGs), but at levels significantly lower than those produced by the mass deposition rates of the cooling flow model. For example, \citep{fraser2014rarity} analyzed a large sample of 245 BCGs at low redshift (z < 0.1) and found that most ($99 \pm 0.6$ \%) have star formation rates (SFR) below $\sim 10 M_\odot / yr$, with a few exceptions such as Cygnus A and Perseus. This result has been confirmed by \cite{mcdonald2018revisiting}, who analyzed a sample of 107 galaxy clusters with z < 0.5 (but mostly z < 0.3) and found that the efficiency of SFR is between $1-10\%$ of that predicted by the cooling flow mass deposition rates.

This situation, at odds with the theoretical predictions, is known as the 'cooling flow problem'. To solve these discrepancies, it is widely accepted that a heating source must be present, preventing further cooling of the ICM. However, the potential solutions have to comply with a number of requirements. In the first place, it has to prevent cooling in the whole core region, not only in the most central parts. Also, it has to operate through 2 orders of magnitude in mass, from the smallest galaxy groups with mass $10^{13} M_\odot$ to the most massive galaxy clusters with mass $10^{15} M_\odot$. Finally, it has to be fine-tuned to effectively quench cooling flows without causing obvious overheating (\cite{fabian2003gravitational}).

One alternative is stellar feedback. However, simulations have shown that there is an excess of cold gas and star formation in the core regions, even if star formation, metal enrichment, and stellar feedback are also included (\cite{nagai2007effects}, \cite{borgani2011cosmological}).

Another option is thermal conductivity as a mechanism to transport heat from outside the core region into it, given the negative temperature gradients seen in the radial direction of the core regions. However, \cite{dolag2004thermal} showed that even a 1/3 value of the Spitzer conductivity \citep{spitzer1962jr}, which would be motivated by randomly oriented magnetic fields, does not significantly alter star formation because most of the cooling and star formation takes place at high redshift, when the temperature of the gas in the ICM is low and therefore thermal conductivity is inefficient. However, it can be an important factor to thermally stabilize the core region and promote mixing once the structure formation process reaches the scales of galaxy clusters and the ICM has temperatures in the order of keVs.

Currently, the most widely accepted mechanism to prevent cooling flows in galaxy clusters is feedback from the central AGN hosted in the brightest cluster galaxy (BCG). This concept works in a straightforward manner at high redshift ($z > 1$), when most of the AGNs are in quasar mode, with luminosities in the range of $10^{44} - 10^{45}$ erg/s, comparable to the ICM X-ray luminosity in the cooling region, and jets at the scale of $\sim 100$ kpc. However, in the modern universe, most of the AGNs hosted in BCGs are in radio mode and have low luminosities of $<10^{43}$ erg/s, clearly below the ICM X-ray luminosities of the cooling region (\cite{russell2013radiative}, \cite{fujita2014agn}).

On the other hand, the radio emission of AGNs in the modern universe is usually coincident with cavities whose enthalpy is in the order of $\sim10^{51}-10^{61}$ erg, which, when divided by typical times of the system, such as the buoyant rise time, the refill time, or the sound crossing time, yields cavity powers in the range of $10^{42} - 10^{45}$ erg/s, comparable with the ICM X-ray luminosity in the cooling region (\cite{rafferty2006feedback}).

To explain the cavity powers in the absence of radiatively efficient AGNs, \cite{churazov2001evolution}, \cite{churazov2005supermassive} proposed mechanical feedback, which thermalizes at almost 100\% efficiency, inflating bubbles that then can rise buoyantly. However, the remaining question is how the energy is isotropically transported to the entire core region.

The first implementations of active galactic nucleus (AGN) feedback in cosmological hydrodynamic simulations, by \cite{di2005energy} and \cite{springel2005modelling}, have demonstrated its effectiveness in reducing the high redshift star formation and reproducing the $M_{\mathrm{BH}}-\sigma$ relation, with a radiative efficiency set to the canonical value of $\epsilon_{\mathrm{r}} = 0.1$, corresponding to a standard Shakura-Sunyaev disk (\cite{shakura1973black}), and a relatively low feedback efficiency value of $\epsilon_{\mathrm{f}} = 0.05$. 

Refinements of the model, by \cite{sijacki2007unified} and \cite{fabjan2010simulating}, aimed at further suppressing star formation at low redshift by introducing a radio mode with higher feedback efficiency at low accretion rates ($\dot{M}_{\mathrm{BH}} / \dot{M}_{\mathrm{Edd}}<10^{-2}$), as suggested by \cite{churazov2005supermassive}, to suppress cooling flows in the modern universe, where most of AGNs are in radio mode and the accretion rates inferred by observations are low.

However, the simulations of \cite{fabjan2010simulating}, which covered a wide range of masses, with $M_{200}$ in the range $[0.52-18.51]10^{14}M_\odot$, showed that the problems at the scale of massive clusters seen in simulations are not solved by increasing the feedback efficiency in radio mode to $\epsilon_{\mathrm{f}} = 0.2$ or even $\epsilon_{\mathrm{f}} = 0.8$. In particular, the stellar mass fractions are still about three times higher than observed. On the other hand, the increased feedback efficiencies introduce differences at the scale of galaxy groups, namely reduced gas fractions and an excess of entropy in the central regions. 

The overheating issues seen in simulations at the scale of galaxy groups have also been pointed out by \cite{gaspari2013solving}, who argues that groups are not scaled-down versions of clusters and require lower feedback efficiencies and gentler feedback to avoid 'catastrophic heating'. Moreover, recent comparisons between simulations and large samples of galaxy groups from the first eROSITA All-Sky survey (eRASS1) by \cite{bahar2024srgerosita} have shown an excess of entropy in the core regions of simulated low-mass galaxy groups.

At this point, it is also important to highlight that, whereas the energetics of AGN feedback in radio mode can halt the development of cooling flows in the modern universe, as inferred from cavities with enthalpy in the order of $\sim10^{51}-10^{61}$ ergs, the energy required to transform a cool-core cluster into a hot-core cluster is orders of magnitude higher, $\sim [1-4] 10^{61}$ ergs, far beyond the most energetic AGN bursts observed in the modern universe \citep{mccarthy2008towards}.

On the other hand, galaxy cluster mergers are the most energetic events since the Big Bang, releasing gravitational binding energies in the order of $\sim10^{64}$ erg \citep{sarazin2002physics}. Actually, cool-core clusters are rarely found among dynamically active systems with strong evidence of merge activity. For example, one indicator of merge activity is the projected offset between the BCG and X-ray centroid, which is under 40 kpc for all low entropy systems of the Chandra ACCEPT sample \citep{hoffer2012infrared} and below <0.02 for the strongest cool-cores of the LoCuSS sample \citep{sanderson2009locuss}.

Moreover, \cite{chen2007statistics} reported a trend whereby the fraction of cool-core clusters decreases towards the most massive, dynamically young systems. This trend has also been reproduced qualitatively by the simulations of \cite{burns2008only} and \cite{planelles2009galaxy}.

However, \cite{burns2008only} showed that although hot-core clusters are characterized by major merges at the early phases of cluster assembly (up to z < 0.5), which destroy the nascent cores, the situation is different for cool-core clusters, which undergo a smoother assembly process at early epochs and become resilient to late mergers. Also, \cite{poole2008impact} studied two-body cluster mergers in simulations and showed that merges between compact cool-cores (CCC) with mass ratios of 1:3 and 1:1 appear disturbed for a period of 4 Gyr, although they recover the CCC state 3 Gyr after the relaxation time.

On the other hand, \cite{rasia2015cool} showed that late mergers can also destroy cool-core systems if AGN feedback is included, whereas the simulations of \cite{burns2008only} and \cite{poole2008impact} included only stellar feedback, not AGN feedback.

These results suggest that cluster growth via accretion and merge processes does have a fundamental role in shaping not only the initial cool-core population but also its evolution. However, pre-heating via both stellar and AGN feedback is required to soften the nascent cool-cores and make them susceptible to disruption by early and late mergers. The pre-heating requirement is also based on previous simulations by \cite{motl2004formation}, which did not include any pre-heating and produced a cool-core in almost all halos.

However, reproducing the fraction of systems hosting a cool-core has been quite challenging \citep{barnes2018census}, due to observational biases and the cool-core classification criteria. From the observational side, \cite{andrade2017fraction} calculated that cool-core systems are overrepresented in observations by a factor of 2.1–2.7 in X-ray selected samples due to the Malmquist bias and obtained a fraction of 44\% cool-core systems in the HIFLUGCS X-ray selected sample using the scaled concentration parameter, versus 28\% in the early Planck survey (ESZ), with a selection based on the thermal Sunyaev-Zeldovich effect (SZ), which approximates mass selection.

Moreover, the fraction of cool-core systems is highly dependent on the indicator and threshold used, which makes it difficult to compare different studies and especially between observations and simulations. For example, \cite{andrade2017fraction} obtained 28\% of cool-core systems using the scaled concentration ratio in the 0.15-1.0 $R_{500c}$ range and a threshold of 0.075, but 36\% of cool-core systems using the concentration ratio in the 40–400 kpc range and a threshold of 0.4. Whereas on the side of simulations \cite{burns2008only} and \cite{planelles2009galaxy} obtained a 16\% of systems hosting a cool-core, defining them as systems where the central temperature drops by at least 20\% in comparison with the viral temperature, and \cite{rasia2015cool} obtained 38\% of cool-core systems using as definition the ratio of pseudo entropy in the 0.05-0.15 $R_{200c}$ range and a threshold of 0.55.

Therefore, in this work, we aim at using the same cool-core classification criteria for observations and simulations to make a more direct comparison. Additionally, we resort to a large observational data sample by combining the recent eROSITA field equatorial deep survey data (\cite{bahar2022erosita}, \cite{chiu2022erosita}) with the Planck SZ-selected sample from \cite{lovisari2020x}. This can be compared to a statistically relevant cluster sample from the large volume simulation {\it Box2b/hr} of the Magneticum Pathfinder simulations \citep{dolag2015magneticum}, allowing to study the mass dependency in cool-core fractions reported by \cite{chen2007statistics}, from the scale of galaxy groups up to the most massive clusters, and to relate this trend to the interplay between dynamical state and AGN feedback. Additionally, we also study which aspects of the AGN feedback model and its numerical implementation relate to the excess of entropy at the scale of galaxy groups reported by \cite{fabjan2010simulating} and \cite{bahar2024srgerosita}, as well as the excess of star formation at the scale of galaxy clusters reported by \cite{fabjan2010simulating}.

\section{The simulations} \label{sec:DescriptionSimulations}

\begin{figure*}[ht!]
    \centering
    \includegraphics[width=0.65\columnwidth]{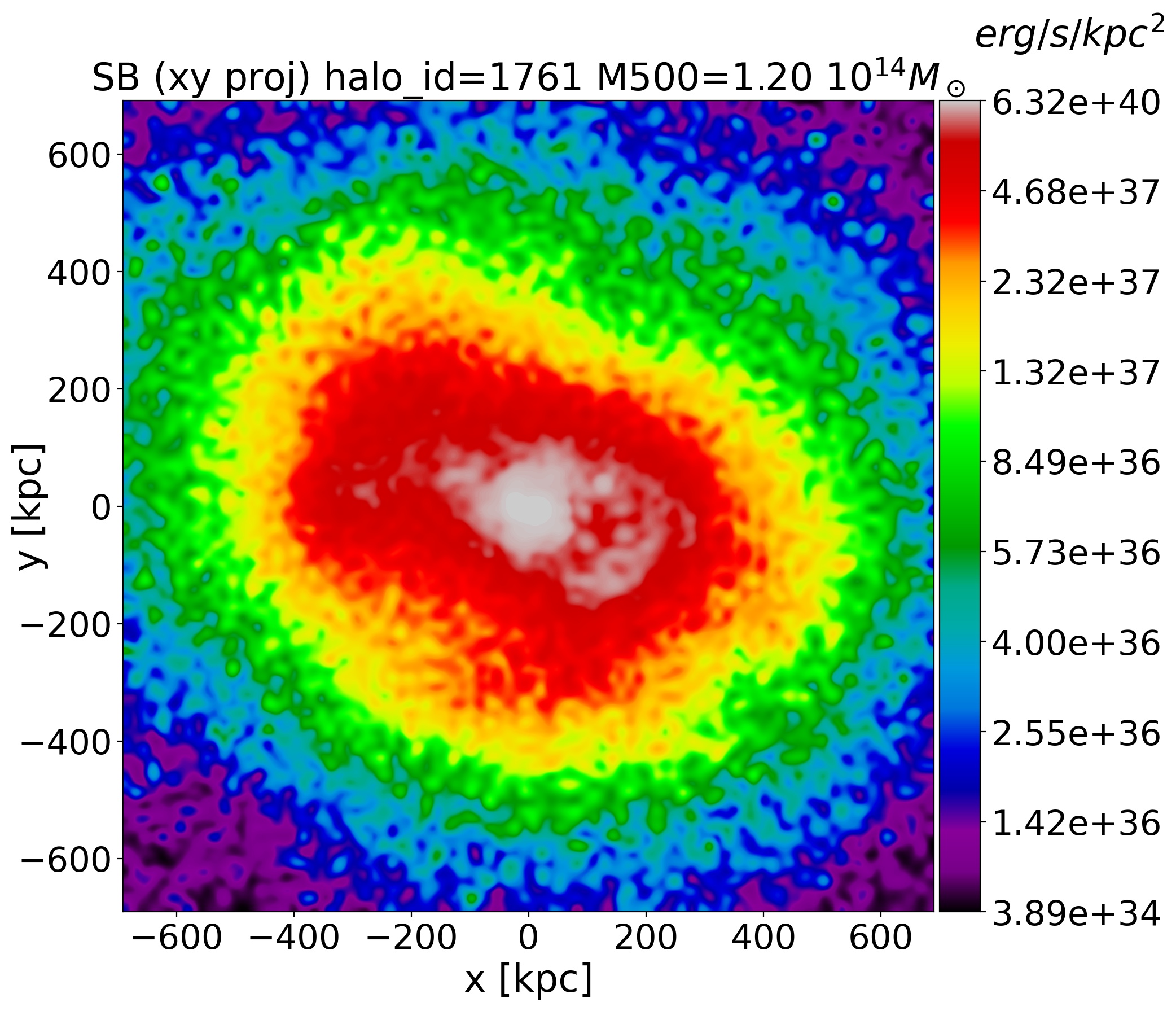}
    \includegraphics[width=0.65\columnwidth]{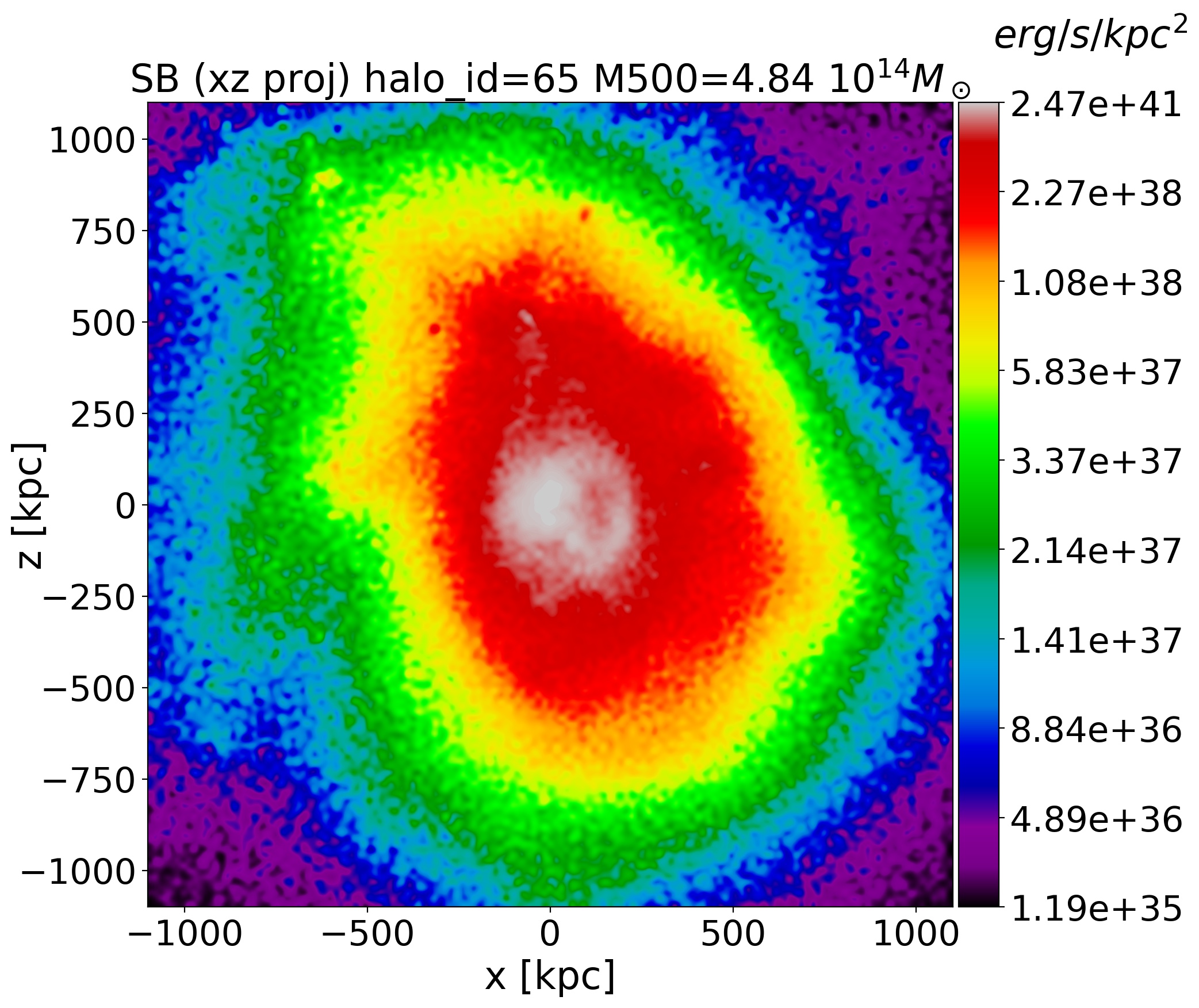}
    \includegraphics[width=0.65\columnwidth]{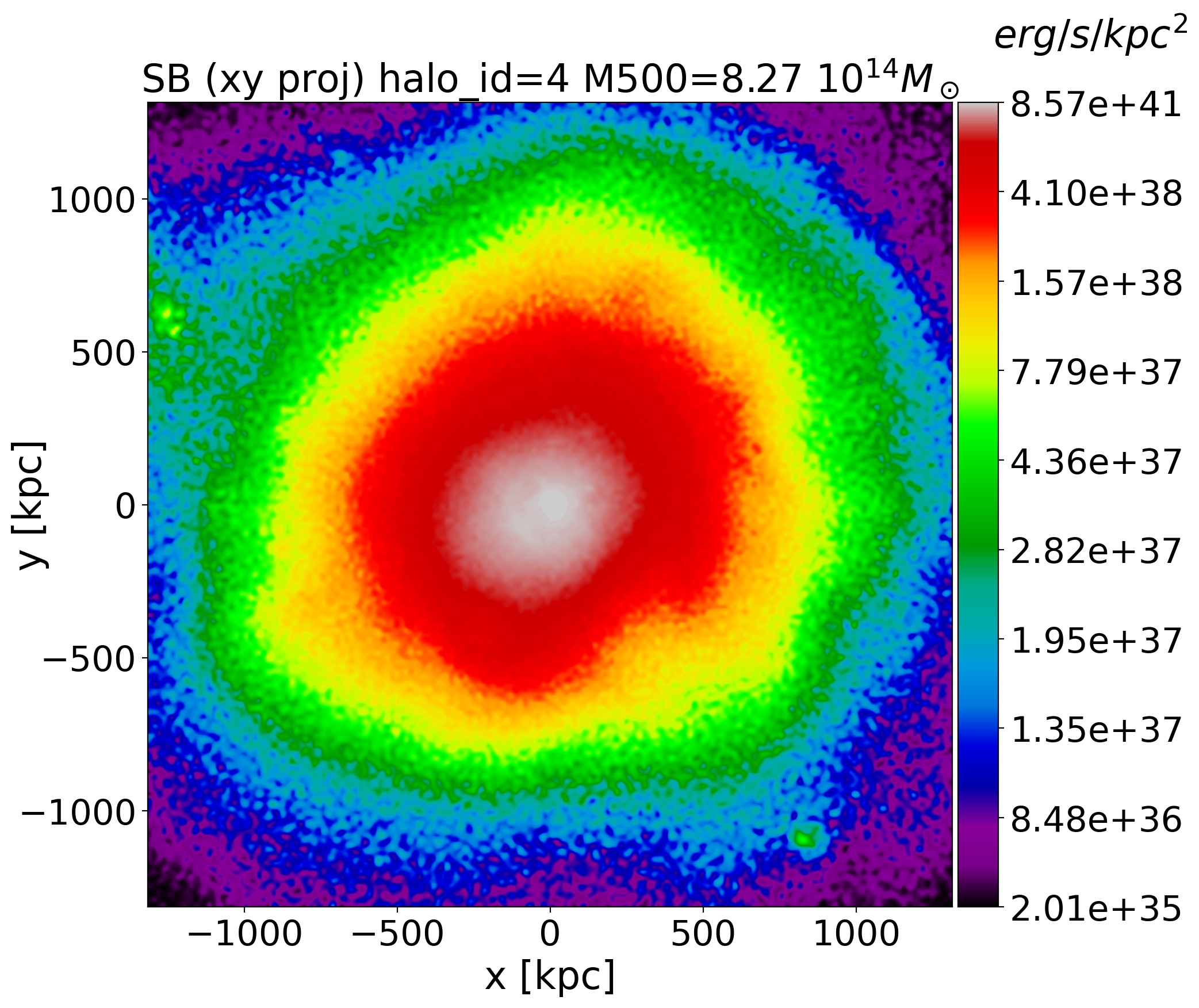}
    \includegraphics[width=0.65\columnwidth]{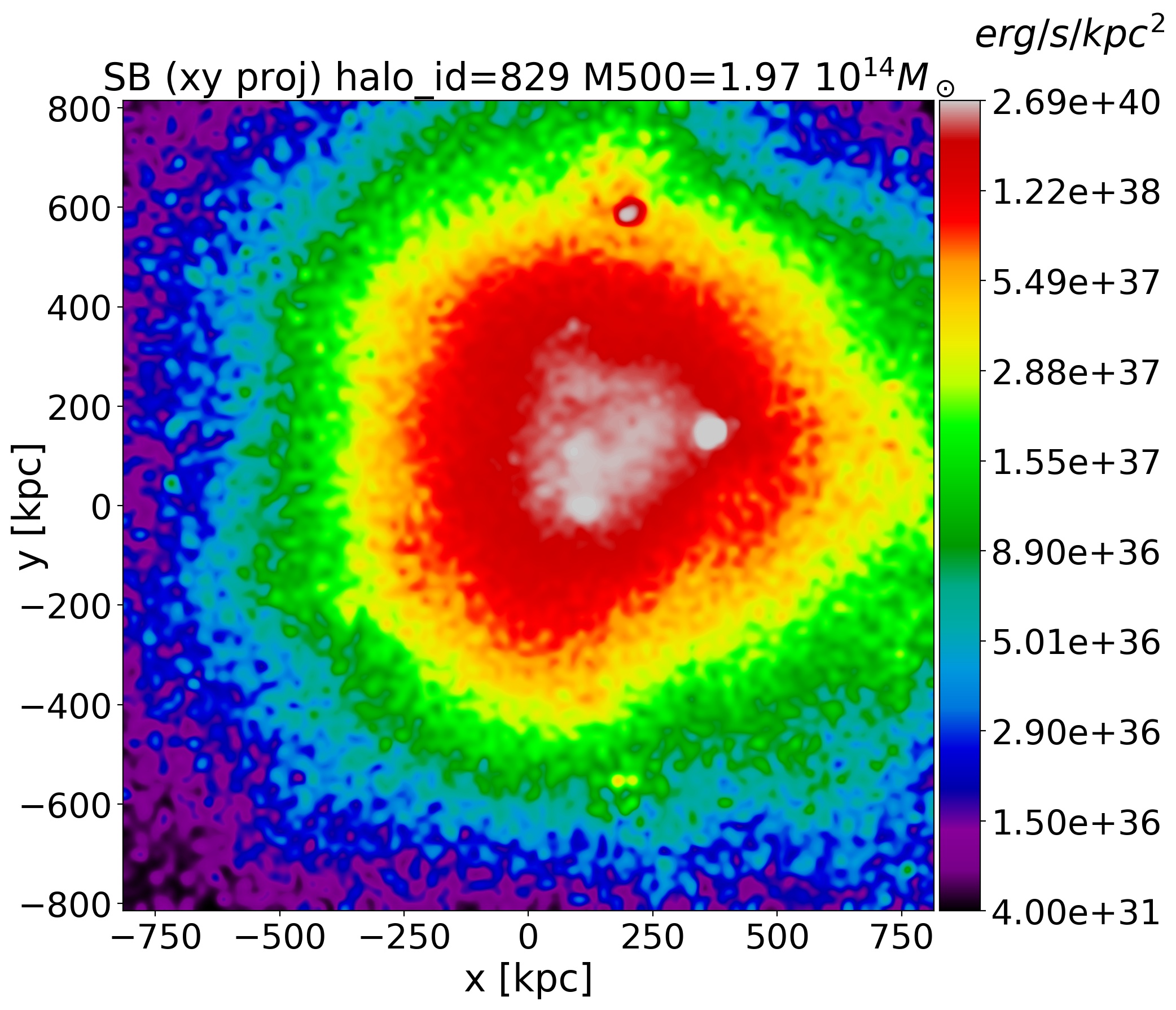}
    \includegraphics[width=0.65\columnwidth]{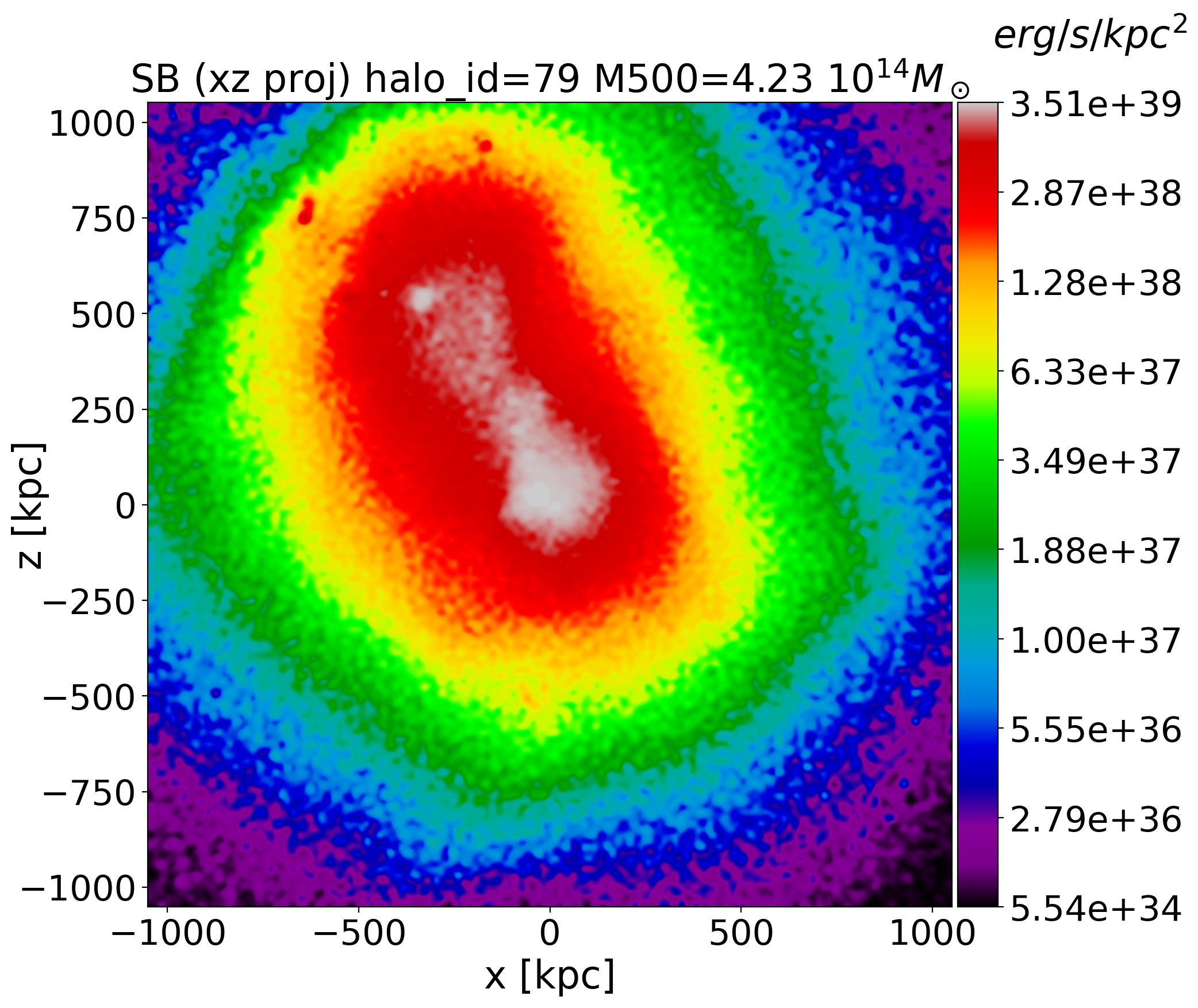}
    \includegraphics[width=0.65\columnwidth]{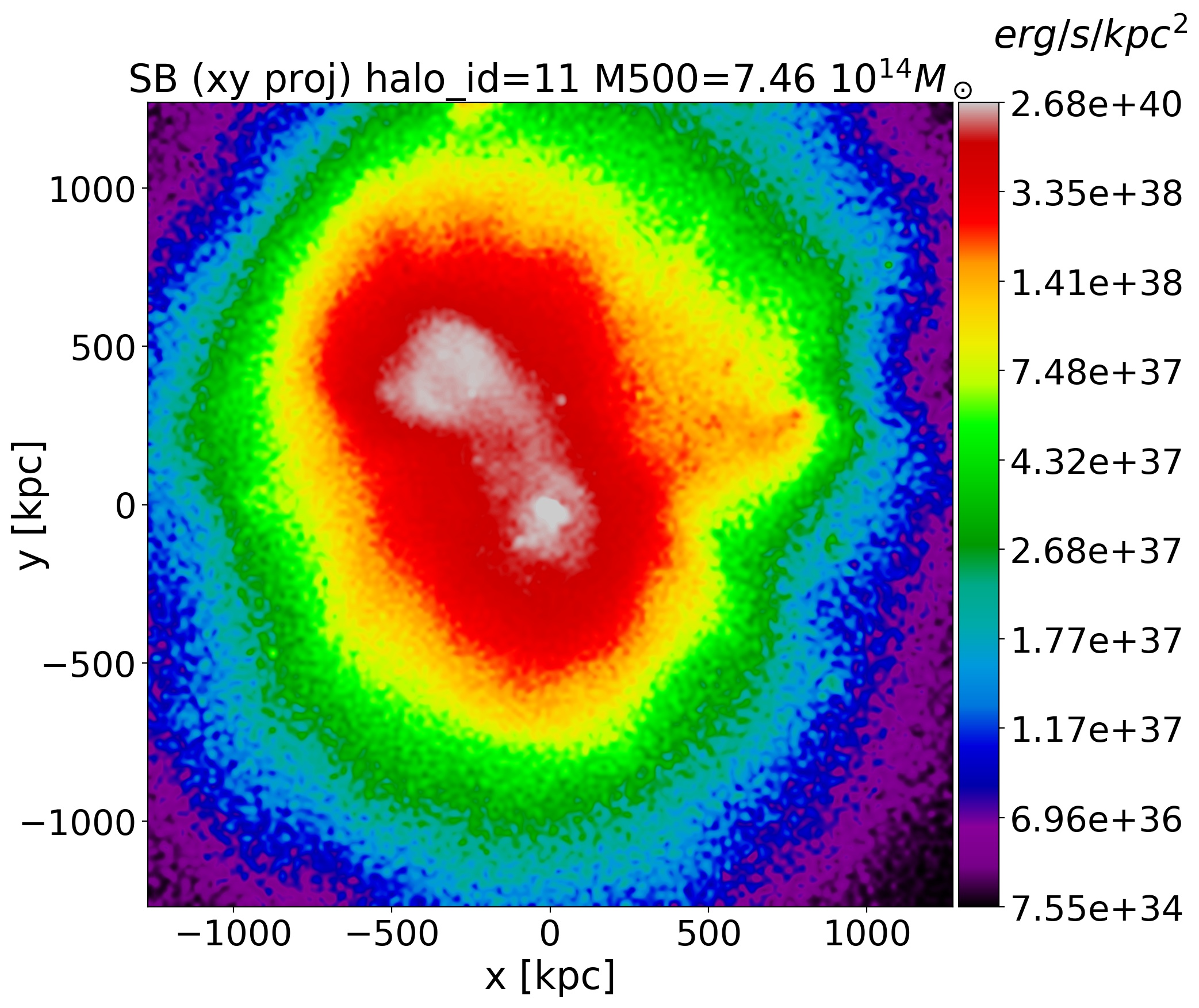}
    \caption{X-ray SB images of three cool-core (upper row) and 3 hot-core (bottom row) clusters selected within the 3 highest mass bins. While the hot-core clusters are highly perturbed and merging systems, the cool-core clusters show more regular shapes but also display some cavity-like features within the SB maps.}
 \label{fig:ClusterExamples}

\end{figure*}

In this work, we compare observational data versus the Magneticum Pathfinder simulations \citep{dolag2015magneticum}, which are based on the parallel cosmological Tree Particle-Mesh (PM) Smoothed-particle Hydrodynamics (SPH) code P-Gadget3, an extended version of P-Gadget2 \citep{springel2005cosmological}, with several new improvements as described by \cite{beck2016improved}: A bias-corrected, sixth order Wendland kernel with 295 neighbours \citep{dehnen2012improving}, an improved locally adaptive treatment for artificial viscosity \citep{dolag2005turbulent,cullen2010inviscid}, and inclusion of locally adaptive artificial conduction \citep{wadsley2008treatment, price2008modelling} among others.

The cosmological model is based on the flat $\Lambda$CDM WMAP7 cosmology \citep{Komatsu_2011}, with dimensionless Hubble constant h=0.704, total matter density $\Omega_m=0.272$, baryonic density $\Omega_b=0.0456$ (baryon fraction 16.76\%), a spectral index of primordial scalar fluctuations of $n_s=0.963$, and an amplitude of the angular power spectrum of $\sigma_8=0.809$.

The most relevant baryonic processes are implemented in different modules, including isotropic thermal conduction \citep{dolag2004thermal} at the level of 1/20 of the Spitzer conductivity \citep{spitzer1962jr}, radiative cooling accounting for the local metallicity \citep{wiersma2009effect}, and heating from a uniform but time-dependent UV/X-ray component modeling the background radiation from quasars and galaxies \citep{haardt2001clusters}.

Star formation is modeled by following a hybrid multiphase star formation model that also provides feedback in the form of galactic winds at a velocity of 350 km/s, produced by 10\% of the massive stars triggering Type II supernova explosions that release $10^{51}$ erg \citep{springel2003cosmological}. Chemical enrichment is also included as in \cite{tornatore2004simulating,tornatore2007simulating}, assuming the \cite{chabrier2003galactic} initial mass function and following stellar evolution models (supernova Type Ia, Type II, and stars in AGB phase).

Thermal AGN feedback is also incorporated, following an updated implementation of the \cite{di2005energy} and \cite{springel2005modelling} model, which includes a radio mode with higher feedback efficiency as described by \cite{fabjan2010simulating}, and some new minor changes by \cite{hirschmann2014cosmological}, such as not enforcing pinning of black hole particles to the minimum of the potential within the smoothing length and a black hole (BH) seeding criterion based on stellar mass instead of dark matter mass.

As shown in previous studies, the physics model implemented in the \textit{Magneticum} simulations leads to an overall successful reproduction of the basic galaxy cluster and group properties. Among many other properties, the simulations reproduce the observable X-ray luminosity-relation \citep{biffi2013scaling} of clusters, the pressure profile of the ICM \citep{gupta2017pressure}, and the chemical composition \citep{dolag2017metals,biffi2018review} of the ICM. The simulations also reproduce the high concentration observed in fossil groups \citep{ragagnin2019concentration} as well as the gas properties within galaxy clusters \citep{angelinelli2022bfrac} and between galaxy clusters \citep{biffi2022bridge}. This accordance also extends to the group regime when compared to eROSITA observations \citep[see][]{bahar2024srgerosita,marini2024groups}. The Magneticum Pathfinder simulations are offered in a variety of resolutions and volumes, which are publicly available on the {\it Cosmological Web Portal}\footnote{https://c2papcosmosim.uc.lrz.de/}\citep{ragagnin2017webportal}. 

In this work, we use data from {\it Box2b/hr}, which is the largest high resolution (hr) simulation, with a side-length of 640 $h^{-1}$cMpc. The simulation is resembled by $2 \cdot 2880^3$ particles with a mass of $6.9\times 10^8h^{-1}M_\odot$ for dark matter particles, $1.4\times 10^8h^{-1}M_\odot$ for gas particles, and $3.5\times 10^7h^{-1}M_\odot$ for star particles. The gravitational softening length is set to 3.75 kpc/h for dark matter and gas particles and to 2 kpc/h for star particles. We are using the output of {\it Box2b/hr} at $z=0.25$, which corresponds to the average redshift of the observational sample we want to compare with. The large volume of this simulation provides a total of 2027 halos in the galaxy cluster regime with $M_{500c} \geq 10^{14} M_\odot$ and 11656 halos in the galaxy group regime with $0.3\times10^{14} M_\odot \leq M_{500c} < 10^{14} M_\odot$ for which we can investigate global properties and internal structures.

In Fig. \ref{fig:ClusterExamples} we show X-ray surface brightness (SB) maps of 3 cool-core and 3 hot-core galaxy clusters selected from the 3 most massive bins to give an impression of the morphologies and internal structures of the simulated galaxy clusters. The cool-core / hot-core classification criteria is described in Sec. \ref{sub:CoolCoreIndicators} and the energy band is XMM-eFEDS as described in Sec. \ref{sec:ComparingObservationsSimulations}. While the hot-core clusters are clearly in an intermediate stage of a merger, the cool-core clusters are showing the effects of the AGN thermal feedback, which is visible as cavity-like features in the SB maps and, although of vastly different origin, seems to effectively resemble some morphological properties of the observed AGN jet-driven cavities in galaxy clusters.

\section{Comparing observational data with simulations} \label{sec:ComparingObservationsSimulations}

Throughout this work, we use detailed cooling functions to obtain luminosity and emission-weighted averages. We have generated the cooling functions using PyAtomDB v0.10.10 \citep{foster2020pyatomdb}, which is based on AtomDB v3.0.9 \citep{foster2018atomdb}, and the APEC model \citep{smith2001collisional}.

This can be obtained by calculating the spectrum, including all radiative processes (line emission, continuum free-free emission from bremsstrahlung, and pseudo-continuum from weak lines) using the PyAtomDB collisional ionization equilibrium (cie) module, and then integrating over the desired band to get the total power. This is done for each metallicity and temperature pair in a grid, which can then be used to interpolate the cooling function for each gas particle from the simulation. 

Once the cooling function has been interpolated to the temperature and metallicity of each gas particle from the simulation, the total luminosity $L_{\mathrm{tot}}$ of a simulated cluster within the region of interest (e.g., $r < R_{500c}$) is computed for all particles within that region through

\begin{equation}
L_{\mathrm{tot}} = \sum_{particles} L_{p} = \sum_{particles} \Lambda\left( T_p,Z_p \right) n_{\mathrm{elec},p} N_{\mathrm{ion},p} 
\label{eq:luminosityWeights}
\end{equation}

where $\Lambda\left( T_p,Z_p \right)$ is the interpolated cooling function using the metallicity $Z_p$ and temperature $T_p$ of each gas particle, $n_{\mathrm{elec},p}$ is the electron number density, and $N_{\mathrm{ion},p}$ is the total number of ions of the individual gas particles, which can be obtained from the atomic masses and total mass of each metal species assuming full ionization. Then the emission-weighted average temperature $T_{\mathrm{spec}}$ can be calculated via

\begin{equation}
T_{\mathrm{spec}} = \frac{1}{L_{\mathrm{tot}}}\sum_{particles} \left( T_p \cdot L_p \right)
\label{eq:spectroscopicTemperature}
\end{equation}

Notice that although \cite{mazzotta2004comparing} and \cite{rasia2004mismatch} reported that emission-weighted temperature tends to be biased towards higher temperatures; they also explained that this was due to the simplification of accounting only for bremsstrahlung in the cooling functions used by cosmological simulations at the time, which is the dominant process only above 3 keV. However, our cooling functions do incorporate line emission; therefore, they should not be biased towards higher temperatures.

We have considered four different bands for this work; for two of them, we don't need to consider rest-frame band corrections because they are used to compare the luminosities versus the energy injection from the central AGN in the simulation. These are bolometric in the [0.01–100] keV band and soft in the [0.1–2.4] keV band.

The other two bands are used to compare the simulated data versus observations from clusters with a redshift of z < 0.3; therefore, we need to consider rest-frame band corrections to be consistent with the observations, these are:

\begin{itemize}
    \item XMM-eFEDS: Using a rest-frame band of [0.55–10.62] keV, corresponding to an average redshift of $z = 0.2$ from the selected clusters of the \cite{lovisari2020x} and \cite{bahar2022erosita} samples originally in the observed bands [0.5-7.0] keV and [0.3-10.0] keV respectively.
    \item ACCEPT: Using a rest-frame band of [0.8–8.0] keV, corresponding to an average redshift of $z = 0.14$ from the selected clusters of the \cite{cavagnolo2009intracluster} sample originally in the observed band [0.7-7.0] keV.
\end{itemize}

To obtain relative abundances, we use \cite{asplund2009chemical} for the XMM-eFEDS cooling function when comparing with data from the \cite{lovisari2020x} and \cite{bahar2022erosita} samples, while we use relative abundances from \cite{anders1989abundances} when comparing with data from the Chandra ACCEPT sample, as used by \cite{cavagnolo2009intracluster}.

As the simulation incorporates star formation physics, we have to carefully consider resolution elements that are describing the multiphase star-forming and interstellar medium (ISM) gas. Therefore, and following \cite{borgani2004x}, we explicitly exclude particles with a temperature below $1\cdot 10^5$ K as well as gas particles that are star-forming (with a cold fraction above 10\%). Additionally, we also exclude particles above 50 keV following \cite{biffi2012observing}, since these represent the cavities filled with relativistic plasma generated by the AGN feedback model.

We also have to consider a number of limitations originating in the underlying, numerical resolution of the cosmological simulations. One is the gravitational smoothing length, which in the case of Magneticum {\it Box2b/hr} is $\epsilon=3.75 \text{kpc/h}$ for gas and dark matter particles and $\epsilon=2\text{kpc/h}$ for star particles. Note that this is the Plummer equivalent, and since the simulation uses a spline interpolation instead, the imposed resolution limit is 2.8 times greater \citep{springel2005cosmological}, which results in a gravitational resolution limit of $2.8 \epsilon=2.8 \cdot 3.75\text{kpc/h} = 14.9 \text{kpc}$.

Additionally, BHs at the center of galaxies excite feedback to their environment via the implemented sub-grid model. Here, analogous to the gas particles, a sphere containing a fixed number of neighboring gas particles defines the so-called 'black hole sphere of influence' (see discussion in Sect. \ref{sub:FeedbackReach}). The thermodynamic properties for particles within this scale can be impacted by the numerical implementation in the form of direct thermal feedback from the central BH.

To minimize the impact caused by these numerical limitations, we follow \cite{borgani2004x} and apply a minimum number of 100 particles in every radial bin. Note that this limit is equivalent, on average, to about two times the gravitational resolution limit. Together with the exclusion of particles representing the ISM, this ensures that the results for the ICM profiles are meaningful.

Finally, for all calculations based on the simulated data, we use as the center the position of the most gravitationally bounded particle equivalent to the deepest point of the gravitational potential located at center of the BCG. Although observational papers typically use the X-ray peak or the X-ray centroid, we argue that our choice is the most adequate to study cooling flows since colder, denser clumps are expected to precipitate in the deeper, central regions of the gravitational potential \citep{voit2015ti}.

\section{Cool / hot-core population statistics}

\begin{figure}[t]
    \centering
    \includegraphics[width=1\columnwidth]{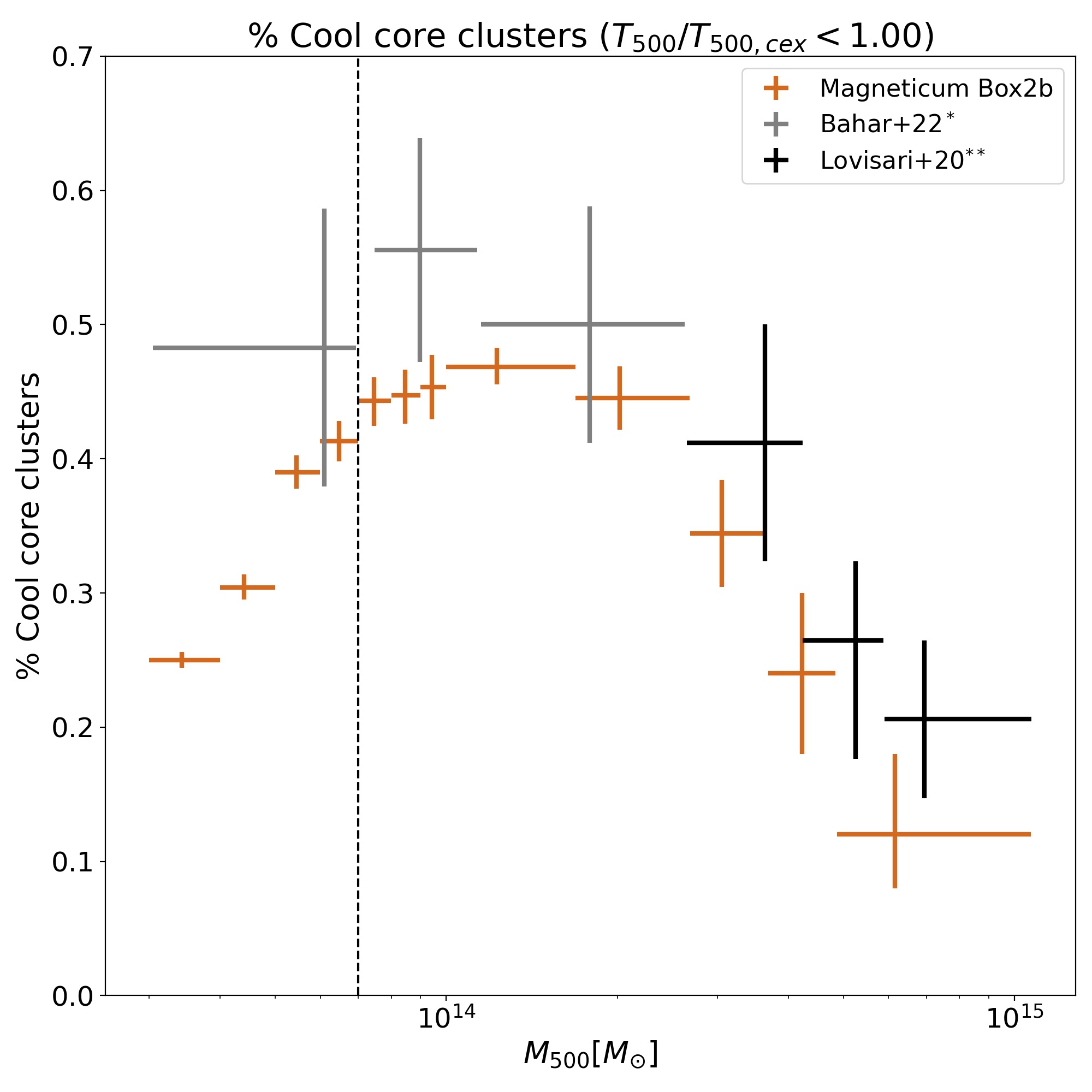}
    \caption{Cool-core fractions determined by the number of clusters for which the total temperature, including the core region, is lower than the core-excised temperature ($Tx_{500}/Tx_{500,cex} < 1$). The gold bars correspond to the simulation, for which the temperature was obtained with emissivity weights in the XMM-eFEDS band. The gray bars for the low-mid mass range correspond to the eROSITA field equatorial deep survey eFEDS data from \cite{bahar2022erosita} and \cite{chiu2022erosita}. ($^*$Also contains 1 mid mass cluster from the \cite{lovisari2020x} sample). The black bars for the high mass range correspond to the Planck SZ-selected sample observed with XMM-Newton from \cite{lovisari2020x}. ($^{**}$Also contains 10 high mass clusters from the \cite{bahar2022erosita} sample). The vertical dashed line corresponds to $M_{500c} = 0.7 \cdot 10^{13} M_\odot$, above which the eFEDS survey is expected to be complete for redshifts below z < 0.3 \citep{comparat2020full}.}
    \label{fig:CCFractions}
\end{figure}

\subsection{cool-core indicators} \label{sub:CoolCoreIndicators}

In order to classify cool/hot-core systems, many studies have resorted to properties in the most central regions, such as the central density, central cooling time, central entropy, and central cuspiness. However, an issue with these classification schemes is how well the innermost radius can be resolved in both observations and simulations.

In observations, the minimum possible scales that can be resolved depend on the angular resolution of the telescope and the redshift of the target object. In general, the average value of central properties, such as entropy or cooling time, decreases with the scale of the innermost radius that can be resolved, as reported by \cite{panagoulia2014volume}, \cite{hogan2017mass} and \cite{sanders2018hydrostatic}. Also, consistent comparisons with simulations can get difficult due to the intrinsic limitations of the simulations and their resolution, as described in Sect. \ref{sec:ComparingObservationsSimulations}.

These shortcomings can be prevented by using indicators that cover the whole core region, typically $r < 0.15 \cdot R_{500c}$, as we use throughout this work. For example, \cite{santos2008searching} and \cite{maughan2012self} proposed the concentration parameter or core flux ratio, defined as the ratio between the bolometric flux in the core region $0.15 \cdot R_{500c}$ and the total flux within $R_{500c}$. However, as \cite{santos2010evolution} explains, the value of the concentration parameter depends on the redshift, due to the K-correction, which can be different for the core emission if it is softer (colder) than the total emission.

\begin{table}[!t]
\begin{center}
\begin{tabular}{||c c c c c c||} 
 \hline
 $M_{500,min}$ & $M_{500,max}$ & $M_{500,med}$ & $N_{tot}$ & $N_{cc}$ & $f_{cc}$\\ 
 \hline
 \hline
 5.91  & 10.71 & 6.94 & 34 & 7 & $0.21\pm0.06$  \\ 
 \hline
 4.24 & 5.88 & 5.25 & 34 & 9 & $0.26\pm0.09$  \\
 \hline
 2.65  & 4.24 & 3.63 & 34 & 14 & $0.41\pm0.09$  \\
 \hline
 1.15 & 2.63 & 1.79 & 34 & 17 & $0.50\pm0.09$  \\
 \hline
 0.75 & 1.13 & 0.90 & 36 & 20 & $0.56\pm0.08$  \\ 
 \hline
 0.30 & 0.69 & 0.61 & 29 & 14 & $0.48\pm0.10$  \\  
 \hline
\end{tabular}
\caption{Cool-core fractions from observational data combining the samples from \cite{lovisari2020x} and \cite{bahar2022erosita} with redshift below z < 0.3. $M_{500,min}$, $M_{500,max}$, and $M_{500,med}$ refer to the minimum, maximum, and median $M_{500c}$ for the clusters included in each bin. $N_{tot}$ refers to the total number of clusters, $N_{cc}$ refers to the number of cool-core systems, and $f_{cc}$ refers to the fraction of cool-core systems, with errors estimated via bootstrapping.}
\label{tab:coolCoreFractionsObservationalData}
\end{center}
\vspace{-1cm}
\end{table}

Alternatively, \cite{rasia2015cool} proposed the ratio of pseudo-entropy between the inner region ($r < 0.05 \cdot R_{180}$) and the outer region ($0.05 \cdot R_{180} < r < 0.15 \cdot R_{180}$). However, the main observational papers, including the large samples used for this work, don't provide entropy measurements directly.

For these reasons, we are using a direct comparison between the total temperature of the cluster, including the core region ($r < R_{500c}$, i.e., $T_{\text{X,500}}$), and the temperature of the cluster, excluding the core region ($0.15 \cdot R_{500c} < r < R_{500c}$, i.e., $T_{\text{X,500cex}}$), and assess the presence of a cool-core cluster by the ratio of these two temperatures ($T_{\text{ratio,500}}$), as shown by Equation \ref{eq:temperatureRatio}:
	
\begin{equation}
    T_{\text{ratio,500}} = \frac{T_{\text{X,500}}}{T_{\text{X,500cex}}} \left\{ 
    \begin{array}{l}
    \geq 1 \rightarrow \text{hot-core} \\ 
    < 1 \rightarrow \text{cool-core}
    \end{array}\right.
    \label{eq:temperatureRatio}
\end{equation}

These two quantities, $T_{\text{X,500}}$ and $T_{\text{X,500cex}}$, are typically provided by observational papers and can also be calculated from simulation data, which facilitates the comparison. Notice that for observations these quantities are obtained by de-projection; therefore, for the simulation we use a sphere and spherical shell, respectively.

This indicator should be free of resolution issues since we consider the whole core region by including or excluding it in the temperature measurement. Also, it does not depend on the K-Correction and, therefore, should be constant over redshift. Moreover, since we don't use absolute values but a ratio, the effect of biases in both the observational measurements and emission-weighted average temperatures obtained from the simulation data should be minimized.

Furthermore, it does not require ad-hoc thresholds depending on the population and has a very direct and simple physical interpretation: If $T_{\text{ratio,500}} < 1$, then the core region is cooler than the average temperature of the cluster, and if $T_{\text{ratio,500}} \geq 1$, then the core region is hotter or has the same average temperature as the rest of the cluster. For this reason we set the temperature ratio threshold to unity for the main part of this study, although in App. \ref{sec:appendixA} we investigate the effect of changing it.

Finally, notice that although we propose to separate clusters into two broad categories of cool and hot-core clusters using the criteria indicated in Equation \ref{eq:temperatureRatio}, this does not imply in itself that we assume the distribution of cool and hot-core clusters to be bimodal. Actually, several observational studies have ruled out a bimodal distribution, such as \cite{santos2010evolution}, \cite{rossetti2017cool}, \cite{yuan2020dynamical} and \cite{ghirardini2022characterization}, as well as studies based on simulations, such as \cite{rasia2015cool} and \cite{barnes2018census}. The lack of a bimodal distribution of cool and hot-core clusters suggests a rather long timescale in the transition from cool to hot-core systems.

\subsection{Cool-core fractions from observational data} 

The next important aspect is to consider how representative an observational sample of galaxy clusters can be. X-ray selected samples are affected by the Malmquist bias because, at a fixed mass, cool-core clusters are brighter than hot-core clusters. In this context, \cite{andrade2017fraction} reported that, for the same mass, cool-core clusters are 1.6–1.8 times more luminous than hot-core clusters, which resulted in an over-representation of cool-core clusters by a factor of 2.1–2.7 in their X-ray selected sample.

\begin{table}[!t]
\begin{center}
\begin{tabular}{||c c c c c c||} 
 \hline
 $M_{500,min}$ & $M_{500,max}$ & $M_{500,med}$ & $N_{tot}$ & $N_{cc}$ & $f_{cc}$\\ 
 \hline
 \hline
 4.88  & 10.68 & 6.15 & 50 & 6 & $0.12\pm0.04$  \\ 
 \hline
 3.69 & 4.84  & 4.23 & 50 & 11 & $0.24\pm0.06$  \\
 \hline
 2.69  & 3.66 & 3.05 & 151 & 51 & $0.34\pm0.04$  \\
 \hline
 1.69 & 2.68 & 2.02 & 467 & 204 & $0.45\pm0.02$  \\
 \hline
 1.00 & 1.69 & 1.23 & 1309 & 609 & $0.47\pm0.01$  \\ 
 \hline
 0.90 & 1.0 & 0.94 & 419 & 189 & $0.45\pm0.02$  \\  
 \hline
 0.80  & 0.90 & 0.85 & 567 & 253 & $0.45\pm0.02$  \\ 
 \hline
 0.70 & 0.80 & 0.74 & 754 & 329 & $0.44\pm0.02$  \\
 \hline
 0.60  & 0.70 & 0.65 & 1071 & 438 & $0.41\pm0.02$  \\
 \hline
 0.50 & 0.60 & 0.54 & 1625 & 642 & $0.39\pm0.01$  \\
 \hline
 0.40 & 0.50 & 0.44 & 2562 & 791 & $0.30\pm0.01$  \\ 
 \hline
 0.30 & 0.40 & 0.34 & 4658 & 1189 & $0.25\pm0.01$  \\  
 \hline
\end{tabular}
\caption{Cool-core fractions from the last snapshot of Magneticum Box2b (z = 0.25). $M_{500,min}$, $M_{500,max}$, and $M_{500,med}$ refer to the minimum, maximum, and median. $M_{500c}$ for the clusters included in each bin. $N_{tot}$ refers to the total number of clusters, $N_{cc}$ refers to the number of cool-core systems, and $f_{cc}$ refers to the fraction of cool-core systems, with errors estimated via bootstrapping.}
\label{tab:coolCoreFractionsMagneticumData}
\end{center}
\vspace{-1cm}
\end{table}

\label{sub:CoolCoreObservationalData}

There are two ways to prevent this problem. On one hand, it is possible to resort to samples selected via the thermal Sunyaev–Zeldovich effect (SZ), which is proportional to the thermal gas pressure integrated along the line of sight and therefore approximates to a mass selection as explained by \cite{lovisari2020x}. Still, the SZ signal can be boosted by shock fronts in disturbed systems. However, Planck measures the SZ signal at scales larger than $R_{500c}$ and therefore should not be significantly affected by small scale physics such as shocks, as pointed out by \cite{rossetti2017cool}.

\begin{figure*}[!t]
    \centering
    \includegraphics[width=0.6\columnwidth]{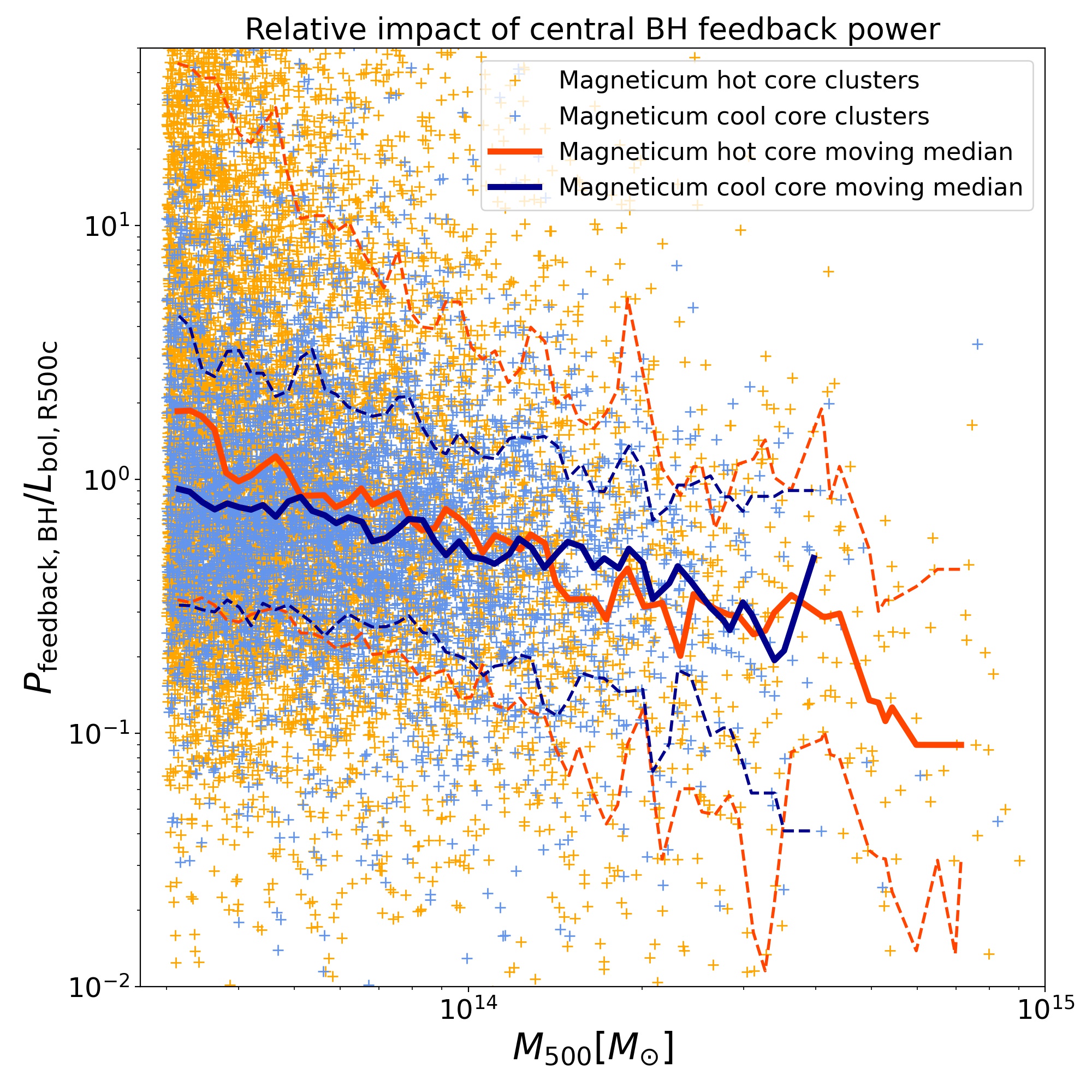}
    \includegraphics[width=0.6\columnwidth]{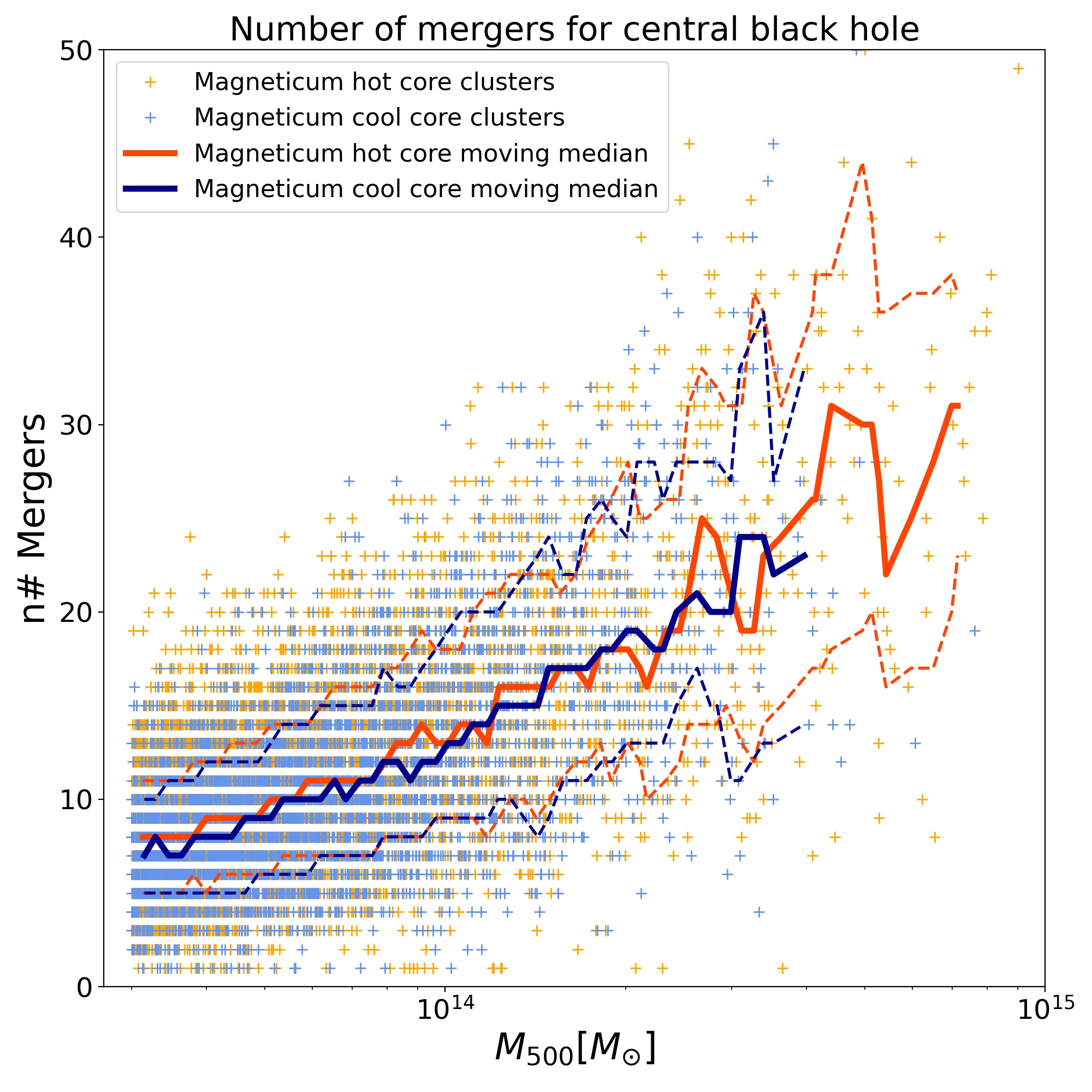}    
    \includegraphics[width=0.75\columnwidth]{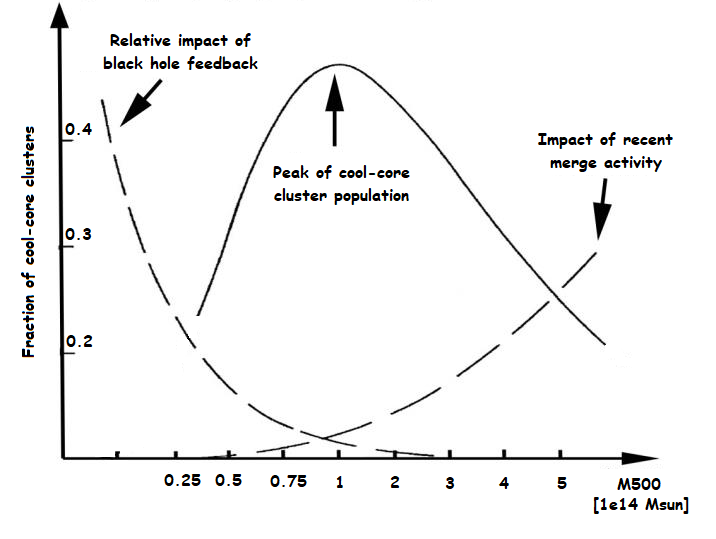}
    
    \caption{The left and central panels correspond to results from the Magneticum simulation ({\it Box2b}), where cool-core clusters are shown in blue and hot-core clusters in orange. Solid lines indicate moving medians, and dashed lines indicate 16\% and 84\% percentiles ($1\sigma$). \textbf{Left panel}: Ratio between the energy injection from the central AGN feedback and the bolometric luminosity in the [0.01–100] keV band for gas particles inside the core region. \textbf{Central panel}: Number of mergers undergone by the central black hole of the brightest cluster galaxy (BCG). \textbf{Right panel}: Sketch of the driving factors behind the characteristic shape of the cool-core fractions for different mass of the system.}
    \label{fig:CCFractionsConcept}
\end{figure*}

On the other hand, the alternative to SZ-selected samples is to still use X-ray selected samples, but reach enough exposure so that the flux limit can be lowered and all objects above a certain redshift and above a certain mass limit can be detected. For example, the full eROSITA survey is expected to be complete for masses greater than $1 \cdot 10^{14} M_\odot$ for redshifts below z < 1, and for masses greater than $0.7 \cdot 10^{13} M_\odot$ for redshifts below z < 0.3 \citep{comparat2020full}.

For these reasons, we use two observational samples to obtain cool/hot-core population statistics. On one hand, we have the sample from \cite{lovisari2020x}, consisting of 120 clusters from the early Planck survey (ESZ) observed with XMM-Newton. On the other hand, we can also use the eROSITA field equatorial deep survey (eFEDS), which comprises 542 groups and clusters and has a similar exposure as the full eROSITA All-Sky \citep{liu2022erosita}. In particular, we use the temperature and luminosity measurements from \cite{bahar2022erosita}, matched to the corresponding weak-lensing masses from \cite{chiu2022erosita}.

Therefore, we combine the samples from \cite{lovisari2020x} and \cite{bahar2022erosita}, both of them providing $T_{\text{X,tot}}$ and $T_{\text{X,cex}}$ measurements. We select clusters with redshift below $z < 0.3$, to match the last snapshot of Magneticum {\it Box2b/hr} (z = 0.25), and to be below the redshift limit for which eROSITA X-ray selected samples are expected to be complete for masses greater than $0.7 \cdot 10^{13} M_\odot$ \citep{comparat2020full}. The redshift cut produces 115 clusters from the \cite{bahar2022erosita} sample with average redshift $z=0.22$ and 93 clusters from the \cite{lovisari2020x} sample with average redshift $z=0.17$, which combined results in 172 clusters with average redshift $z=0.20$, quite close to $z=0.25$, which is the redshift of the Magnetium {Box2b/hr} snapshot used for this work.

Then, we divide the clusters in the combined sample with $M_{500c} > 0.7 \cdot 10^{13} M_\odot$ into 5 bins with approximately an equal number of clusters in each bin. Additionally, we produce a bin for the groups from eFEDS with $M_{500c} < 0.7 \cdot 10^{13} M_\odot$. For each bin, we calculate directly the number of cool-core systems using the definition given by Equation \ref{eq:temperatureRatio} and the corresponding fraction of cool-core systems. Additionally, we estimate the cool-core fraction uncertainty via bootstrapping as the standard deviation in the cool-core fraction from 1000 new random samples created from the original cluster sample for each bin, with replacement. The results are shown as black data points in Fig. \ref{fig:CCFractions} and the values are also reported in Table \ref{tab:coolCoreFractionsObservationalData}.

\subsection{Cool-core fractions from Magneticum Box2b} \label{sub:CoolCoreFractionsMagneticum}

Now that we have obtained an unbiased observational reference for the cool-core fractions, we would like to estimate the cool-core fractions from the last snapshot of Magneticum {\it Box2b/hr} (z = 0.25) in a similar way so that they are comparable to the observational data.

As explained in Sect. \ref{sec:ComparingObservationsSimulations}, we use emission-weighted averages for each cluster to obtain $T_{\text{X,tot}}$ in the $r < R_{500c}$ region, $T_{\text{X,cex}}$ in the $0.15 \cdot R_{500c} < r < R_{500c}$ region, and the corresponding $T_{\text{ratio}}$ as described by Equation \ref{eq:temperatureRatio}.

Then we bin the clusters, starting with the most massive ones. In the high mass range, we have 100 clusters between $M_{500c}=[10.68-3.69] \cdot 10^{14} M_\odot$, which we split into 2 bins of 50 clusters each. Below $M_{500c}=3.69 \cdot 10^{14} M_\odot$, we have many more clusters, which allows to construct bins with width $1 \cdot 10^{14} M_\odot$ down to $M_{500c}= 1 \cdot 10^{14} M_\odot$ in the middle mass range and with width $0.1 \cdot 10^{14} M_\odot$ down to $M_{500c}= 0.3 \cdot 10^{14} M_\odot$ in the low mass range.

As done for the observational data, we calculate directly the number and fraction of cool-core systems using the definition given by Equation \ref{eq:temperatureRatio} and estimate the cool-core fraction uncertainty via bootstrapping, as the standard deviation in the cool-core fraction from 1000 new random samples created from the original cluster sample for each bin, with replacement. The results are shown as orange data points in Fig. \ref{fig:CCFractions} and the values are reported in Table \ref{tab:coolCoreFractionsMagneticumData}.

\subsection{Comparison and interpretation of cool-core fractions} \label{sub:CoolCoreFractionComparison}

\begin{figure*}[h!t]
    \centering
    \includegraphics[width=2\columnwidth]{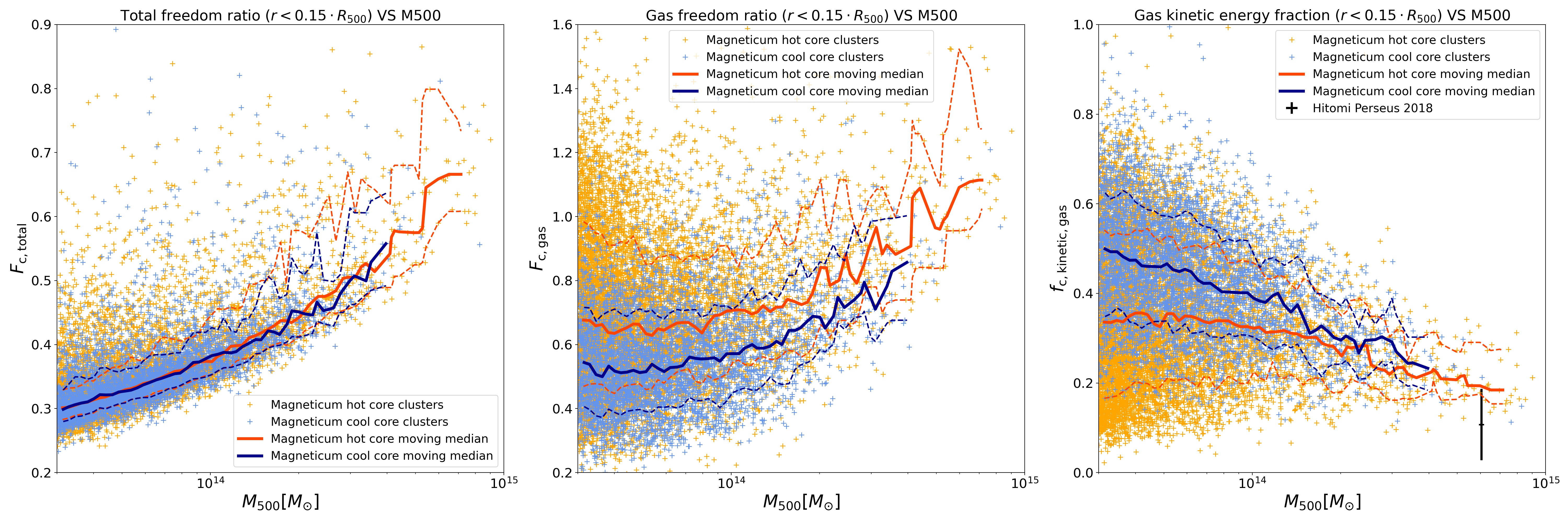}
    \caption{Cool-core clusters are shown in blue, and hot-core clusters in orange. Solid lines indicate moving medians, and dashed lines 16\% and 84\% percentiles ($2\sigma$) from the Magneticum simulation ({\it Box2b/hr}). \textbf{Left panel}: Freedom ratio for all particles inside the core region ($F_\text{c,total}$) \textbf{Central panel}: Freedom ratio for gas particles inside the core region ($F_\text{c,gas}$) \textbf{Right panel}: Kinetic energy fraction $f_\text{c,gas,kinetic}$ for gas particles inside the core region. The black bar indicates the first and only measurement of kinetic energy fraction in galaxy clusters, of 10\% measured for the Perseus cluster by Hitomi \citep{hitomi2018atmospheric}. }
    \label{fig:FreedomRatio}
\end{figure*}

In Fig. \ref{fig:CCFractions} we compare the fraction of cool-core systems obtained from the observational data by combining the samples of \cite{lovisari2020x} and \cite{bahar2022erosita} for clusters below $z < 0.3$ with the fraction of cool-core systems obtained from the last snapshot of Magneticum {\it Box2b/hr} at z = 0.25. Simulation and observations coincide within the error bars. Both show a characteristic curve for the fraction of cool-core systems, which peaks around a mass scale of $M_{500c}\approx 10^{14}M_\odot$ and decreasing towards lower mass groups as well as towards the very massive systems in both cases. The position of the peak in the simulation data, which lies in the range $M_{500c}=[1.00-1.69]. 10^{14}M_\odot$ is quite compatible (within the error bars) with the peak position found observational data, which is around $M_{500c}=[0.75-1.13] 10^{14}M_\odot$.

The transition from cool-core systems to hot-core systems is a very complex interplay between the cooling of the ICM, the additional energy source from AGN, heat transport, and mixing, together with the merger process \citep[see discussions in][]{rasia2015cool,barns2019ccaniso}. Therefore, it is not straightforward to interpret the shape of the obtained cool-core fractions. This is complicated by the large difference in timescales between the different processes, where the current BH accretion rate and the associated energy imprint reflect a timescale much shorter than the timescale of transition, while the overall timescale of a merger most likely reflects a much larger timescale. With this in mind, we propose an interpretation that reflects this shape through the interplay of two distinct factors, as illustrated in Fig. \ref{fig:CCFractionsConcept}.

On one hand, the relative impact of the central AGN feedback increases generally, when going from more massive, bigger clusters towards less massive, smaller groups. This can be seen in the left panel of Fig. \ref{fig:CCFractionsConcept}, which shows the ratio between the feedback power injected by the central AGN and the bolometric luminosity of the core region computed from the simulation. Therefore the probability of transitioning from cool-core to hot-core is increased at the scale of smaller galaxy groups due to the extra energy available in the system. 

The other factor is the impact of merger activity. From the simulation, we can use the number of mergers that the BH in the central galaxy of the cluster underwent, shown in the central panel of Fig. \ref{fig:CCFractionsConcept}, as a proxy for the number of mergers of the overall cluster. This shows a clear trend increasing towards the large mass end, thereby increasing the probability of transitioning from cool-core to hot-core at the scale of massive galaxy clusters.

Therefore, we propose that the combination of these two factors can produce a characteristic curve, peaking at around the middle mass range ($M_{500c} \approx 1\cdot10^{14}M_\odot$), where both factors minimize the probability of a transition from cool-core to hot-core systems, thus increasing the fraction of cool-core systems, as sketched in the right panel of Fig. \ref{fig:CCFractionsConcept}.

Although the simulated cool-core fractions shown in Fig. \ref{fig:CCFractions} coincide with the observations within the error bars, there are indications of a small, systematic bias of the cool-core fraction in the simulation to be smaller across the entire mass range. For the sample of massive systems from \cite{lovisari2020x}, the so-called hydrostatic mass bias could potentially affect the comparison between observational and simulated galaxy clusters. However, accounting for such bias, the observed data points would shift by 10\%-20\% towards larger masses \citep{biffi2016nature}, which would actually increase the difference between simulated and observed cool-core fractions for the massive systems. Moreover, this should not affect the eFEDS data at low mass, since these masses have been obtained via weak gravitational lensing \citep{chiu2022erosita}. 

At the scale of smaller groups, the simulated cool-core fractions decline sharply in comparison with the observations. Possible observational explanations for this discrepancy can be lack of data or undetected hot-core systems in the lower mass group regime \citep[see discussion in][]{marini2024groups}. On the other hand, the increased impact of AGN feedback in small mass groups, as seen in the simulation, is consistent with the excess of entropy in the core region ($r < 0.15 \cdot R_{500c}$) reported in the first eROSITA All-Sky survey \citep{bahar2024srgerosita} which accounted for selection effects.

These findings suggest that the details in the numerical implementations of the coupling of the AGN energy to the surrounding ICM could affect the predicted cool-core fractions from the simulation, especially for small groups, since at this mass range the injected energy can significantly exceed the cooling losses, as can be seen in the left panel of Fig. \ref{fig:CCFractionsConcept}. We look into further insights and possible ways to improve the numerical implementation of AGN feedback in cosmological simulations in Sec. \ref{sub:FeedbackEfficiencies}.

Finally, in App. \ref{sec:appendixA}, we study how the cool-core fractions depend on the choice of the threshold adopted for the temperature ratio, which we set to unity for the main part of the study. Changing the temperature ratio (lowering it) provides a gauge for the mid-to-strong cool-core population, which is more sensitive to the fine-tuning of the simulation and also is less robust in statistical terms due to the reduced population of strong cool cores but still provides a valuable insight to interpret the results overall.

\subsection{Dynamical analysis}

\begin{figure*}[ht!]
    \centering
    \includegraphics[width=1.33\columnwidth]{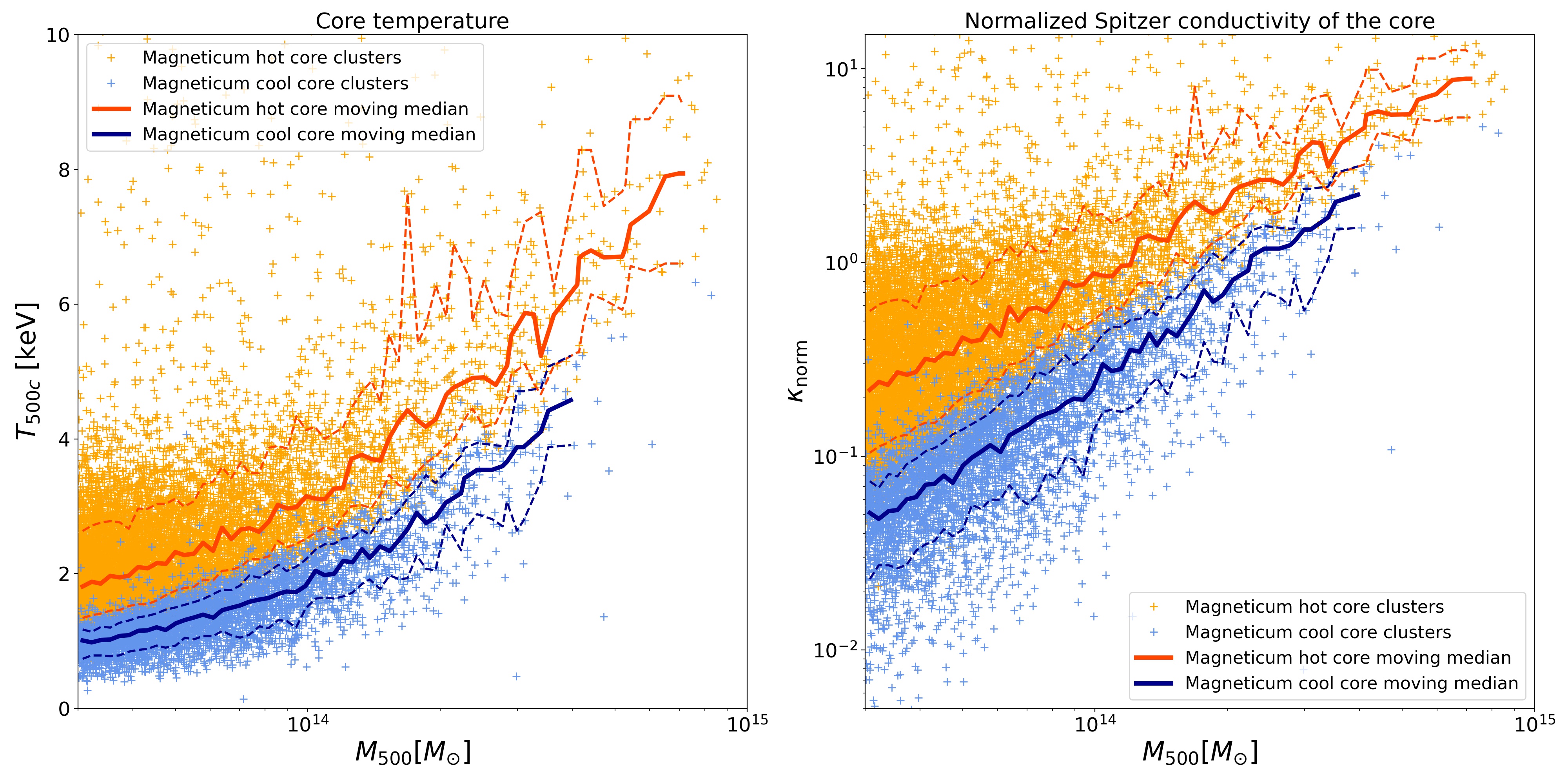}
    \caption{Cool-core clusters are shown in blue, and hot-core clusters in orange. Solid lines indicate moving medians, and dashed lines 16\% and 84\% percentiles ($2\sigma$) from the Magneticum simulation ({\it Box2b/hr}). \textbf{Left panel}: Temperature of the core region obtained with emissivity weights in the XMM-eFEDS band. \textbf{Right panel}: Effective Spitzer conductivity coefficient for the core temperature, normalized to the value for a system at 1 keV. }
    \label{fig:SpitzerConductivity}
\end{figure*}

Having a set of simulated groups and clusters which show good agreement with the observed cool-core fractions, we can now investigate the impact that merger activity has on cool-core and hot-core groups and clusters in the simulation and how that contributes to the observed mass trend. 

To gauge the impact of recent merge activity, we consider the relation between the kinetic ($K_\text{c,component}$), internal ($U_\text{c,component}$), and gravitational potential ($W_\text{c,component}$) energies, which we compute as a sum over the contributions of the individual particles. Here the sub-index c refers to the core region ($r < R_c = 0.15 \cdot R_{500c}$) and the sub-index component refers to either total, which means all particles (stellar, dark matter, gas, and BH) or only gas particles. We use the potential and internal energy provided directly by the snapshot output for each particle, but for the kinetic energy, we define the velocities w.r.t. the velocity of the center of mass for all particles inside $R_{200c}$, which is expected to be close to the viral radius of the system.

We define the freedom ratio as the ratio between the sum of kinetic and internal energy divided by the potential energy. We calculate two different freedom ratios, one including all particles $F_\text{c,total}=(K_\text{c,all} + U_\text{c,all})/\vert W_\text{c,all} \vert$ and one including only gas particles $F_\text{c,gas}$. Notice that this is not the classical ratio from the Virial theorem, which can only be obtained by either accounting for all particles in the system (not a specific region) or by accounting for boundary conditions around the region of interest (e.g., external potential, surface pressure), which are difficult to estimate (\cite{davis2011virialization}, \cite{klypin2016multidark}). On the other hand, the core freedom ratio can be easily obtained and has a clear interpretation in terms of the average probability of particles in the core region to move to the outer regions of the halo since, for collisionless, non-gas particles, the freedom ratio represents the square of the particle velocity divided by the escape velocity.

Additionally, we compute the kinetic energy fraction, defined as the ratio between the kinetic and the total (kinetic + internal) energy $f_\text{c,kinetic,gas}=K_\text{c,gas}/(K_\text{c,gas} + U_\text{c,gas})$.

We show the results in Fig. \ref{fig:FreedomRatio}. The left panel shows the core freedom ratio for all particles, which allows us to gauge the effect of merge (dynamical) activity independent from the AGN feedback. We see that the core freedom ratio increases towards the most massive clusters, indicating an increasing presence of high-energy particles injected from the merge activity. This confirms that more massive systems are also dynamically young systems \citep[see also discussion in][]{chen2007statistics}. However, there is no difference between cool-core and hot-core systems regarding the total composition inside the cores of groups and clusters.

This situation changes when only considering the gas particles within the core region shown in the central panel of Fig. \ref{fig:FreedomRatio}. While the gas particles show a very similar trend with the mass of the system, there is a clear offset of hot-core clusters having larger total energies than cool-core systems. Interestingly, especially in low-mass systems, the hot-core clusters show a large scatter towards high total energy. This is mainly due to a larger internal energy fraction and again points towards the AGN feedback driving the transition from cool-core to hot-core for lower mass systems.

On the other hand, this offset is of similar strength at all masses, from which we can conclude that the reduced cool-core fraction at higher masses does not directly come from increased merger activity. A clarification of the situation can be seen in the right panel of Fig. \ref{fig:FreedomRatio}, where we show the kinetic energy fraction decreases towards the higher mass end, thus indicating that the thermalization of kinetic energy injected by merge processes is more efficient towards higher mass clusters, thus establishing the first step to increase the core entropy. Then the implementation of thermal conduction in the Magneticum simulations, which has a strong dependency on temperature following the \cite{spitzer1962jr} description, provides a more efficient energy transport at larger masses due to the tight connection between cluster mass and temperature as explained in Sec. \ref{subsec:ThermalConductivity}. Therefore, the combination of these factors can effectively reduce the cool-core fractions towards massive clusters.

However, the prediction for kinetic energy fractions from the Magneticum simulations is difficult to verify since current X-ray telescopes cannot measure the kinetic energy support (bulk velocity and velocity dispersion), although future missions like Athenea are designed for this \citep{roncarelli2018measuring}. So far, we only have the Hitomi observations from Perseus ($M_{500c} \sim 6 \cdot 10^{14} M_\odot$), from which \citep{hitomi2018atmospheric} derived a ratio between the kinetic and thermal energy in a range of [11–13]\%, accounting for geometry corrections to the velocity dispersion as suggested by \cite{zuhone2012turbulence}. This is equivalent to a kinetic energy fraction of $f_\text{kinetic,gas} = 0.11 \pm 0.08$ using our definition of $f_\text{c,kinetic,gas}$ and as shown by the black bar in the right panel of Fig. \ref{fig:FreedomRatio}, which is slightly lower than the predictions of the simulations, although overlapping within the error bars. This is in line with previous findings that simulations over-predict the amount of kinetic energy in the ICM when compared to observations of relaxed galaxy clusters \citep{eckert2019turb}. Soon, XRISM will largely increase such measures for massive clusters; however, extending this to the group regime to investigate the mass trend found in the Magneticum simulations will only be possible with possible future instruments like Athena.

\subsection{Effect of thermal conductivity} \label{subsec:ThermalConductivity}

As mentioned in the previous section, the kinetic energy injected by the merge processes has to be first thermalized but also mixed through the core of the ICM. In this sense, the Magneticum simulations provide an implementation of isotropic physical thermal conductivity following the \cite{spitzer1962jr} description:

\begin{equation}
    \kappa=4.6 \cdot 10^{13}\left(\frac{T_e}{10^8 K}\right)^{5 / 2} \frac{40}{\ln \Lambda} \quad \frac{\text{erg}}{\text{s cm K}}
    \label{eq:spitzerConductivity}
\end{equation}

Where $T_e$ is the electron temperature and $\Lambda$ the Coulomb logarithm set to 37.8, which is an appropriate value for core regions of galaxy clusters (see \cite{arth2014anisoconduction}). Although the effective conductivity coefficient is reduced to 1/20th of the total value obtained from Eq. \ref{eq:spitzerConductivity}, to account for suppression of the conductivity due to magnetic fields, still the effective value of the conductivity coefficient can be significantly larger for massive clusters due to the strong temperature dependence ($T^{5/2}$) and the tight connection between cluster mass and temperature according to scaling relations ($T \sim M^{2/3}$, \cite{boehringer2011self}). 

In this sense, we show in Fig. \ref{fig:SpitzerConductivity} the emission-weighted average temperature of the cool-core and hot-core groups and clusters (left panel) and the corresponding value of the effective Spitzer conductivity coefficient for the core temperature, normalized to the value for a system at 1 keV ($\kappa_{\text{norm}} = \kappa [T_{500c}] / \kappa [\text{1keV}]$, right panel). We see that the median effective value of the conductivity can be almost one order of magnitude higher for the most massive clusters in comparison with smaller mass clusters and groups, thus increasing the mixing of the energy injected by mergers through the core regions and facilitating the conversion of cool-core clusters to hot-core clusters towards the higher mass end.

\section{Radial profiles} \label{sec:RadialProfiles}

Since the simulated cool-core fractions are consistent with the observed ones, we can now investigate the temperature, density, and entropy profiles. To ensure that profiles are resolved properly, we restrict here to clusters with $M_{500c} > 10^{14}M_\odot$ and divide the 2027 clusters in this sample into 3 broad mass ranges: low: $10^{14}M_\odot < M_{500c} \leq 2.69\times10^{14}M_\odot$, medium: $2.69\times 10^{14}M_\odot < M_{500c} \leq 4.88\times10^{14}M_\odot$ and high: $4.88\times 10^{14}M_\odot < M_{500c} \leq 9.02\times10^{14}M_\odot$, corresponding to the 3 most massive bins of the characteristic cool-core fraction curve presented in Sect. \ref{sub:CoolCoreFractionsMagneticum}. For computing the simulated profiles, we use a series of radial annuli in increments of $0.01 \cdot R_{500c}$, but for the innermost annulus, we impose a minimum physical radius of 15 kpc above the gravitational resolution limit and a minimum of 100 particles as described in Sect. \ref{sec:ComparingObservationsSimulations}.

We compare the simulated profiles with median profiles obtained from the Chandra ACCEPT sample \citep{cavagnolo2009intracluster}. For this, we first categorize the Chandra ACCEPT clusters into the previously mentioned mass bins using the corresponding cross-matched masses from the M2C Galaxy Cluster Database. This results in 35, 51, and 50 clusters in the low, median, and high mass bins, respectively, where the observed clusters in these 3 bins have a median redshift of $z = 0.08$, $z = 0.14$, and $z = 0.20$, respectively. The total resulting sample consists of 136 clusters with an average redshift of $z = 0.14$. To reduce the noise for the constructed median profiles, we draw 100 random instances of each profile, assuming a gaussian distribution for the errors, and construct a collection of random instances from all clusters included in each mass bin, from which we can calculate the overall median and 1-sigma percentiles for each radial bin. In this way, we account for the dispersion in the population but also take into account the measurement errors. Additionally we also consider a correction of +20\% in the masses retrieved from the M2C Galaxy Cluster Database to account for the hydrostatic mass bias \citep{biffi2016nature} and construct two sets of median profiles with the original and corrected observational masses.

\begin{figure*}[!ht]
    \centering
    \includegraphics[width=2\columnwidth]{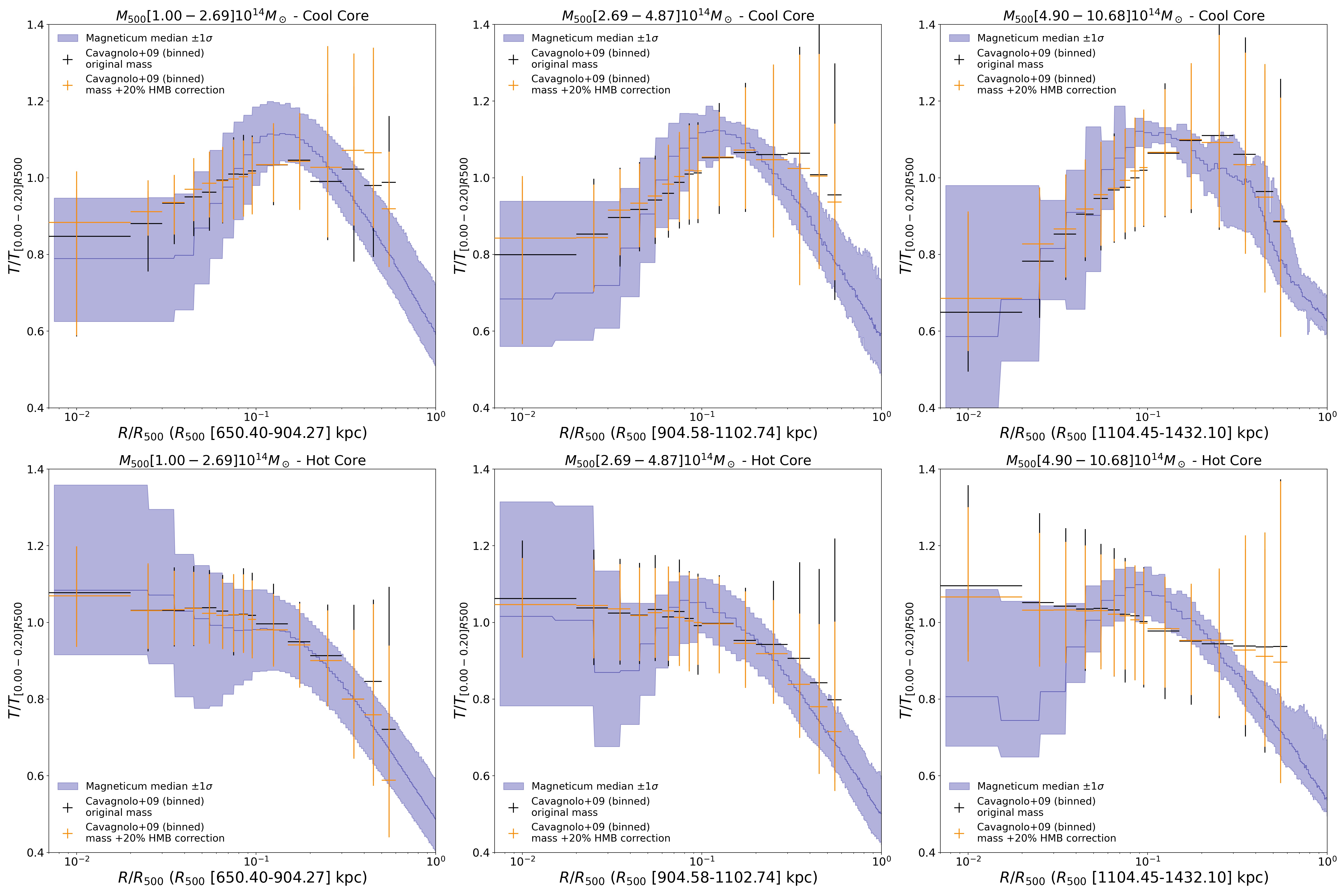}
    \caption{Projected X-ray temperature profiles in the [0.7-7.0]keV band, normalised by the mean X-ray temperature in the range $[0.1-0.2]R_{500c}$. The error bars correspond to the median Chandra ACCEPT sample profiles and $\pm 1\sigma$ intervals using the original masses from the M2C Galaxy Cluster Database (black) and a +20\% hydrostatic mass bias correction (orange). The blue line and shaded areas correspond to the Magneticum simulation (Box2b) median profiles and $\pm 1\sigma$ intervals. Columns are sorted by mass range, as indicated on top of each panel (increasing mass range from left to right). The upper row panels correspond to cool-core clusters, and the lower row panels to hot-core clusters.}
    \label{fig:TemperatureProfile}
\end{figure*}

The Chandra ACCEPT profiles don't typically extend up to $R_{500c}$ for the low-redshift clusters, as this would be outside of the field of view (FoV) of the Chandra Advanced CCD Imaging Spectrometer (ACIS). Therefore, we compute the temperature ratio in two radial ranges corresponding to $r<0.1 \cdot R_{500c}$ and $0.1 \cdot R_{500c}<r<0.2 \cdot R_{500c}$ to classify the clusters from the Chandra ACCEPT sample as cool-core or hot-core.

Provided that we compare the simulated profiles with the Chandra ACCEPT sample, we follow in general the same procedures and definitions for the individual quantities described by \cite{cavagnolo2009intracluster}. However, note that our centering procedure always chooses the position of the most gravitationally bounded particle, equivalent to the deepest point of the gravitational potential, whereas the centering procedure used for ACCEPT uses the X-ray peak as the default center but switches to the X-ray centroid if it is separated by more than 70 kpc from the peak \citep{cavagnolo2008bandpass}.

\begin{figure*}[!ht]
    \raggedleft
    \includegraphics[width=2\columnwidth]{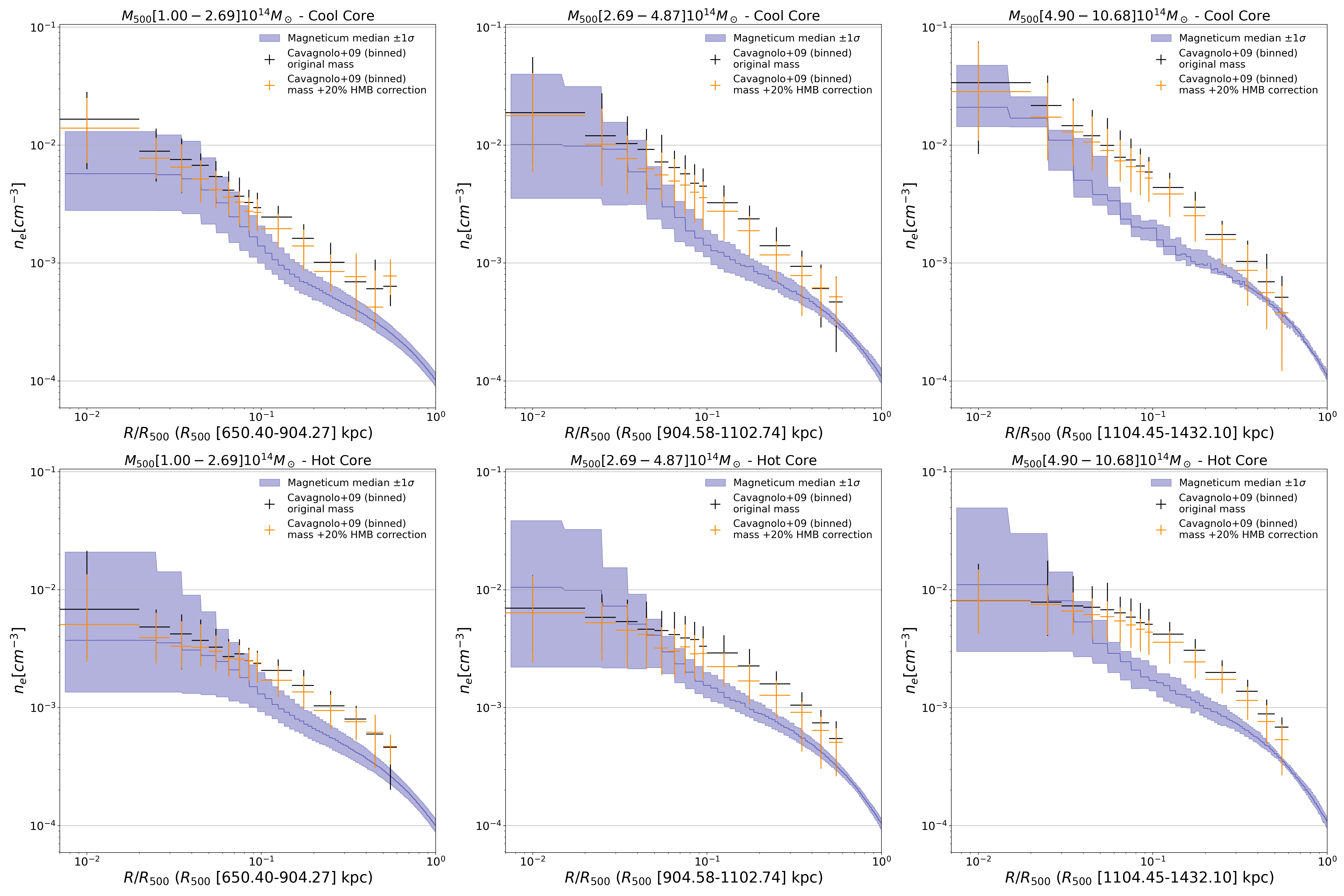}
    \caption{Electron number density profiles. The error bars correspond to the median Chandra ACCEPT sample profiles and $\pm 1\sigma$ intervals using the original masses from the M2C Galaxy Cluster Database (black) and a +20\% hydrostatic mass bias correction (orange). The blue line and shaded areas correspond to the Magneticum simulation (Box2b) median profiles and $\pm 1\sigma$ intervals. Columns are sorted by mass range, as indicated on top of each panel (increasing mass range from left to right). The upper row panels correspond to cool-core clusters, and the lower row panels to hot-core clusters.}
    \label{fig:RhoProfile}
\end{figure*}

Since hot-core clusters are typically disturbed, the ACCEPT centering procedure switches to the X-Ray centroid, whereas cool-core clusters are typically relaxed, and it defaults to the X-Ray peak. This procedure results in two very distinct types of profiles for cool-core and hot-core clusters and can be behind the bimodal distribution reported for ACCEPT. On the other hand, our centering procedure produces more similar profiles for cool-core and hot-core clusters but can better track the development of cooling flows, which are expected to precipitate in the deepest regions of the potential.

\subsection{Radial temperature profiles} \label{sub:RadialTemperatureProfiles}

For the temperature profile, we use emission-weighted average temperatures as before, where the emissivity weights are adapted to the energy range of the observations. Following the approach used in the observations, we compute the projected radial temperature profiles out to $R_{500c}$ in cylindrical shells as done in \cite{cavagnolo2009intracluster}, where we use a depth in the z direction corresponding to $R_{200c}$. 

The comparison of the constructed profiles from the observations with the results from the simulations is shown in Fig. \ref{fig:TemperatureProfile}. In general, there is a good agreement within the intrinsic error bars between them. Here the simulations reproduce the same characteristic shapes, which differ for cool-core and hot-core systems. Especially the simulated cool-core systems show the same characteristic temperature drop towards the center to roughly halve of the maximum temperature, while the hot-core systems show an almost isotherm core albeit at the low- and mid-mass range, they still have traces of a peak/turnover similar to the cool-core systems.

\begin{figure*}[!ht]
    \raggedleft
    \includegraphics[width=2\columnwidth]{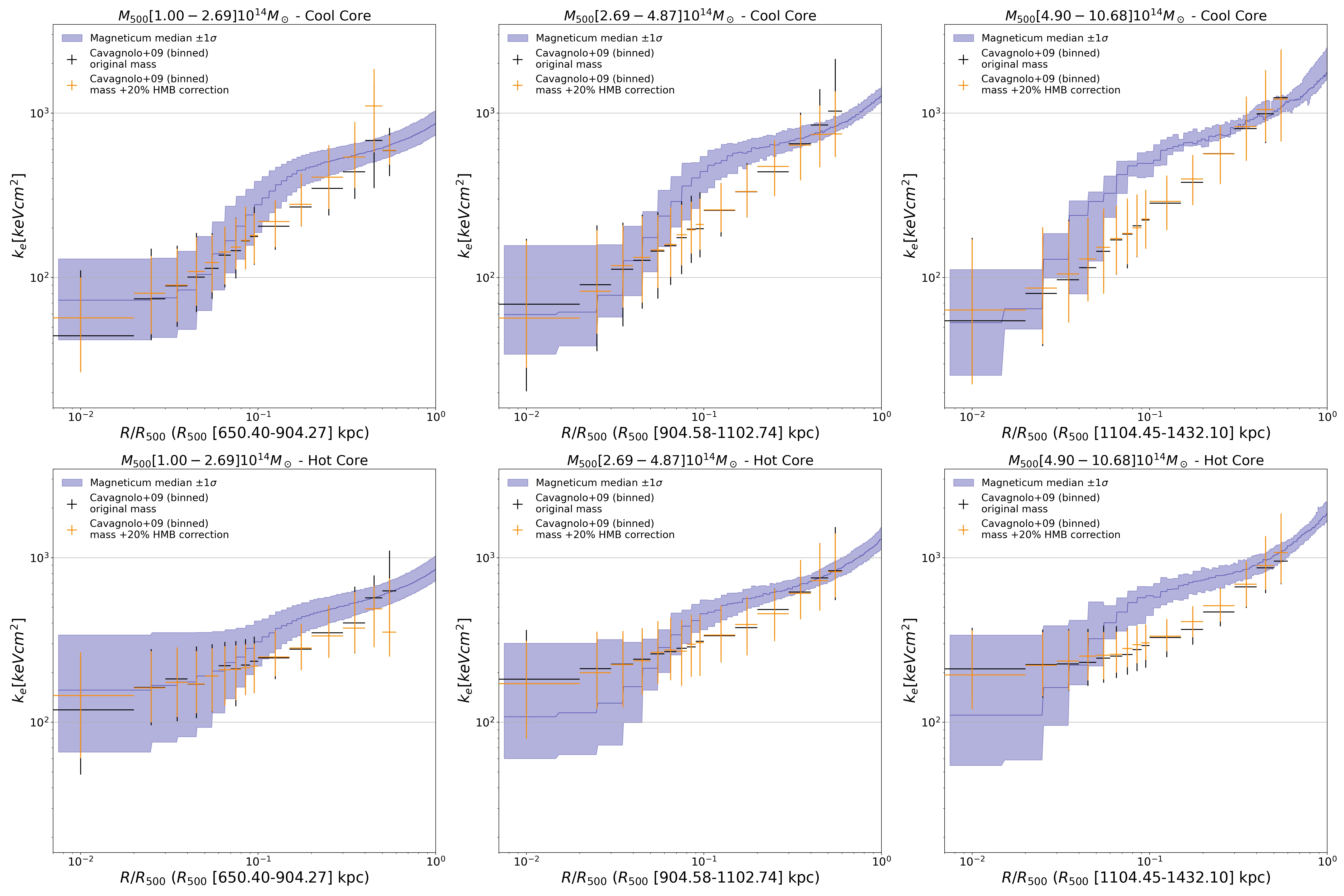}
    \caption{Electron entropy profiles. The error bars correspond to the median Chandra ACCEPT sample profiles and $\pm 1\sigma$ intervals using the original masses from the M2C Galaxy Cluster Database (black) and a +20\% hydrostatic mass bias correction (orange). The blue line and shaded areas correspond to the Magneticum simulation (Box2b) median profiles and $\pm 1\sigma$ intervals. Columns are sorted by mass range, as indicated on top of each panel (increasing mass range from left to right). The upper row panels correspond to cool-core clusters, and the lower row panels to hot-core clusters.}
    \label{fig:EntropyProfile}
\end{figure*}

The hot-core temperature profiles from the ACCEPT sample show a stronger isothermal core; however, this roots in the previously mentioned centering procedure used for ACCEPT sample that selects the X-ray centroid as the center if it is separated by more than 70 kpc from the peak, which is typically the case for hot-core clusters. The X-ray centroid is usually located in between the dominant galaxies and corresponds to a rather shallow part of the gravitational potential, resulting in no temperature gradients (isothermal) structure. On the other hand, our simulated clusters are centered at the deepest point of the gravitational potential, which enhances temperature gradients and the accumulation of denser cold clumps. In App. \ref{sec:appendixB} we show how the temperature profiles would look like when following the ACCEPT centering procedure.

Additionally, the hot-core clusters of the low and mid mass range appear to be slightly overheated in the innermost regions, which exhibit a 'hot bubble' as a remnant of the AGN feedback implementation in the simulations. This hot bubble is preceded by a peak/turnover similar to the cool-core systems and indicates that the AGN feedback energy is not transported effectively away from the immediate surroundings of the AGN where the feedback is initially injected. In the case of Magneticum, this problem is partially addressed by the physical conductivity model; however, it strongly depends on the overall temperature of the system, which decreases towards low-mass systems as explained in Sec. \ref{subsec:ThermalConductivity}, and moreover the current model applies a strong suppression down to 1/20 of the Spitzer value assuming turbulent magnetic fields \citep[see][]{arth2014anisoconduction} which we plan to review in future works. Also, the implemented AGN feedback could be too strong, especially in low mass systems, as already reported by \cite{vogelsberger2013model}, who argued that the residual black hole accretion, when there is no star-forming gas in the vicinity of the black hole, can create artificial hot bubbles around the black hole as an artifact of the imperfections of the sub-grid black hole accretion and feedback model.

Finally, we clearly see that for the cool-core clusters, the median temperature profiles from the ACCEPT sample peak at a larger radii than the simulated clusters. This is due to the significant presence in the ACCEPT sample of clusters with increasing temperatures outwards beyond the core region, such as Abell 1835, Abell 2142, and PKS 0745-191. 

\subsection{Radial density profiles} \label{sub:RadialDensityProfiles}

While the differences in the cool-core and hot-core temperature profiles are not fully independent from the cool-core criteria used, it is worth while to check also the gas density profile and their systematic differences between the two cluster populations. This is especially important, as the density is the fundamental thermodynamic property of the ICM, which is directly linked with the development of cooling flows, star formation, and black hole accretion. 

As the density profiles from the Chandra ACCEPT sample are de-projected profiles \citep{cavagnolo2009intracluster} we are using here directly spherical, radial density profiles from the clusters selected in the Magneticum {\it Box2b/hr} simulation but keep the cool-core classification as before. In addition, to minimize the impact of sub-structures, which would be typically masked in the observations, we build the mean density in each radial bin by summing the mass of all gas particles, dividing by the total volume of the spherical radial bin shell instead of averaging the individual densities and assuming full ionization of the gas to obtain the electron number density.

The results of this comparison are shown in Fig. \ref{fig:RhoProfile}. Interestingly, the simulated clusters which are classified as cool-core systems show systematically higher central densities compared to the hot-core clusters, well in agreement with the observations.

Generally, the simulated profiles of both cool and hot-core clusters show increasingly lower densities towards the high mass range in comparison with the median profiles from the ACCEPT sample. This discrepancy is reduced when considering the observational masses corrected by the hydrostatic mass bias but not entirely removed. A possible explanation for this behavior can be associated with excessive star formation towards more massive systems, which has not been effectively quenched by the central AGN and has therefore led to lower gas fractions in the baryonic component. We elaborate more on this scenario in Sect. \ref{sec:GasStellarMassFractions}.

\subsection{Radial entropy profiles} \label{sub:RadialEntropyProfiles}

Finally, we can construct gas entropy profiles in the same way as for the Chandra ACCEPT sample by combining the projected temperature profiles with the spherical density profiles \citep[see][]{cavagnolo2009intracluster}. The results are shown in Fig. \ref{fig:EntropyProfile}, where we see that the simulated profiles for the cool-core systems typically decline towards the center with a central entropy of less than 100kev/cm$^2$, while the hot-core systems typically have a flat entropy profile with a central entropy of larger than 100kev/cm$^2$. This compares very well with the observational trend. 

Driven by the trends in the density profiles, the simulated cool and hot-core clusters show increasingly higher entropy profiles, especially for the high mass bins. As for the densities, assuming a 20\% hydrostatic mass bias in the observations brings the simulated and observational profiles closer, although not entirely overlapping. In particular, the simulations consistently produce a flatter profile at the intermediate radii before steepening in the core, diverging from the observed power-law-like shape.

\section{Gas and stellar mass fractions} \label{sec:GasStellarMassFractions}

It is well established that the AGN feedback not only suppresses the star formation in very massive systems, it also impacts the gas mass fraction in clusters \citep[see for example][]{planelles2013baryons}. Here, the overall gas fraction within $R_{500c}$ of the clusters and groups as predicted by the Magneticum simulations agrees with observational derived fractions from x-ray observations, both in absolute numbers and also in the trend of having lower gas mass fractions in lower mass systems \citep{angelinelli2022bfrac,angelinelli2023bfrac}.

In contrast, stellar masses typically are still significantly larger than observational inferred stellar masses, especially at the high mass end, pointing towards an inefficient suppression of star formation by the AGN feedback implementation in simulations for very massive systems. A similar result was reported by \cite{fabjan2010simulating}, who compared the stellar mass fractions at the scale of $R_{500c}$ and found them 2-3 times higher than the observational measurements from \cite{gonzalez2007census}, which also included the contribution from the intra-cluster light (ICL).

To investigate this point in more detail, we evaluate the gas mass fraction, the stellar mass fraction, and the total baryonic mass fraction within $R_{2500c}$\footnote{Note that typically $R_{2500c} \sim 0.4 \cdot R_{500c}$. See \cite{ragagnin2021mass} for the conversions between different over-densities.} to compare to \cite{lagana2013comprehensive}, who presented the gas and stellar mass fractions at $R_{2500c}$ for a sample of 27 clusters with average redshift $z = 0.22$, quite close to $z = 0.25$ of our sample.

\cite{lagana2013comprehensive} obtained the gas mass ($M_{\text{2500,gas}}$) by modeling the X-ray emission with a modified $\beta$-model profile \citep{maughan2012self}, which is then integrated up to $R_{2500c}$. The total mass ($M_{\text{2500,total}}$) is calculated assuming hydrostatic equilibrium and isothermality. The stellar mass ($M_{\text{2500,stellar}}$) is calculated by converting apparent magnitudes to absolute magnitudes using K-corrections depending on the morphological type and using two different mass-to-light ratios for early-type and late-type galaxies. The corresponding gas, stellar, and baryon mass fractions ($f_{\text{2500,gas}}$, $f_{\text{2500,stellar}}$ and $f_{\text{2500,bar}}$ respectively) are obtained as the ratios between the mass of each component and the total mass, and we calculate the errors by standard error propagation of the $1\sigma$ uncertainty in quadrature. 

\begin{figure*}[!ht]
    \centering
    \includegraphics[width=2\columnwidth]{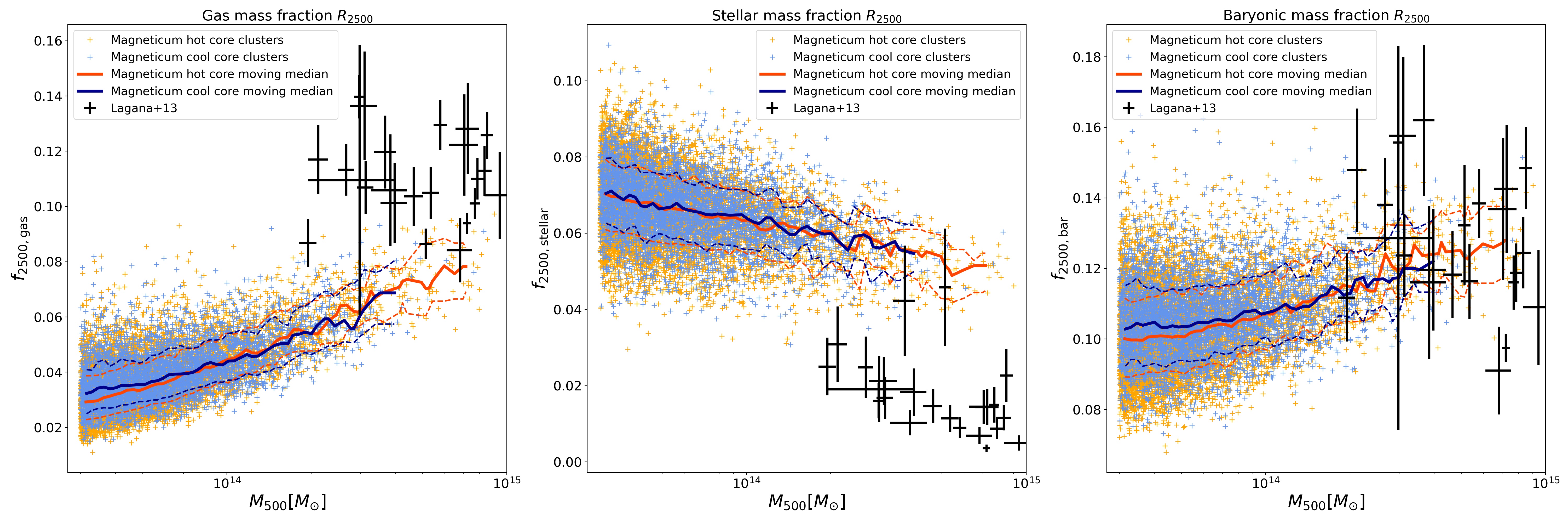}
    \caption{Cool-core clusters are shown in blue, and hot-core clusters are shown in orange. Solid lines indicate moving medians, and dashed lines represent 16\% and 84\% percentiles ($2\sigma$) from the Magneticum simulation (Box2b). Observational data is shown in black, with error bars at the $1 \sigma$ level, based on XMM-Newton, Chandra, and SDSS from \cite{lagana2013comprehensive}. \textbf{Left panel}: Gas mass fraction inside $R_{2500c}$. \textbf{Central panel}: Stellar mass fraction inside $R_{2500c}$, \textbf{Right panel}: Baryonic mass fraction inside $R_{2500c}$}
    \label{fig:MassFractionsR2500}
\end{figure*}

From the simulation, we can directly obtain the gas, stellar, and total mass of the components contained in the sphere of $R_{2500c}$. The results are shown in Fig. \ref{fig:MassFractionsR2500}, where we see that the simulated gas fractions are systematically lower than the observed gas fractions, as we would have expected from the comparison of the radial density profiles. Interestingly, similar to the findings at $R_{500c}$ there is a clear mass trend predicted by the simulations, which however is not reflected in the observations. On the other hand, the simulated stellar mass fractions within $R_{2500c}$ are much higher than the observed one. Being 5 times larger than the observational stellar mass fraction at the very massive end, this strongly deviates from the observations more than the overall stellar mass fraction within $R_{500c}$. Surprisingly, the total baryonic mass fraction seems to not be biased in comparison with observations, although the observational results present a much wider scatter than the simulated data. Also worth noting is that the simulation does not show any significant difference between cool-core and hot-core systems.

While there are still uncertainties in the observations \citep[see detailed discussion in][]{lagana2013comprehensive}, both in the gas mass based on the assumption of hydrostatic equilibrium and isothermality, as well as the stellar mass, where undetected ICL could contribute an additional 10\% to 40\% to the total stellar mass, this would not solve the observed differences between the simulation and observations.

The fact that the total baryonic mass fraction within $R_{2500c}$ aligns with the observations but there are too many stars formed within the central regions of clusters points towards a deficit in the detailed coupling of the AGN feedback within the simulations rather than a vastly different energy injection by the AGN, which would lift more gas to larger distances. This is also in agreement with the fact that the AGN luminosity function, which reflects the general energy available for the AGN feedback, in the Magneticum simulations agrees well with observations \citep{hirschmann2014cosmological,biffi2018agn}. Therefore, this points towards the coupling of the feedback from the central AGN within the simulations to the surrounding medium, which is not able to fully suppress star formation at the scale of clusters, in line with the previous findings reported by \cite{fabjan2010simulating}.

\section{Implications for the AGN feedback model} \label{sec:AGNFeedbackModel}

The results presented so far are consistent with earlier findings \citep{fabjan2010simulating}, showing that the AGN feedback model implemented in the simulations is significantly suppressing star formation at the scale of groups and clusters. However, it only partially prevents catastrophic cooling and star formation at the scale of massive galaxy clusters, despite increased AGN feedback efficiencies in radio mode. To investigate the possible origin of this problem, we can compare the effective implementation within the simulations to the observed signatures of AGN feedback in galaxy groups and clusters.

\begin{figure*}[!ht]
    \centering
    \includegraphics[width=2\columnwidth]{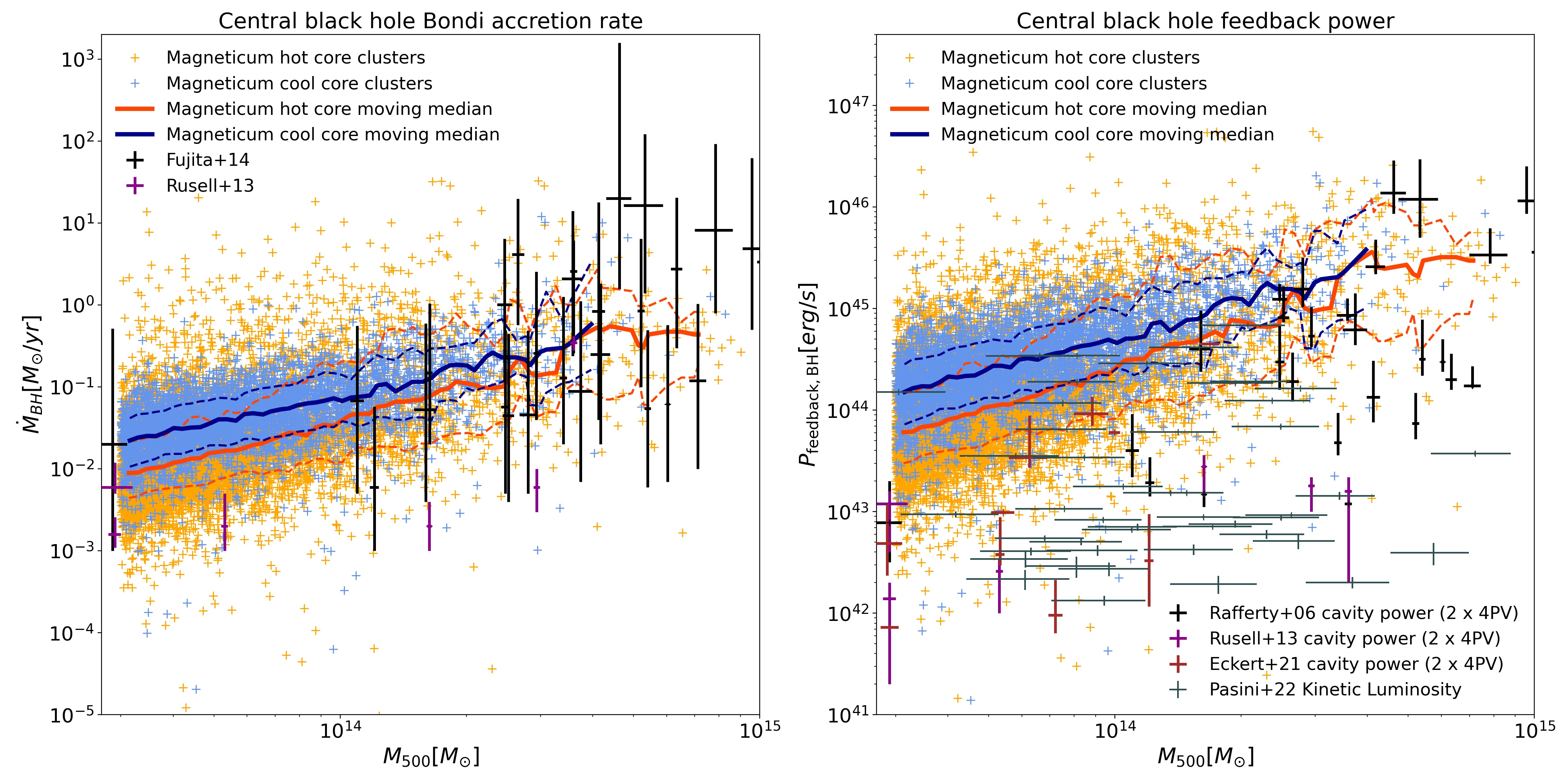}
    \caption{Accretion rates and energy injection of the AGN hosted in the centers on BCGs. Cool-core clusters are shown in blue, and hot-core clusters are shown in orange. Solid lines indicate moving medians, and dashed lines represent 16\% and 84\% percentiles ($1\sigma$) from the Magneticum simulation ({\it Box2b/hr}). Observational data is shown with error bars at the $1 \sigma$ level. \textbf{Left panel}: Comparison of the accretion rates from the AGN hosted in the centers on BCGs, with the estimations from \cite{fujita2014agn} and \cite{russell2013radiative}. \textbf{Right panel}: Comparison of black hole energy injection with the observational estimations from \cite{rafferty2006feedback}, \cite{russell2013radiative}, \cite{eckert2021feedback} and \cite{pasini2022erosita}.}
    \label{fig:AGNFeedback}
\end{figure*}

\subsection{AGN accretion and energy output} \label{sub:BondiAccretion}

There are currently two models for how cool-core systems are formed. One is described as precipitation, where the deposition of cold gas onto the core is driven by thermal instability \citep[TI,][]{mccourt2012ti,voit2015ti}. In the other one, the raining of cold gas onto the core is driven by chaotic cold accretion \citep[CCA,][]{gaspari2018cca}. The onset of these mechanisms is driven by the ratio of cooling time to free-fall time below a typical threshold of $t_\mathrm{cool}/t_\mathrm{ff} \lesssim 10$ or the ratio of cooling time over the eddy turnover time of $t_\mathrm{cool}/t_\mathrm{eddy} \lesssim 1$, respectively. In practice, these two criteria are almost identical, and observations indicate that they correspond to a central entropy threshold of $\approx 35$keV/cm$^2$ across a wide range of redshifts \citep{mcdonald2013ccevol}. This is in line with the entropy profiles of simulated cool-core clusters to be decreasing towards the center, while the entropy profiles of hot-core clusters show more flattening towards the center at values above 100 keV/cm$^2$. In cool-core clusters, we then expect that a feedback loop is established that prevents low-entropy cooling flows from developing further \citep{churazov2005supermassive}.

As most of the AGN treatment in cosmological simulation, the AGN accretion model used by Magneticum \citep{hirschmann2014cosmological} is based on the Bondi-Hoyle-Lyttleton model \citep{hoyle1939effect,bondi1944mechanism,bondi1952spherically}, with an implementation that follows \cite{di2005energy}, \cite{springel2005modelling} and \cite{fabjan2010simulating} and has the form

\begin{equation}
    \dot{M}_{\text{BH}}=\alpha \frac{4 \pi  G^2 M_{\text{BH}}^2 \rho_g}{\left(c_{\text{s}}^2+v^2\right)^{3 / 2}}.    
    \label{eq:Bondi–Hoyle–Lyttleton}
\end{equation}

Here, G is the gravitational constant, $M_{\text{BH}}$ is the mass of the BH, $\rho_g$ is the gas density, $c_s$ is the speed of sound, $v$ is the black hole velocity with respect to the surrounding gas, and $\alpha$ is a dimensionless parameter (boost factor) used by the simulation, typically set to 100 to account for the unresolved increase in density towards the central regions surrounding the black hole. Since $c_s \sim T^{1/2}$ and the entropy is defined as $S \sim T \rho^{-2/3}$, the $\dot{M}_\text{B}$ predicted by the Bondi-Hoyle-Lyttleton accretion formula inversely depends on the entropy to the power ${3/2}$, implying that aside from the gas velocity, AGN accretion is larger in the simulation when the entropy of the surrounding medium is low.

How well this implementation describes the general accretion onto the central BH in groups and clusters is shown in the left panel of Fig. \ref{fig:AGNFeedback}, where we compare observational inferred $\dot{M}_{\text{BH}}$ as function of halo mass with the results from the simulations. To do so, we resort to the measurements from \cite{fujita2014agn}, who obtained the Bondi accretion rate of a sample of BCGs in the near universe ($z < 0.35$) by modeling the temperature and gravitational contribution from the dark matter halo, galaxy, and central black hole, to then integrate the hydrostatic equation in order to derive the gas density down to the Bondi radius. The obtained accretion of the AGNs hosted in the centers on BCGs agrees very well with the ones predicted by the simulations, indicating that the use of the Bondi-Hoyle-Lyttleton model in the simulations is an appropriate approximation. Also clear to see is that in the simulation the accretion onto the central BHs is systematically larger than in hot-core systems.

In the simulation, the accretion rate of the BH is converted into the energy output (feedback power) $P_{\text{feedback,BH}} = \dot{M}_{\text{BH}} c^2 \epsilon_r \epsilon_f$ by assuming a radiative efficiency ($\epsilon_r$), which determines what fraction of the accreted mass is converted into energy, and a feedback efficiency ($\epsilon_f$), which determines how much energy released by the black hole accretion is deposited in the surrounding medium. In our case, they are chosen to be $0.2$ and $0.15$ in the quasar mode, while the later is 4 times large in the radio mode \citep{fabjan2010simulating}. The radio mode is assumed whenever the actual accretion rate is below one hundredth of the Eddington accretion rate. Therefore, in radio mode the total efficiency reaches $0.12$, close to the 10\% as inferred in galaxy clusters by \cite{churazov2005supermassive} to compensate the cooling losses of the ICM.

In the right panel of Fig. \ref{fig:AGNFeedback}, we compare the power in observed cavities obtained by \cite{rafferty2006feedback}, \cite{russell2013radiative}, and \cite{eckert2021feedback} within galaxy clusters with the feedback energy released from the central AGN in the simulation. The observed cavity power ($P_{\text{cavity}}$) is derived by dividing the cavity enthalpy

\begin{equation}
    E_{\text{cav}}=\frac{\gamma_{c}}{\gamma_{c}-1} p_{s} V_{c}
    \label{eq:cavity-enthalpy}
\end{equation}

by the buoyancy time scale. Here $\gamma_{c}$ is the adiabatic index of the material filling the cavity, $V_{c}$ is the cavity volume, and $p_{s}$ is the pressure at the cavity surface. As these cavities are filled with relativistic material, $\gamma_{c}=4/3$ and $E_{\text{cav}}=4V_c p_s$. Additionally, in this work, we apply a x2 factor to account for the shock energy as estimated by \cite{rafferty2006feedback} and confirmed by the simulations of \cite{guo2010simulating}. We also add to the comparison the kinetic luminosity sample inferred by \cite{pasini2022erosita} based on LOFAR 144MHz radio power observations of the BCG in the eFEDS sample. We can see that the energy injected in the simulation agrees well with the observations for massive clusters; however, at the scale of galaxy groups, the injected energy is significantly larger than the observed values. The simulation also show a clear trend that cool-core systems inject a larger amount of energy than hot-core systems.

A possible explanation for the divergency at the scale of galaxy groups could be that the previously mentioned x2 factor to account for the shock energy is underestimated at the scale of groups; however, the fact that this divergency is also visible for the kinetic luminosity sample from \cite{pasini2022erosita} indicates that the observed cavity powers cannot be significantly underestimated at the scales of galaxy groups since we would expect at least higher radio emission even if the cavities are not detectable.

In addition, we show in the left panel of Fig. \ref{fig:AGNvsLICM} the soft band [0.1–2.4] keV luminosity of the ICM ($L_{500,\text{soft}}$) from the observational samples of \cite{lovisari2020x} and \cite{bahar2022erosita}\footnote{We use a factor of 1.64 to convert the luminosity from the observed [0.5–2.0] keV band to the [0.1–2.4] keV band}, as well as the cavity powers previously mentioned. We see that the observational cavity powers fluctuate around the ICM luminosity, but are not systematically above. On the other hand, the ICM luminosity of the simulation matches the observations, but the energy input by the AGN feedback of the simulation is systematically above the ICM luminosity, matches the cavity powers only at the scales of very massive clusters, and is much larger at low mass clusters and group scales. This bias is consistent with the bias towards lower cool-core fractions in simulations at the scale of galaxy groups, as shown in Fig. \ref{fig:CCFractions}.

Since in the simulation as well as in observations, almost all AGNs in the centers of groups and clusters are in radio mode in the modern universe, this indicates that the AGN feedback efficiency in groups must be significantly smaller than in clusters at the radio mode regime.

\subsection{A new model for AGN feedback efficiency in radio mode} \label{sub:FeedbackEfficiencies}

\begin{figure*}[!ht]
    \centering
    \includegraphics[width=1\textwidth]{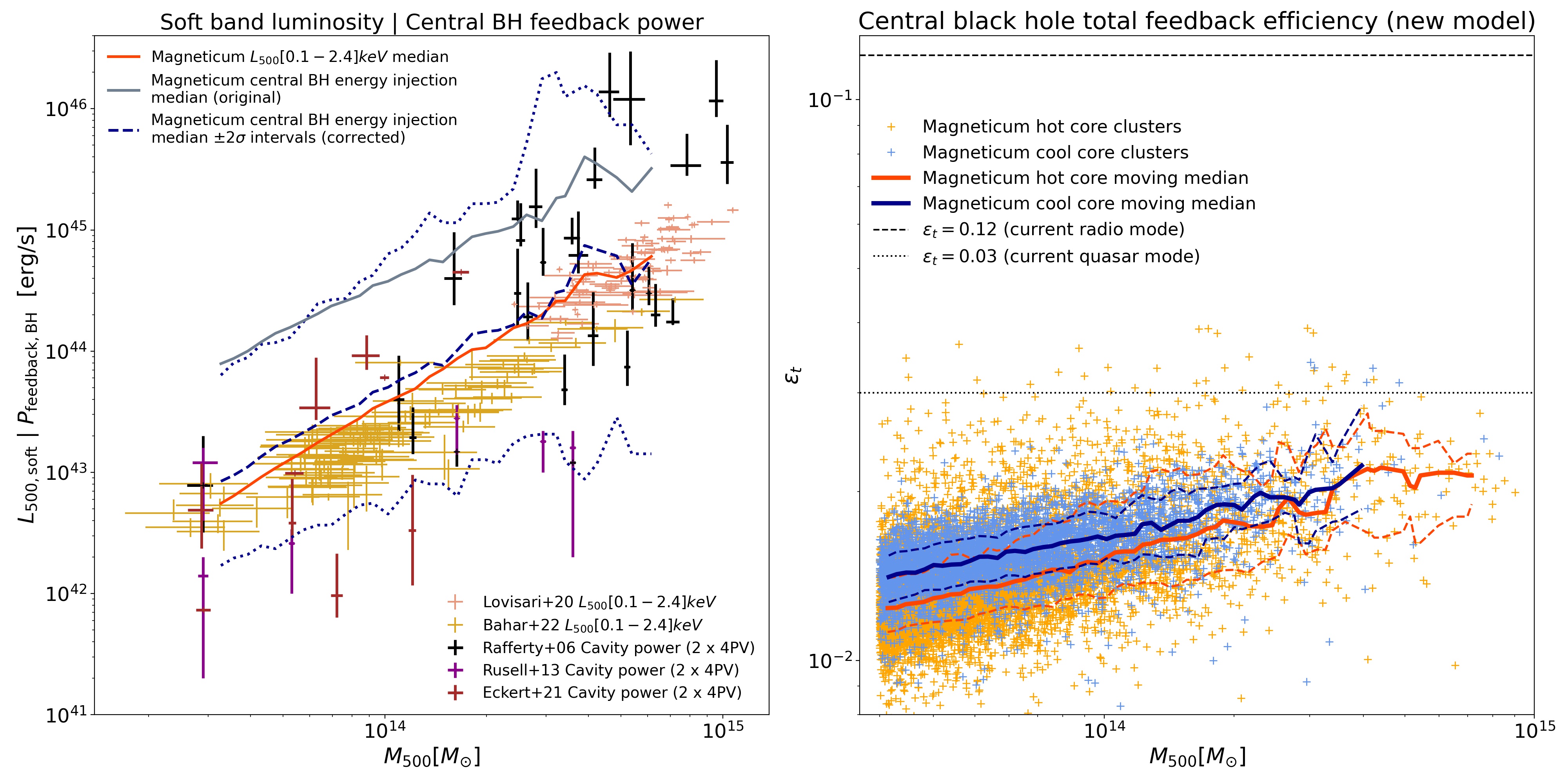}
    \caption{\textbf{Left panel}: Alignment of soft band luminosity with the central AGN feedback. The golden and salmon bars correspond to the ICM $L_{500}$ luminosity in the [0.1-2.4] keV (soft) band from the \cite{bahar2022erosita} and \cite{lovisari2020x} samples, respectively, whereas black, magenta, and brown bars correspond to the cavity powers from \cite{rafferty2006feedback}, \cite{russell2013radiative}, and the kinetic luminosity sample from \cite{pasini2022erosita}, respectively. Observational error bars are at the $1\sigma$ level. The solid orange-red and grey lines correspond to the luminosity and original AGN feedback model obtained from the simulation, whereas the blue dashed line corresponds to the 'corrected' AGN feedback model obtained from the simulation using Eq. \ref{eq:PjetMbondi}, and the blue dotted lines are the corresponding $2\sigma$ level intervals. \textbf{Right panel}: Total black hole efficiency in radio mode following the new model described by Eq. \ref{eq:EfficiencyTotal} based on the original accretion rates from the simulation shown in the left panel of Fig. \ref{fig:AGNFeedback}. Cool-core clusters are shown in blue, and hot-core clusters are shown in orange. Solid lines indicate moving medians, and dashed lines represent 16\% and 84\% percentiles ($1\sigma$) from the Magneticum simulation ({\it Box2b/hr}). For comparison, the total black hole efficiencies in radio and quasar mode are shown with a dashed and doted line, respectively.}
    \label{fig:AGNvsLICM}
\end{figure*}

Given that the accretion rates produced by the simulation seem to be well aligned with the observations as shown in the left panel of Fig. \ref{fig:AGNFeedback}, we can check if the total feedback efficiency used in radio mode, especially for groups, can be adjusted. For this, we can interpret the observational relation between Bondi power defined as $P_\text{Bondi} = 0.1 c^2 \dot{M}_{\text{B}}$ and cavity power ($P_{\text{cavity}}$) presented in \cite{fujita2014agn} as implied total feedback efficiencies in radio mode.

Consistently with the computation of the cavity powers from before and also motivated by \cite{fujita2016agn}, we introduce two adjustments to the $P_{\text{cavity}}-P_{\text{Bondi}}$ relation originally presented by \cite{fujita2014agn}, to assume that the jet cavities are filled with relativistic cosmic rays, therefore $\gamma_{c}=4/3$ and $E_\text{cav}=4V_c p_s$, and to apply a x2 factor to account for the shock energy as estimated by \cite{rafferty2006feedback} and \cite{guo2010simulating}.

\begin{figure*}[!ht]
    \centering
    \includegraphics[width=1\textwidth]{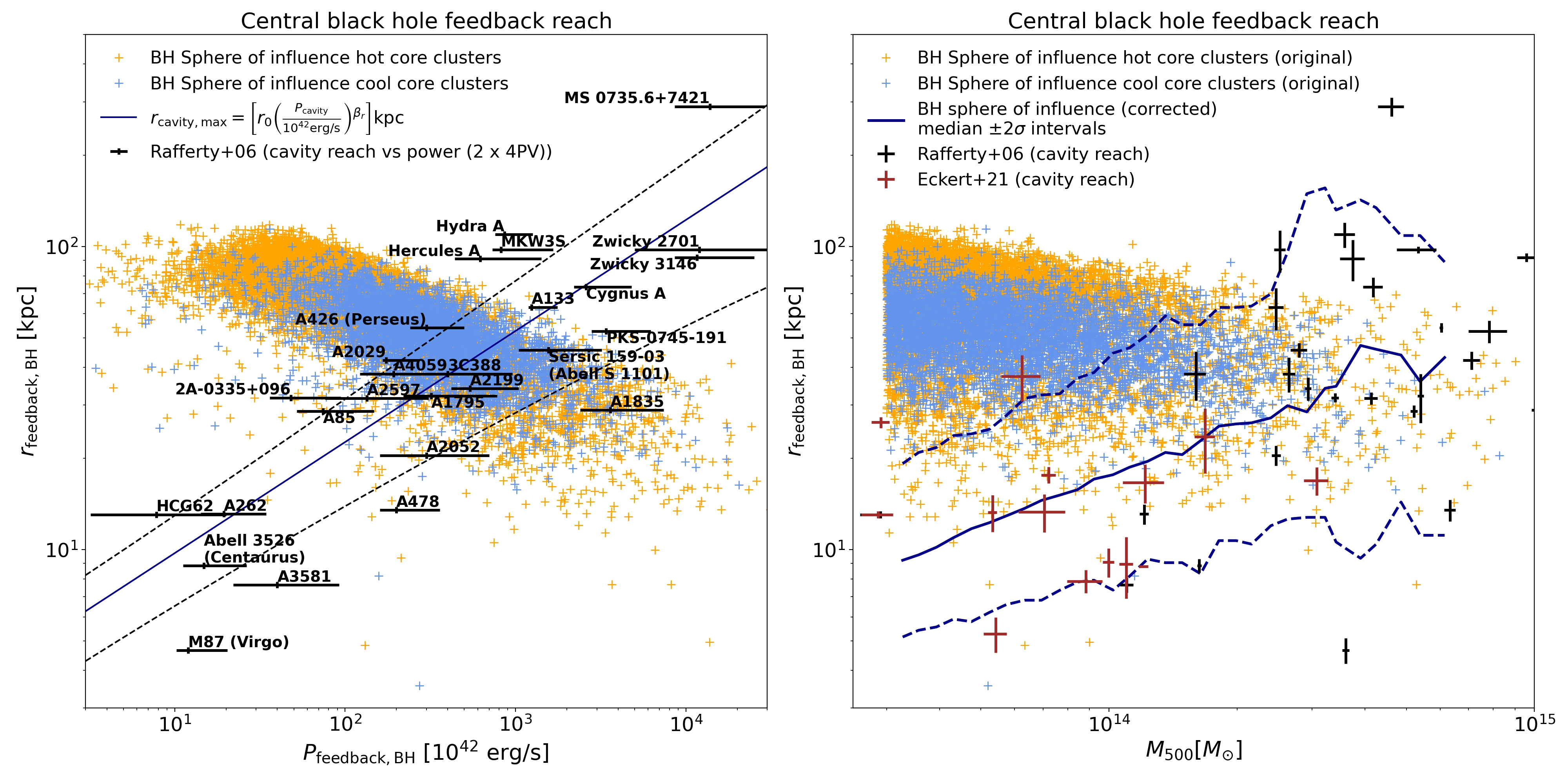}
    \caption{Size of the sphere of influence of the simulated AGNs in the centers of groups and clusters. Cool-core clusters are shown in blue, and hot-core clusters are shown in orange. Observational data from \cite{rafferty2006feedback} and \cite{eckert2021feedback} is shown with error bars at the $1 \sigma$ level. 
    \textbf{Left panel}: Size vs power. Shown is the maximum size or distance of cavities in observed groups and clusters versus cavity power, whereas for the simulation we show the 'black hole sphere of influence' versus the AGN energy injection. The dark blue solid line shows the power law fit described by Equation \ref{eq:CavityDistanceVsPower}, and the dashed lines represent $\pm 1 \sigma$ confidence intervals. \textbf{Right panel}: Same as the left panel, but plotting versus $M_{500c}$ in the X-axis. }
    \label{fig:AGNReach}
\end{figure*}

We now can express the observed cavity power directly as a function of the inferred Bondi accretion rate $\dot{M}_{\text{B}}$ where the final parameters are $P_0 = 13.60^{+55.56}_{-35.74}$ and $\beta_P = 1.14^{+0.09}_{-0.22}$:

\begin{align}
        P_{\text{cavity}} = \left[ P_0 \left( \frac{\dot{M_B}}{M_{\odot}\text{/yr}} \right)^{\beta_P} \right] 10^{44} \text{erg/s}
    \label{eq:PjetMbondi}
\end{align}

We can further divide both sides by $\dot{M_B}c^2$ to obtain the implied total efficiency where $\epsilon_0 = 0.024^{+0.10}_{-0.06}$ and $\beta_P - 1 = 0.14^{+0.09}_{-0.22}$, thus indicating a weak dependency of the total feedback efficiency with the accretion rate:

\begin{equation}
    \epsilon_t = \frac{P_{\text{cavity}} }{\dot{M_B} c^2} = \epsilon_0 \left( \frac{\dot{M_B}}{M_{\odot}\text{/yr}} \right)^{(\beta_P-1)}
    \label{eq:EfficiencyTotal}
\end{equation}

Recomputing the feedback power from the simulation using the above total efficiency depending on the BH accretion rate instead of a fixed the value of $0.12$ as currently used by the simulation in the radio mode, we obtain that the injected feedback energy in the simulations matches the observed cavity power over the full range of masses, from massive clusters down to groups, as illustrated by the dashed line in the left panel of Fig. \ref{fig:AGNvsLICM}. Therefore , the reduced feedback efficiency at lower masses can potentially alleviate the excess of feedback at group scales currently seen in the simulation.

Note that the reduced values of the total feedback efficiency obtained by this scaling are at a similar level as the mechanical efficiencies of the best cold accretion model presented by \cite{gaspari2013solving}, who found values in the order of $5 \cdot 10^{-3}-10^{-2}$ for galaxy clusters. The right panel of Fig. \ref{fig:AGNvsLICM} shows a comparison of the total black hole efficiencies in radio mode following the new 
model described by Eq. 8 based on the original accretion rates from the simulation shown in the left panel of Fig. \ref{fig:AGNFeedback}.

\subsection{Limitations of AGN feedback models in simulations} \label{sub:FeedbackReach}

In the Magneticum simulations, the AGN feedback is always injected purely as thermal energy within the surrounding gas. Following the original prescription from \cite{springel2005cosmological}, this region is defined by requiring a fixed number of resolution elements, similar to the smooth particle hydrodynamics (SPH) treatment. The energy inside this so-called 'black hole sphere of influence' (H) is distributed to the neighboring gas particles by the same kernel weighting scheme as used for the hydrodynamic treatment.

The mixing of this energy received by the particles with the rest of the ICM is mainly promoted by turbulent motions within the cluster core as well as through thermal conductivity, which is implemented in an isotropic manner at 1/20 of the Spitzer level. However, buoyancy of the heated gas in the center is largely suppressed by the centered, spherical injection of the AGN feedback energy as well as by the smoothed gravitational potential under the gravitational resolution limit of $\sim 15$ kpc.

To verify if the implemented AGN feedback model of the simulation somehow effectively mimics the situation in galaxy clusters and groups, we can compare this 'black hole sphere of influence' with the region where the energy of the radio mode feedback by jets is thermalized in observed galaxy clusters and groups. For this, we obtain the cavity maximum radius $r_{\text{cavity,max}} = R_{\text{cavity,center}} + 0.5(a+b)$ from the samples of \cite{rafferty2006feedback} and \cite{eckert2021feedback} where a is the projected semi-major axis, b is the projected semi-minor axis, $R_{\text{cavity,center}}$ is the distance to the cavity center and we estimate the corresponding error to be $0.5(b-a)$.

In Fig. \ref{fig:AGNReach} we show $r_{\text{cavity,max}}$ against the cavity power (left panel) as well as against $M_{500c}$ (right panel). For comparison, we show the 'black hole sphere of influence' of the central BHs from the simulation against their feedback power. While the simulation shows a completely different scaling in the size vs. power relation at the resolution of the Magneticum {\it Box2b/hr} simulation, the two distributions have a significant overlap and similar mean values on both axes. Therefore, the current implementation of the AGN feedback model gives a reasonable, effective description for the resolution of the simulation, despite the different scaling. The same is visible in the size vs. mass relation. Although the scatter and trend in the observations differ largely from the model in the simulations, at massive cluster scale the distributions largely overlap, so that the implemented AGN feedback model gives a reasonable, effective description. 
 
However, the difference increases with increasing feedback power, which is produced by stronger accretion rates in higher gas density environments that typically harbor star formation; therefore, it can be related to the problems in suppressing star formation previously mentioned. Also, towards the group scale, the model in the simulation clearly injects the AGN feedback within a larger region.

Also, note that the differences will increase when increasing the simulation resolution, which would scale the 'black hole sphere of influence' down with increasing resolution. In addition, alternative AGN feedback implementations used in other cosmological hydrodynamical simulations, even when based on injecting kinetic feedback, do not guarantee that the effective thermalization of the feedback energy coincides with these observations.

To improve future AGN feedback models, we can quantify the observational relation of cavity reach vs. power. A fit to the observational data from \cite{rafferty2006feedback} with an orthogonal distance regression (ODR), accounting for the errors in both variables\footnote{We where using the Python package SciPy.odr, which implements the algorithm proposed by \cite{boggs1990orthogonal}}, results in $\beta_r = 0.37 \pm 0.05$ and $r_0 = 4.2 \pm 1.3$ for a power law relation in the form:

\begin{equation}
    r_{\text{cavity,max}} = \left[ r_0  \left( \frac{P_{\text{cavity}}} {10^{42} \text{erg/s} } \right)^{\beta_r} \right] \text{kpc}.
    \label{eq:CavityDistanceVsPower}
\end{equation}

Using this relation, we can correct the 'black hole sphere of influence' used in the simulation to overlap with the observations of cavity reach, as shown by the blue solid and dashed lines in the right panel of Fig. \ref{fig:AGNReach}, representing the median and $\pm 2 \sigma$ intervals, respectively. Therefore, this relation can be used to improve future, effective models for AGN feedback in simulations and make them more resolution independent, although the accretion rate in itself still depends on the resolution and requires adjusting the boost factor in Eq. \ref{eq:Bondi–Hoyle–Lyttleton} accordingly. Additionally, to fully reproduce the AGN feedback signatures, it is necessary to track the BH spin and inject the feedback along its axis in a bipolar pattern, as shown by the simulations of \cite{sala2023supermassive}.

\subsection{Interpretation}

While the size of the 'black hole sphere of influence' is purely given by a technical choice in the implementation and therefore has no physical meaning, the original choice of the feedback efficiency in radio mode was motivated by observations of massive galaxy clusters. Given the tight relation between halo mass and cavity power and the weak dependency between BH accretion rate and total feedback efficiencies presented in Eq. \ref{eq:EfficiencyTotal}, which results in significantly smaller total efficiencies at the scale of groups, it is worth while to see if there could be a theoretical argument for such scaling.

The most established theoretical model to produce jets (mechanical feedback) is the \cite{blandford1977electromagnetic} (BZ) model, in which a spinning BH drags the magnetic field lines, producing an electric field by induction that accelerates charged particles. Therefore, in this scenario, the energy extraction efficiency is proportional to the BH spin parameter, which means that slowly spinning BHs should be less efficient at generating jets. 

However, according to observational constraints by \cite{reynolds2021observational}, there is a decreasing tendency in the spin parameter towards higher BH masses typically hosted in the most massive systems. This trend has been confirmed by simulations of spin evolution \citep{sala2023supermassive} and should be expected according to the isotropic principle, since mergers of BHs with spins oriented in random directions should produce higher-mass BHs with an increasingly lower spin parameter. As a consequence, and according to the BZ model, higher-mass black holes should be less efficient at generating jets. This scenario would be in tension with observations since stronger jets and cavity powers are actually observed towards the most massive systems, as shown by the right panel of Fig. \ref{fig:AGNFeedback}.

On the other hand, an alternative route to generate mechanical feedback even for non-spinning black holes can be collisions between particles orbiting the inner most stable orbits ($R_\text{ISCO}$) and free falling particles. In this sense, theoretical advancements by \cite{frolov2012weakly} have demonstrated that even with a weak uniform magnetic field, the radius of the inner most circular orbit of a non-rotating BH approaches the event horizon thanks to the additional support of a repulsive Lorentz force, which can potentially boost the mass-energy conversion efficiency for the accreted material. 

The effect is characterized by the ratio b between the Lorentz force and the surface gravity $\kappa$ at the event horizon \citep{baker2023charged} as shown by Equation \ref{eq:FrolovParameter}, where q and m are the charge and mass of the particle orbiting $R_\text{ISCO}$, B the magnetic field at the event horizon, $M_\text{BH}$ is the mass of the BH, G the gravitational constant, and c the speed of light:

\begin{equation}
    b= \frac{1}{4} \frac{qBc}{\kappa} =\frac{q B G M_\text{BH}}{m c^3} \sim  10^{15}\left(\frac{q}{e}\right)\left(\frac{m_e}{m}\right)\left(\frac{B}{10^4 G}\right)\left(\frac{M_\text{BH}}{10^9 M_{\odot}}\right)
    \label{eq:FrolovParameter}
\end{equation}

This ratio can indeed become very large even for a weak magnetic field in the order of $[1-30] G$, as measured by the Event Horizon Telecospe for M87 \citep{event2021first}. Moreover, \cite{frolov2012weakly} showed that the energy of a collision between a particle at $R_\text{ISCO}$ and a free falling particle is directly proportional to $E_\text{collision} \sim mc^2b^{1/4}$.

To quantify the magnetic field, we have to consider that the BHs hosted in the center of BCGs generally have a low Eddington ratio and luminosity, characterized by the hot advection-dominated accretion flow model (ADAF, \cite{narayan1995advection}), which is usually associated with jets, according to observations. In the ADAF model, the magnetic field strength at the Schwarzschild radius is proportional to the accretion rate $\dot{M}_\text{BH}$, as shown by Equation \ref{eq:B_ADAC} \citep{yuan2014hot}, where $\alpha$ is the viscosity parameter, $\beta = \frac{P_{\text{gas}}}{P_{\text{mag}}}$ the magnetization parameter, and $M_\text{BH}$ the mass of the BH:

\begin{equation}
    B = 0.44 \left( 1 + \beta \right)^{-\frac{1}{2}} \alpha^{-\frac{1}{2}} \left( \frac{M_\text{BH}}{10^9 M_{\odot}}\right)^{-1} \left( \frac{\dot{M}_\text{BH}}{M_\odot / \text{yr}}\right)^{\frac{1}{2}} 10^{4} G
    \label{eq:B_ADAC}
\end{equation}

Then, if we combine Eqs. \ref{eq:FrolovParameter} and \ref{eq:B_ADAC} and consider the energy extraction efficiency ($\epsilon$) of the \cite{frolov2012weakly} process as the ratio between the collision energy ($E_{collision}$) and the rest mass energy of the free falling particle ($mc^2$) we obtain that the energy extraction efficiency depends on $\dot{M}_\text{BH}^{1/8}$ as shown by Equation \ref{eq:CollisionEnergy}, and quite close to the $0.14$ exponent obtained from the observational relation shown in Equation \ref{eq:EfficiencyTotal}.

\begin{equation}
    \epsilon \sim \frac{E_{collision}}{mc^2} \sim b^{\frac{1}{4}} \sim \left( BM_\text{BH} \right)^{\frac{1}{4}} \sim \dot{M}_\text{BH}^{1/8}
    \label{eq:CollisionEnergy}
\end{equation}

This process can provide a channel independent of BH spin to produce jets (mechanical feedback), which can be a key ingredient to explain increasing jet and cavity powers towards the most massive systems and decreasing towards low mass systems to reduce the overall mechanical feedback efficiency at the scale of groups and address the overheating problems described in Sect. \ref{sub:CoolCoreFractionComparison} and also reported by \cite{gaspari2013solving} and suggested by the entropy excess at the cores of simulated galaxy groups shown by \cite{bahar2024srgerosita}.

\section{Comparison with the results of other simulations}

Looking at the results from other cosmological simulations, we find that the decreasing trend in cool-core fractions towards higher mass clusters initially reported by \cite{chen2007statistics} and confirmed by the simulations of \cite{burns2008only}, \cite{planelles2009galaxy}, and this work, is however not systematically reproduced by all cool-core indicators for the TNG-Cluster simulation; \cite{lehle2024heart} reports that the non-cool core fraction increases with mass only when using the central cooling time and entropy as indicators, whereas it decreases when using the central density and cuspiness. Here we argue that as explained in Sec. \ref{sec:ComparingObservationsSimulations} and Sec. \ref{sub:CoolCoreIndicators} the central properties are subject to resolution issues in both observations and simulations, and also, for the case of simulations, are affected by the direct AGN feedback, thus subject to the limitations in the sub-grid model affecting the distribution of the AGN feedback as described in Section \ref{sub:FeedbackReach}.

Regarding the problems with overheating and excess of entropy in the central regions at the scale of galaxy groups initially reported by \cite{fabjan2010simulating}, and confirmed by \cite{bahar2024srgerosita} and this work, \cite{barnes2017cluster} reports a similar problem in the temperature profiles already at the scale of galaxy clusters (median $ M_{500c} = 2.1 \cdot 10^{14}M_\odot$) for the Cluster-EAGLE simulations (C-Eagle), also confirmed by \cite{altamura2023eagle} in the entropy profiles at the scales of groups ($ M_{500c} = 8.8 \cdot 10^{12}M_\odot$) and clusters ($ M_{500c} = 2.9 \cdot 10^{14}M_\odot$). Similarly we also see some traces of overheating in our hot-core clusters at the mid-mass range ($10^{14}M_\odot < M_{500c} \leq 4.88 \cdot 10^{14}M_\odot$), although is possible that this problem is partially mitigated for the Magneticum simulations at the scale of massive galaxy clusters due to the implementation of physical conductivity at 1/20 of the Spitzer value, which nevertheless we plan to review in future works. At the galaxy group regime, our conclusions are aligned with results obtained by the simulations of \cite{gaspari2013solving} in that lower mass halos require reduced AGN feedback efficiencies to prevent overheating.

Regarding the distribution of AGN feedback, \cite{le2014towards} reports that the best feedback model for the OverWhelmingly Large Simulations (OWLS) requires injecting the feedback only in one gas particle per feedback loop and only if the injected energy is above $10^8$ K (8.6 keV) to prevent that the feedback energy gets quickly radiated away if distributed among all neighboring particles. This contrasts with the approach presented in this work, to distribute the feedback through the thermal channel directly in the region where the signatures of AGN feedback (radio emission, cavities) are observed. Here we point out that a key aspect to prevent catastrophic cooling is to inject the energy into the ICM particles that have not yet been cooled, which can be a more efficient process given the longer cooling times of the ICM. Moreover, although we agree that the AGN feedback in real physical scenarios is generated in the surroundings of the AGN, it is expelled at relativistic speeds and thermalizes at a much larger distance corresponding to the observed excavated cavities that overlap with radio emission.

Finally, our conclusion that merge activity does not directly turn cool-cores into hot-cores since it first requires to be thermalized coincides with the conclusions of \cite{poole2008impact} based on a suite of idealized mergers with different mass ratios and impact parameters and \cite{hahn2017rhapsody} based on the RHAPSODY-G simulations, who reported that only direct collisions (low angular momentum mergers) can effectively turn cool-cores into hot-cores. Additionally, we also agree with \cite{rasia2015cool} and \cite{hahn2017rhapsody} in that thermal conduction is required to effectively redistribute the energy injected by mergers and AGN feedback across the ICM in the core region.

\section{Conclusions}

We conducted a study to compare the halo mass dependency of cool-core fractions for a large sample of groups and clusters from observational data combining the samples from \cite{lovisari2020x} and \cite{bahar2022erosita} at low redshift ($z < 0.3$) with a large scale, cosmological hydrodynamical simulation {\it Box2b/hr} from the Magneticum set. This covers a mass range of $0.3\times10^{14}M_\odot < M_{500c} < 10^{15}M_\odot$ with a sample size of 201 groups and clusters for the observational dataset and 13683 for the simulation dataset. For the set of 2027 simulated clusters above a mass of $M_{500c} = 10^{14}M_\odot$ we compared in detail the radial profiles of density, temperature, and entropy with the results obtained from X-ray observations. We investigate the possible physical mechanism that drives the observed trends of the cool-core fractions in simulations and observations across this mass range. We also compare the assumptions and predictions of the AGN feedback model implemented within the simulations with the observed properties of the central AGN and their inflated radio bubbles in clusters and groups. Our main findings can be summarized as follows:

\begin{itemize}
    \item In general, the simulation is able to reproduce the overall cool-core fractions at the scale of galaxy clusters ($M_{500c} > 1 \cdot 10^{14}M_\odot$) and the mid-to-strong cool-core fractions at the scale of massive clusters ($M_{500c} > 2.7 \cdot 10^{14} M_\odot$). When splitting the simulated clusters in cool-core and hot-core systems, the radial profiles of the two classes reproduce the different shapes observed for the two different classes, where hot-core systems show a more cored entropy profile with larger values compared to cool-core systems and hot-core systems showing a more isothermal temperature profile in the center compared to the drop of the temperature of cool-core systems towards the center. However, some traces of overheating and inefficient energy transport are visible in the temperature profiles of the hot-core clusters at the mid-mass range ($10^{14}M_\odot < M_{500c} \leq 4.88 \cdot 10^{14}M_\odot$), calling for a review of the physical conductivity model currently set at 1/20 of the Spitzer value.
    \item The cool-core fractions clearly decrease towards high-mass galaxy clusters; this trend is observed in both simulations and observations and confirms the results reported by \cite{chen2007statistics}, \cite{burns2008only}, and \cite{planelles2009galaxy}. However, the relative contribution of AGN feedback to this process also decreases at the high mass end, indicating that an additional factor is required to suppress cool-cores. While within the core, the amount of internal plus kinetic energy compared to the potential energy for the total system, including dark matter, stellar, and gas components, is increasing with mass, there is no visible difference for cool-core and hot-core systems. However, when only considering the gas component, we find a clear separation of cool-core and hot-core systems, with hot-core systems having a larger internal plus kinetic energy compared to cool-core systems as well as higher internal energy fractions, both increasing with mass. This indicates that the thermalization of the kinetic energy induced by mergers is more efficient for hot-core systems and also increases in the high-mass regime. This factor, combined with the implementation of physical conductivity in the Magneticum simulations, which is also more efficient as the halo mass increases due to its strong temperature dependency, can effectively reduce the cool-core fractions at the scale of massive clusters.
    \item On the other hand, the cool-core fractions decrease much more sharply in the simulation than in observations for the low-mass galaxy groups. This divergence is equivalent to the excess of entropy in the cores of simulated galaxy groups shown by \cite{bahar2024srgerosita}. This can be associated with the relatively high impact that the AGN feedback injected in radio mode has in comparison with the luminosity at the scale of galaxy groups. While the simulation produces Bondi accretion rates compatible with the refined estimations by \cite{fujita2014agn} for the AGNs hosted in the centers of BCGs, the usage of a relatively large AGN feedback efficiency as found in inferred in massive clusters might be questionable in galaxy groups and might be much lower and in the range as inferred from the cold accretion model presented in \cite{gaspari2013solving}.
    \item The AGN feedback efficiencies in the simulations at group scales could be aligned when using the relation between Bondi power and cavity power presented by \cite{fujita2014agn}. This relation is equivalent to a weak dependency of the total feedback efficiency in radio mode on the accretion rate, with an exponent close to 1/8. Interestingly, the same dependency and exponent are obtained considering an AGN energy injection mechanism based on collisions between free-falling particles and charged particles orbiting the inner most stable orbits in the presence of a weak magnetic field, following the theoretical model of \cite{frolov2012weakly}. This process also provides an AGN energy injection mechanism independent of spin, which decreases towards the supermassive BH hosted in the centers of galaxy clusters, as suggested by the observations of \cite{reynolds2021observational} and confirmed by the simulations of \cite{sala2023supermassive}.
    \item Finally, we point out that current implementation of AGN feedback models in simulations often scales the size of the region where the AGN feedback energy is injected inversely with the local density, which is in contrast to the scaling of the observed size and reach of cavities. This inverted scaling can reduce the effectiveness of the AGN feedback to suppress star formation in high-density environments. Although for the current simulation there is still a significant overlap, which makes such effective models at least work partially, future simulations with increased resolution need to adapt the sub-grid models accordingly. Taken the observational data from \cite{rafferty2006feedback}, we have obtained a power law relation of the AGN energy injection and the expected reach that can be used to bring the AGN feedback distribution schema from simulations into agreement with observations.
\end{itemize}
\begin{acknowledgements}

This work was supported by the COMPLEX project from the European Research Council (ERC) under the European Union’s Horizon 2020 research and innovation program grant agreement ERC-2019-AdG 882679.

VB acknowledges support by the DFG project nr. 415510302. 

The calculations for the hydrodynamical simulations were carried out at the Leibniz Supercomputer Center (LRZ) under the project pr83li.

We thank the developers of the following software packages that we have used for the data analysis: Python 3.6.15 \citep{van1995python}, numpy 1.19.5 \citep{harris2020array}, g3read, g3matcha \citep{ragagnin2017webportal}, astropy 4.1 (\cite{robitaille2013astropy}, \cite{astropy2022astropy}), scipy 1.5.4 \citep{virtanen2020scipy}, AtomDB v3.0.9 \citep{foster2018atomdb}, PyAtomDB v0.10.10 \citep{foster2020pyatomdb}, statsmodels  0.12.2 \citep{seabold2010statsmodels}, asymmetric uncertainty 0.2.1 \citep{gobat2022asymmetric}, and matplotlib 3.3.4 (\cite{hunter2007matplotlib}, \cite{caswell2023matplotlib})

This research has made use of the M2C Galaxy Cluster Database, constructed as part of the ERC project M2C (The Most Massive Clusters across cosmic time, ERC-Adv grant No. 340519).

We thank Adam Foster for his support and insights regarding the APEC model, AtomDB and PyAtomDB; Thomas Pasini for kindly providing the LOFAR kinetic luminosity data; and Lucas Sala for the many useful discussions and feedback regarding accretion flows and AGN feedback.

\end{acknowledgements}

%
\bibliographystyle{aa} 
\bibliography{bibliography} 

\begin{thebibliography}{135}
\expandafter\ifx\csname natexlab\endcsname\relax\def\natexlab#1{#1}\fi

\bibitem[{Altamura {et~al.}(2023)Altamura, Kay, Bower, Schaller, Bah{\'e},
  Schaye, Borrow, \& Towler}]{altamura2023eagle}
Altamura, E., Kay, S.~T., Bower, R.~G., {et~al.} 2023, Monthly Notices of the
  Royal Astronomical Society, 520, 3164

\bibitem[{Anders \& Grevesse(1989)}]{anders1989abundances}
Anders, E. \& Grevesse, N. 1989, Geochimica et Cosmochimica acta, 53, 197

\bibitem[{Andrade-Santos {et~al.}(2017)Andrade-Santos, Jones, Forman, Lovisari,
  Vikhlinin, Van~Weeren, Murray, Arnaud, Pratt, D{\'e}mocl{\`e}s,
  {et~al.}}]{andrade2017fraction}
Andrade-Santos, F., Jones, C., Forman, W.~R., {et~al.} 2017, The Astrophysical
  Journal, 843, 76

\bibitem[{{Angelinelli} {et~al.}(2022){Angelinelli}, {Ettori}, {Dolag},
  {Vazza}, \& {Ragagnin}}]{angelinelli2022bfrac}
{Angelinelli}, M., {Ettori}, S., {Dolag}, K., {Vazza}, F., \& {Ragagnin}, A.
  2022, \aap, 663, L6

\bibitem[{{Angelinelli} {et~al.}(2023){Angelinelli}, {Ettori}, {Dolag},
  {Vazza}, \& {Ragagnin}}]{angelinelli2023bfrac}
{Angelinelli}, M., {Ettori}, S., {Dolag}, K., {Vazza}, F., \& {Ragagnin}, A.
  2023, \aap, 675, A188

\bibitem[{{Arth} {et~al.}(2014){Arth}, {Dolag}, {Beck}, {Petkova}, \&
  {Lesch}}]{arth2014anisoconduction}
{Arth}, A., {Dolag}, K., {Beck}, A.~M., {Petkova}, M., \& {Lesch}, H. 2014,
  arXiv e-prints, arXiv:1412.6533

\bibitem[{Asplund {et~al.}(2009)Asplund, Grevesse, Sauval, \&
  Scott}]{asplund2009chemical}
Asplund, M., Grevesse, N., Sauval, A.~J., \& Scott, P. 2009, Annual review of
  astronomy and astrophysics, 47, 481

\bibitem[{Astropy {et~al.}(2022)Astropy, Price-Whelan, Lim, Earl, Starkman, \&
  B{\'o}di}]{astropy2022astropy}
Astropy, C., Price-Whelan, A.~M., Lim, P.~L., {et~al.} 2022, Astrophysical
  Journal, 935

\bibitem[{Bahar {et~al.}(2022)Bahar, Bulbul, Clerc, Ghirardini, Liu, Nandra,
  Pacaud, Chiu, Comparat, Ider-Chitham, {et~al.}}]{bahar2022erosita}
Bahar, Y.~E., Bulbul, E., Clerc, N., {et~al.} 2022, Astronomy \& Astrophysics,
  661, A7

\bibitem[{Bahar {et~al.}(2024)Bahar, Bulbul, Ghirardini, Sanders, Zhang, Liu,
  Clerc, Artis, Balzer, Biffi, Bose, Comparat, Dolag, Garrel, Hadzhiyska,
  Hernández-Aguayo, Hernquist, Kluge, Krippendorf, Merloni, Nandra, Pakmor,
  Popesso, Ramos-Ceja, Seppi, Springel, Weller, \&
  Zelmer}]{bahar2024srgerosita}
Bahar, Y.~E., Bulbul, E., Ghirardini, V., {et~al.} 2024, The SRG/eROSITA
  All-Sky Survey: Constraints on AGN Feedback in Galaxy Groups

\bibitem[{Baker \& Frolov(2023)}]{baker2023charged}
Baker, N.~P. \& Frolov, V.~P. 2023, Physical Review D, 108, 024045

\bibitem[{{Barnes} {et~al.}(2019){Barnes}, {Kannan}, {Vogelsberger},
  {Pfrommer}, {Puchwein}, {Weinberger}, {Springel}, {Pakmor}, {Nelson},
  {Marinacci}, {Pillepich}, {Torrey}, \& {Hernquist}}]{barns2019ccaniso}
{Barnes}, D.~J., {Kannan}, R., {Vogelsberger}, M., {et~al.} 2019, \mnras, 488,
  3003

\bibitem[{Barnes {et~al.}(2017)Barnes, Kay, Bah{\'e}, Dalla~Vecchia, McCarthy,
  Schaye, Bower, Jenkins, Thomas, Schaller, {et~al.}}]{barnes2017cluster}
Barnes, D.~J., Kay, S.~T., Bah{\'e}, Y.~M., {et~al.} 2017, Monthly Notices of
  the Royal Astronomical Society, 471, 1088

\bibitem[{Barnes {et~al.}(2018)Barnes, Vogelsberger, Kannan, Marinacci,
  Weinberger, Springel, Torrey, Pillepich, Nelson, Pakmor,
  {et~al.}}]{barnes2018census}
Barnes, D.~J., Vogelsberger, M., Kannan, R., {et~al.} 2018, Monthly Notices of
  the Royal Astronomical Society, 481, 1809

\bibitem[{Beck {et~al.}(2016)Beck, Murante, Arth, Remus, Teklu, Donnert,
  Planelles, Beck, F{\"o}rster, Imgrund, {et~al.}}]{beck2016improved}
Beck, A.~M., Murante, G., Arth, A., {et~al.} 2016, Monthly Notices of the Royal
  Astronomical Society, 455, 2110

\bibitem[{Biffi {et~al.}(2016)Biffi, Borgani, Murante, Rasia, Planelles,
  Granato, Ragone-Figueroa, Beck, Gaspari, \& Dolag}]{biffi2016nature}
Biffi, V., Borgani, S., Murante, G., {et~al.} 2016, The Astrophysical Journal,
  827, 112

\bibitem[{Biffi {et~al.}(2012)Biffi, Dolag, Boehringer, \&
  Lemson}]{biffi2012observing}
Biffi, V., Dolag, K., Boehringer, H., \& Lemson, G. 2012, Monthly Notices of
  the Royal Astronomical Society, 420, 3545

\bibitem[{{Biffi} {et~al.}(2013){Biffi}, {Dolag}, \&
  {B{\"o}hringer}}]{biffi2013scaling}
{Biffi}, V., {Dolag}, K., \& {B{\"o}hringer}, H. 2013, \mnras, 428, 1395

\bibitem[{{Biffi} {et~al.}(2018{\natexlab{a}}){Biffi}, {Dolag}, \&
  {Merloni}}]{biffi2018agn}
{Biffi}, V., {Dolag}, K., \& {Merloni}, A. 2018{\natexlab{a}}, \mnras, 481,
  2213

\bibitem[{{Biffi} {et~al.}(2022){Biffi}, {Dolag}, {Reiprich}, {Veronica},
  {Ramos-Ceja}, {Bulbul}, {Ota}, \& {Ghirardini}}]{biffi2022bridge}
{Biffi}, V., {Dolag}, K., {Reiprich}, T.~H., {et~al.} 2022, \aap, 661, A17

\bibitem[{{Biffi} {et~al.}(2018{\natexlab{b}}){Biffi}, {Mernier}, \&
  {Medvedev}}]{biffi2018review}
{Biffi}, V., {Mernier}, F., \& {Medvedev}, P. 2018{\natexlab{b}}, \ssr, 214,
  123

\bibitem[{Blandford \& Znajek(1977)}]{blandford1977electromagnetic}
Blandford, R.~D. \& Znajek, R.~L. 1977, Monthly Notices of the Royal
  Astronomical Society, 179, 433

\bibitem[{Boehringer {et~al.}(2011)Boehringer, Dolag, \&
  Chon}]{boehringer2011self}
Boehringer, H., Dolag, K., \& Chon, G. 2011, arXiv preprint arXiv:1112.5035

\bibitem[{Boggs \& Rogers(1990)}]{boggs1990orthogonal}
Boggs, P.~T. \& Rogers, J.~E. 1990, Contemporary mathematics, 112, 183

\bibitem[{Bondi(1952)}]{bondi1952spherically}
Bondi, H. 1952, Monthly Notices of the Royal Astronomical Society, 112, 195

\bibitem[{Bondi \& Hoyle(1944)}]{bondi1944mechanism}
Bondi, H. \& Hoyle, F. 1944, Monthly Notices of the Royal Astronomical Society,
  104, 273

\bibitem[{Borgani \& Kravtsov(2011)}]{borgani2011cosmological}
Borgani, S. \& Kravtsov, A. 2011, Advanced Science Letters, 4, 204

\bibitem[{Borgani {et~al.}(2004)Borgani, Murante, Springel, Diaferio, Dolag,
  Moscardini, Tormen, Tornatore, \& Tozzi}]{borgani2004x}
Borgani, S., Murante, G., Springel, V., {et~al.} 2004, Monthly Notices of the
  Royal Astronomical Society, 348, 1078

\bibitem[{Burns {et~al.}(2008)Burns, Hallman, Gantner, Motl, \&
  Norman}]{burns2008only}
Burns, J.~O., Hallman, E.~J., Gantner, B., Motl, P.~M., \& Norman, M.~L. 2008,
  The Astrophysical Journal, 675, 1125

\bibitem[{Caswell {et~al.}(2023)}]{caswell2023matplotlib}
Caswell, T. {et~al.} 2023, matplotlib/matplotlib: REL: v3. 7.0 rc1, doi:
  10.5281/zenodo. 7570264

\bibitem[{Cavagnolo {et~al.}(2008)Cavagnolo, Donahue, Voit, \&
  Sun}]{cavagnolo2008bandpass}
Cavagnolo, K.~W., Donahue, M., Voit, G.~M., \& Sun, M. 2008, The Astrophysical
  Journal, 682, 821

\bibitem[{Cavagnolo {et~al.}(2009)Cavagnolo, Donahue, Voit, \&
  Sun}]{cavagnolo2009intracluster}
Cavagnolo, K.~W., Donahue, M., Voit, G.~M., \& Sun, M. 2009, The Astrophysical
  Journal Supplement Series, 182, 12

\bibitem[{Chabrier(2003)}]{chabrier2003galactic}
Chabrier, G. 2003, Publications of the Astronomical Society of the Pacific,
  115, 763

\bibitem[{Chen {et~al.}(2007)Chen, Reiprich, B{\"o}hringer, Ikebe, \&
  Zhang}]{chen2007statistics}
Chen, Y., Reiprich, T., B{\"o}hringer, H., Ikebe, Y., \& Zhang, Y.-Y. 2007,
  Astronomy \& Astrophysics, 466, 805

\bibitem[{Chiu {et~al.}(2022)Chiu, Ghirardini, Liu, Grandis, Bulbul, Bahar,
  Comparat, Bocquet, Clerc, Klein, {et~al.}}]{chiu2022erosita}
Chiu, I.-N., Ghirardini, V., Liu, A., {et~al.} 2022, Astronomy \& Astrophysics,
  661, A11

\bibitem[{Churazov {et~al.}(2001)Churazov, Br{\"u}ggen, Kaiser, B{\"o}hringer,
  \& Forman}]{churazov2001evolution}
Churazov, E., Br{\"u}ggen, M., Kaiser, C., B{\"o}hringer, H., \& Forman, W.
  2001, The Astrophysical Journal, 554, 261

\bibitem[{Churazov {et~al.}(2005)Churazov, Sazonov, Sunyaev, Forman, Jones, \&
  B{\"o}hringer}]{churazov2005supermassive}
Churazov, E., Sazonov, S., Sunyaev, R., {et~al.} 2005, Monthly Notices of the
  Royal Astronomical Society: Letters, 363, L91

\bibitem[{Collaboration {et~al.}(2021)}]{event2021first}
Collaboration, E. H.~T. {et~al.} 2021, arXiv preprint arXiv:2105.01173

\bibitem[{Collaboration {et~al.}(2018)Collaboration, Aharonian, Akamatsu,
  Akimoto, Allen, Angelini, Audard, Awaki, Axelsson, Bamba,
  {et~al.}}]{hitomi2018atmospheric}
Collaboration, H., Aharonian, F., Akamatsu, H., {et~al.} 2018, Publications of
  the Astronomical Society of Japan, 70, 9

\bibitem[{Comparat {et~al.}(2020)Comparat, Eckert, Finoguenov, Schmidt,
  Sanders, Nagai, Lau, Kaefer, Pacaud, Clerc, {et~al.}}]{comparat2020full}
Comparat, J., Eckert, D., Finoguenov, A., {et~al.} 2020, arXiv preprint
  arXiv:2008.08404

\bibitem[{Cullen \& Dehnen(2010)}]{cullen2010inviscid}
Cullen, L. \& Dehnen, W. 2010, Monthly Notices of the Royal Astronomical
  Society, 408, 669

\bibitem[{Davis {et~al.}(2011)Davis, D’Aloisio, \&
  Natarajan}]{davis2011virialization}
Davis, A.~J., D’Aloisio, A., \& Natarajan, P. 2011, Monthly Notices of the
  Royal Astronomical Society, 416, 242

\bibitem[{Dehnen \& Aly(2012)}]{dehnen2012improving}
Dehnen, W. \& Aly, H. 2012, Monthly Notices of the Royal Astronomical Society,
  425, 1068

\bibitem[{Di~Matteo {et~al.}(2005)Di~Matteo, Springel, \&
  Hernquist}]{di2005energy}
Di~Matteo, T., Springel, V., \& Hernquist, L. 2005, nature, 433, 604

\bibitem[{Dolag(2015)}]{dolag2015magneticum}
Dolag, K. 2015, IAU General Assembly, 29, 2250156

\bibitem[{Dolag {et~al.}(2004)Dolag, Jubelgas, Springel, Borgani, \&
  Rasia}]{dolag2004thermal}
Dolag, K., Jubelgas, M., Springel, V., Borgani, S., \& Rasia, E. 2004, The
  Astrophysical Journal, 606, L97

\bibitem[{{Dolag} {et~al.}(2017){Dolag}, {Mevius}, \&
  {Remus}}]{dolag2017metals}
{Dolag}, K., {Mevius}, E., \& {Remus}, R.-S. 2017, Galaxies, 5, 35

\bibitem[{Dolag {et~al.}(2005)Dolag, Vazza, Brunetti, \&
  Tormen}]{dolag2005turbulent}
Dolag, K., Vazza, F., Brunetti, G., \& Tormen, G. 2005, Monthly Notices of the
  Royal Astronomical Society, 364, 753

\bibitem[{Eckert {et~al.}(2021)Eckert, Gaspari, Gastaldello, Le~Brun, \&
  O’Sullivan}]{eckert2021feedback}
Eckert, D., Gaspari, M., Gastaldello, F., Le~Brun, A.~M., \& O’Sullivan, E.
  2021, Universe, 7, 142

\bibitem[{Eckert {et~al.}(2019)Eckert, Ghirardini, Ettori, Rasia, Biffi,
  Pointecouteau, Rossetti, Molendi, Vazza, Gastaldello, Gaspari, De~Grandi,
  Ghizzardi, Bourdin, Tchernin, \& Roncarelli}]{eckert2019turb}
Eckert, D., Ghirardini, V., Ettori, S., {et~al.} 2019, Astronomy and
  Astrophysics, 621, A40

\bibitem[{Fabian(2003)}]{fabian2003gravitational}
Fabian, A. 2003, Monthly Notices of the Royal Astronomical Society, 344, L27

\bibitem[{Fabian(2002)}]{fabian2002cooling}
Fabian, A.~C. 2002, in Lighthouses of the Universe: The Most Luminous Celestial
  Objects and Their Use for Cosmology: Proceedings of the MPA/ESO/MPE/USM Joint
  Astronomy Conference Held in Garching, Germany, 6-10 August 2001, Springer,
  24--36

\bibitem[{Fabjan {et~al.}(2010)Fabjan, Borgani, Tornatore, Saro, Murante, \&
  Dolag}]{fabjan2010simulating}
Fabjan, D., Borgani, S., Tornatore, L., {et~al.} 2010, Monthly Notices of the
  Royal Astronomical Society, 401, 1670

\bibitem[{Foster {et~al.}(2018)Foster, Smith, Brickhouse, Mullen, Cumbee,
  Stancil, \& Cui}]{foster2018atomdb}
Foster, A., Smith, R., Brickhouse, N.~S., {et~al.} 2018, in American
  Astronomical Society Meeting Abstracts\# 231, Vol. 231, 253--03

\bibitem[{Foster \& Heuer(2020)}]{foster2020pyatomdb}
Foster, A.~R. \& Heuer, K. 2020, Atoms, 8, 49

\bibitem[{Fraser-McKelvie {et~al.}(2014)Fraser-McKelvie, Brown, \&
  Pimbblet}]{fraser2014rarity}
Fraser-McKelvie, A., Brown, M.~J., \& Pimbblet, K.~A. 2014, Monthly Notices of
  the Royal Astronomical Society: Letters, 444, L63

\bibitem[{Frolov(2012)}]{frolov2012weakly}
Frolov, V.~P. 2012, Physical Review D, 85, 024020

\bibitem[{Fujita {et~al.}(2014)Fujita, Kawakatu, \& Shlosman}]{fujita2014agn}
Fujita, Y., Kawakatu, N., \& Shlosman, I. 2014, arXiv preprint arXiv:1406.6366

\bibitem[{Fujita {et~al.}(2016)Fujita, Kawakatu, \& Shlosman}]{fujita2016agn}
Fujita, Y., Kawakatu, N., \& Shlosman, I. 2016, Publications of the
  Astronomical Society of Japan, 68, 26

\bibitem[{Gaspari {et~al.}(2013)Gaspari, Brighenti, \&
  Ruszkowski}]{gaspari2013solving}
Gaspari, M., Brighenti, F., \& Ruszkowski, M. 2013, Astronomische Nachrichten,
  334, 394

\bibitem[{{Gaspari} {et~al.}(2018){Gaspari}, {McDonald}, {Hamer}, {Brighenti},
  {Temi}, {Gendron-Marsolais}, {Hlavacek-Larrondo}, {Edge}, {Werner}, {Tozzi},
  {Sun}, {Stone}, {Tremblay}, {Hogan}, {Eckert}, {Ettori}, {Yu}, {Biffi}, \&
  {Planelles}}]{gaspari2018cca}
{Gaspari}, M., {McDonald}, M., {Hamer}, S.~L., {et~al.} 2018, \apj, 854, 167

\bibitem[{Ghirardini {et~al.}(2022)Ghirardini, Bahar, Bulbul, Liu, Clerc,
  Pacaud, Comparat, Liu, Ramos~Ceja, Hoang,
  {et~al.}}]{ghirardini2022characterization}
Ghirardini, V., Bahar, Y.~E., Bulbul, E., {et~al.} 2022, AAS/High Energy
  Astrophysics Division, 54, 107

\bibitem[{Gobat(2022)}]{gobat2022asymmetric}
Gobat, C. 2022, Astrophysics Source Code Library, ascl

\bibitem[{Gonzalez {et~al.}(2007)Gonzalez, Zaritsky, \&
  Zabludoff}]{gonzalez2007census}
Gonzalez, A.~H., Zaritsky, D., \& Zabludoff, A.~I. 2007, The Astrophysical
  Journal, 666, 147

\bibitem[{Guo \& Mathews(2010)}]{guo2010simulating}
Guo, F. \& Mathews, W.~G. 2010, The Astrophysical Journal, 712, 1311

\bibitem[{{Gupta} {et~al.}(2017){Gupta}, {Saro}, {Mohr}, {Dolag}, \&
  {Liu}}]{gupta2017pressure}
{Gupta}, N., {Saro}, A., {Mohr}, J.~J., {Dolag}, K., \& {Liu}, J. 2017, \mnras,
  469, 3069

\bibitem[{Haardt {et~al.}(2001)Haardt, Madau, Neumann, \&
  Tran}]{haardt2001clusters}
Haardt, F., Madau, P., Neumann, D., \& Tran, J. 2001, Clusters of galaxies and
  the high redshift universe observed in X-rays

\bibitem[{Hahn {et~al.}(2017)Hahn, Martizzi, Wu, Evrard, Teyssier, \&
  Wechsler}]{hahn2017rhapsody}
Hahn, O., Martizzi, D., Wu, H.-Y., {et~al.} 2017, Monthly Notices of the Royal
  Astronomical Society, 470, 166

\bibitem[{Harris {et~al.}(2020)Harris, Millman, Van Der~Walt, Gommers,
  Virtanen, Cournapeau, Wieser, Taylor, Berg, Smith,
  {et~al.}}]{harris2020array}
Harris, C.~R., Millman, K.~J., Van Der~Walt, S.~J., {et~al.} 2020, Nature, 585,
  357

\bibitem[{Hirschmann {et~al.}(2014)Hirschmann, Dolag, Saro, Bachmann, Borgani,
  \& Burkert}]{hirschmann2014cosmological}
Hirschmann, M., Dolag, K., Saro, A., {et~al.} 2014, Monthly Notices of the
  Royal Astronomical Society, 442, 2304

\bibitem[{Hoffer {et~al.}(2012)Hoffer, Donahue, Hicks, \&
  Barthelemy}]{hoffer2012infrared}
Hoffer, A.~S., Donahue, M., Hicks, A., \& Barthelemy, R. 2012, The
  Astrophysical Journal Supplement Series, 199, 23

\bibitem[{Hogan {et~al.}(2017)Hogan, McNamara, Pulido, Nulsen, Russell,
  Vantyghem, Edge, \& Main}]{hogan2017mass}
Hogan, M., McNamara, B., Pulido, F., {et~al.} 2017, The Astrophysical Journal,
  837, 51

\bibitem[{Hoyle \& Lyttleton(1939)}]{hoyle1939effect}
Hoyle, F. \& Lyttleton, R.~A. 1939, Mathematical Proceedings of the Cambridge
  Philosophical Society, 35, 405–415

\bibitem[{Hunter \& Dale(2007)}]{hunter2007matplotlib}
Hunter, J. \& Dale, D. 2007, Matplotlib 0.90. 0 user’s guide

\bibitem[{Klypin {et~al.}(2016)Klypin, Yepes, Gottl{\"o}ber, Prada, \&
  Hess}]{klypin2016multidark}
Klypin, A., Yepes, G., Gottl{\"o}ber, S., Prada, F., \& Hess, S. 2016, Monthly
  Notices of the Royal Astronomical Society, 457, 4340

\bibitem[{Komatsu {et~al.}(2011)Komatsu, Smith, Dunkley, Bennett, Gold,
  Hinshaw, Jarosik, Larson, Nolta, Page, Spergel, Halpern, Hill, Kogut, Limon,
  Meyer, Odegard, Tucker, Weiland, Wollack, \& Wright}]{Komatsu_2011}
Komatsu, E., Smith, K.~M., Dunkley, J., {et~al.} 2011, The Astrophysical
  Journal Supplement Series, 192, 18

\bibitem[{Lagan{\'a} {et~al.}(2013)Lagan{\'a}, Martinet, Durret, Neto, Maughan,
  \& Zhang}]{lagana2013comprehensive}
Lagan{\'a}, T., Martinet, N., Durret, F., {et~al.} 2013, Astronomy \&
  Astrophysics, 555, A66

\bibitem[{Le~Brun {et~al.}(2014)Le~Brun, McCarthy, Schaye, \&
  Ponman}]{le2014towards}
Le~Brun, A.~M., McCarthy, I.~G., Schaye, J., \& Ponman, T.~J. 2014, Monthly
  Notices of the Royal Astronomical Society, 441, 1270

\bibitem[{Lehle {et~al.}(2024)Lehle, Nelson, Pillepich, Truong, \&
  Rohr}]{lehle2024heart}
Lehle, K., Nelson, D., Pillepich, A., Truong, N., \& Rohr, E. 2024, Astronomy
  \& Astrophysics

\bibitem[{Liu {et~al.}(2022)Liu, Bulbul, Ghirardini, Liu, Klein, Clerc,
  {\"O}zsoy, Ramos-Ceja, Pacaud, Comparat, {et~al.}}]{liu2022erosita}
Liu, A., Bulbul, E., Ghirardini, V., {et~al.} 2022, Astronomy \& Astrophysics,
  661, A2

\bibitem[{Lovisari {et~al.}(2020)Lovisari, Schellenberger, Sereno, Ettori,
  Pratt, Forman, Jones, Andrade-Santos, Randall, \& Kraft}]{lovisari2020x}
Lovisari, L., Schellenberger, G., Sereno, M., {et~al.} 2020, The Astrophysical
  Journal, 892, 102

\bibitem[{{Marini} {et~al.}(2024){Marini}, {Popesso}, {Lamer}, {Dolag},
  {Biffi}, {Vladutescu-Zopp}, {Dev}, {Toptun}, {Bulbul}, {Comparat},
  {Malavasi}, {Merloni}, {Mroczkowski}, {Ponti}, {Seppi}, {Shreeram}, \&
  {Zhang}}]{marini2024groups}
{Marini}, I., {Popesso}, P., {Lamer}, G., {et~al.} 2024, arXiv e-prints,
  arXiv:2404.12719

\bibitem[{Maughan {et~al.}(2012)Maughan, Giles, Randall, Jones, \&
  Forman}]{maughan2012self}
Maughan, B., Giles, P., Randall, S., Jones, C., \& Forman, W. 2012, Monthly
  Notices of the Royal Astronomical Society, 421, 1583

\bibitem[{Mazzotta {et~al.}(2004)Mazzotta, Rasia, Moscardini, \&
  Tormen}]{mazzotta2004comparing}
Mazzotta, P., Rasia, E., Moscardini, L., \& Tormen, G. 2004, Monthly Notices of
  the Royal Astronomical Society, 354, 10

\bibitem[{McCarthy {et~al.}(2008)McCarthy, Babul, Bower, \&
  Balogh}]{mccarthy2008towards}
McCarthy, I.~G., Babul, A., Bower, R.~G., \& Balogh, M.~L. 2008, Monthly
  Notices of the Royal Astronomical Society, 386, 1309

\bibitem[{{McCourt} {et~al.}(2012){McCourt}, {Sharma}, {Quataert}, \&
  {Parrish}}]{mccourt2012ti}
{McCourt}, M., {Sharma}, P., {Quataert}, E., \& {Parrish}, I.~J. 2012, \mnras,
  419, 3319

\bibitem[{{McDonald} {et~al.}(2013){McDonald}, {Benson}, {Vikhlinin},
  {Stalder}, {Bleem}, {de Haan}, {Lin}, {Aird}, {Ashby}, {Bautz}, {Bayliss},
  {Bocquet}, {Brodwin}, {Carlstrom}, {Chang}, {Cho}, {Clocchiatti}, {Crawford},
  {Crites}, {Desai}, {Dobbs}, {Dudley}, {Foley}, {Forman}, {George},
  {Gettings}, {Gladders}, {Gonzalez}, {Halverson}, {High}, {Holder},
  {Holzapfel}, {Hoover}, {Hrubes}, {Jones}, {Joy}, {Keisler}, {Knox}, {Lee},
  {Leitch}, {Liu}, {Lueker}, {Luong-Van}, {Mantz}, {Marrone}, {McMahon},
  {Mehl}, {Meyer}, {Miller}, {Mocanu}, {Mohr}, {Montroy}, {Murray},
  {Nurgaliev}, {Padin}, {Plagge}, {Pryke}, {Reichardt}, {Rest}, {Ruel}, {Ruhl},
  {Saliwanchik}, {Saro}, {Sayre}, {Schaffer}, {Shirokoff}, {Song},
  {{\v{S}}uhada}, {Spieler}, {Stanford}, {Staniszewski}, {Stark}, {Story}, {van
  Engelen}, {Vanderlinde}, {Vieira}, {Williamson}, {Zahn}, \&
  {Zenteno}}]{mcdonald2013ccevol}
{McDonald}, M., {Benson}, B.~A., {Vikhlinin}, A., {et~al.} 2013, \apj, 774, 23

\bibitem[{McDonald {et~al.}(2018)McDonald, Gaspari, McNamara, \&
  Tremblay}]{mcdonald2018revisiting}
McDonald, M., Gaspari, M., McNamara, B., \& Tremblay, G. 2018, The
  Astrophysical Journal, 858, 45

\bibitem[{Motl {et~al.}(2004)Motl, Burns, Loken, Norman, \&
  Bryan}]{motl2004formation}
Motl, P.~M., Burns, J.~O., Loken, C., Norman, M.~L., \& Bryan, G. 2004, The
  Astrophysical Journal, 606, 635

\bibitem[{Nagai {et~al.}(2007)Nagai, Kravtsov, \& Vikhlinin}]{nagai2007effects}
Nagai, D., Kravtsov, A.~V., \& Vikhlinin, A. 2007, The Astrophysical Journal,
  668, 1

\bibitem[{Narayan \& Yi(1995)}]{narayan1995advection}
Narayan, R. \& Yi, I. 1995, Astrophysical Journal v. 452, p. 710, 452, 710

\bibitem[{Panagoulia {et~al.}(2014)Panagoulia, Fabian, \&
  Sanders}]{panagoulia2014volume}
Panagoulia, E., Fabian, A., \& Sanders, J. 2014, Monthly Notices of the Royal
  Astronomical Society, 438, 2341

\bibitem[{Pasini {et~al.}(2022)Pasini, Br{\"u}ggen, Hoang, Ghirardini, Bulbul,
  Klein, Liu, Shimwell, Hardcastle, Williams, {et~al.}}]{pasini2022erosita}
Pasini, T., Br{\"u}ggen, M., Hoang, D., {et~al.} 2022, Astronomy \&
  Astrophysics, 661, A13

\bibitem[{Peterson {et~al.}(2003)Peterson, Kahn, Paerels, Kaastra, Tamura,
  Bleeker, Ferrigno, \& Jernigan}]{peterson2003high}
Peterson, J., Kahn, S., Paerels, F., {et~al.} 2003, The Astrophysical Journal,
  590, 207

\bibitem[{{Planelles} {et~al.}(2013){Planelles}, {Borgani}, {Dolag}, {Ettori},
  {Fabjan}, {Murante}, \& {Tornatore}}]{planelles2013baryons}
{Planelles}, S., {Borgani}, S., {Dolag}, K., {et~al.} 2013, \mnras, 431, 1487

\bibitem[{Planelles \& Quilis(2009)}]{planelles2009galaxy}
Planelles, S. \& Quilis, V. 2009, Monthly Notices of the Royal Astronomical
  Society, 399, 410

\bibitem[{Poole {et~al.}(2008)Poole, Babul, McCarthy, Sanderson, \&
  Fardal}]{poole2008impact}
Poole, G.~B., Babul, A., McCarthy, I.~G., Sanderson, A., \& Fardal, M.~A. 2008,
  Monthly Notices of the Royal Astronomical Society, 391, 1163

\bibitem[{Price(2008)}]{price2008modelling}
Price, D.~J. 2008, Journal of Computational Physics, 227, 10040

\bibitem[{Rafferty {et~al.}(2006)Rafferty, McNamara, Nulsen, \&
  Wise}]{rafferty2006feedback}
Rafferty, D.~A., McNamara, B., Nulsen, P., \& Wise, M. 2006, The Astrophysical
  Journal, 652, 216

\bibitem[{{Ragagnin} {et~al.}(2017){Ragagnin}, {Dolag}, {Biffi}, {Cadolle Bel},
  {Hammer}, {Krukau}, {Petkova}, \& {Steinborn}}]{ragagnin2017webportal}
{Ragagnin}, A., {Dolag}, K., {Biffi}, V., {et~al.} 2017, Astronomy and
  Computing, 20, 52

\bibitem[{{Ragagnin} {et~al.}(2019){Ragagnin}, {Dolag}, {Moscardini},
  {Biviano}, \& {D'Onofrio}}]{ragagnin2019concentration}
{Ragagnin}, A., {Dolag}, K., {Moscardini}, L., {Biviano}, A., \& {D'Onofrio},
  M. 2019, \mnras, 486, 4001

\bibitem[{{Ragagnin} {et~al.}(2021){Ragagnin}, {Saro}, {Singh}, \&
  {Dolag}}]{ragagnin2021mass}
{Ragagnin}, A., {Saro}, A., {Singh}, P., \& {Dolag}, K. 2021, \mnras, 500, 5056

\bibitem[{Rasia {et~al.}(2015)Rasia, Borgani, Murante, Planelles, Beck, Biffi,
  Ragone-Figueroa, Granato, Steinborn, \& Dolag}]{rasia2015cool}
Rasia, E., Borgani, S., Murante, G., {et~al.} 2015, The Astrophysical Journal
  Letters, 813, L17

\bibitem[{Rasia {et~al.}(2004)Rasia, Mazzotta, Borgani, Moscardini, Dolag,
  Tormen, Diaferio, \& Murante}]{rasia2004mismatch}
Rasia, E., Mazzotta, P., Borgani, S., {et~al.} 2004, The Astrophysical Journal,
  618, L1

\bibitem[{Reynolds(2021)}]{reynolds2021observational}
Reynolds, C.~S. 2021, Annual Review of Astronomy and Astrophysics, 59, 117

\bibitem[{Robitaille {et~al.}(2013)Robitaille, Tollerud, Greenfield,
  Droettboom, Bray, Aldcroft, Davis, Ginsburg, Price-Whelan, Kerzendorf,
  {et~al.}}]{robitaille2013astropy}
Robitaille, T.~P., Tollerud, E.~J., Greenfield, P., {et~al.} 2013, Astronomy \&
  Astrophysics, 558, A33

\bibitem[{Roncarelli {et~al.}(2018)Roncarelli, Gaspari, Ettori, Biffi,
  Brighenti, Bulbul, Clerc, Cucchetti, Pointecouteau, \&
  Rasia}]{roncarelli2018measuring}
Roncarelli, M., Gaspari, M., Ettori, S., {et~al.} 2018, Astronomy \&
  Astrophysics, 618, A39

\bibitem[{Rossetti {et~al.}(2017)Rossetti, Gastaldello, Eckert, Della~Torre,
  Pantiri, Cazzoletti, \& Molendi}]{rossetti2017cool}
Rossetti, M., Gastaldello, F., Eckert, D., {et~al.} 2017, Monthly Notices of
  the Royal Astronomical Society, 468, 1917

\bibitem[{Russell {et~al.}(2013)Russell, McNamara, Edge, Hogan, Main, \&
  Vantyghem}]{russell2013radiative}
Russell, H., McNamara, B., Edge, A., {et~al.} 2013, Monthly Notices of the
  Royal Astronomical Society, 432, 530

\bibitem[{Sala {et~al.}(2023)Sala, Valentini, Biffi, \&
  Dolag}]{sala2023supermassive}
Sala, L., Valentini, M., Biffi, V., \& Dolag, K. 2023, arXiv preprint
  arXiv:2312.07657

\bibitem[{Sanders {et~al.}(2010)Sanders, Fabian, Frank, Peterson, \&
  Russell}]{sanders2010deep}
Sanders, J., Fabian, A., Frank, K., Peterson, J., \& Russell, H. 2010, Monthly
  Notices of the Royal Astronomical Society, 402, 127

\bibitem[{Sanders {et~al.}(2018)Sanders, Fabian, Russell, \&
  Walker}]{sanders2018hydrostatic}
Sanders, J., Fabian, A., Russell, H., \& Walker, S. 2018, Monthly Notices of
  the Royal Astronomical Society, 474, 1065

\bibitem[{Sanderson {et~al.}(2009)Sanderson, Edge, \&
  Smith}]{sanderson2009locuss}
Sanderson, A.~J., Edge, A.~C., \& Smith, G.~P. 2009, Monthly Notices of the
  Royal Astronomical Society, 398, 1698

\bibitem[{Santos {et~al.}(2008)Santos, Rosati, Tozzi, B{\"o}hringer, Ettori, \&
  Bignamini}]{santos2008searching}
Santos, J.~S., Rosati, P., Tozzi, P., {et~al.} 2008, Astronomy \& Astrophysics,
  483, 35

\bibitem[{Santos {et~al.}(2010)Santos, Tozzi, Rosati, \&
  B{\"o}hringer}]{santos2010evolution}
Santos, J.~S., Tozzi, P., Rosati, P., \& B{\"o}hringer, H. 2010, Astronomy \&
  Astrophysics, 521, A64

\bibitem[{Sarazin(2002)}]{sarazin2002physics}
Sarazin, C.~L. 2002, Merging Processes in Galaxy Clusters, 1

\bibitem[{Seabold \& Perktold(2010)}]{seabold2010statsmodels}
Seabold, S. \& Perktold, J. 2010, SciPy, 7

\bibitem[{Shakura \& Sunyaev(1973)}]{shakura1973black}
Shakura, N.~I. \& Sunyaev, R.~A. 1973, Astronomy and Astrophysics, Vol. 24, p.
  337-355, 24, 337

\bibitem[{Sijacki {et~al.}(2007)Sijacki, Springel, Di~Matteo, \&
  Hernquist}]{sijacki2007unified}
Sijacki, D., Springel, V., Di~Matteo, T., \& Hernquist, L. 2007, Monthly
  Notices of the Royal Astronomical Society, 380, 877

\bibitem[{Smith {et~al.}(2001)Smith, Brickhouse, Liedahl, \&
  Raymond}]{smith2001collisional}
Smith, R.~K., Brickhouse, N.~S., Liedahl, D.~A., \& Raymond, J.~C. 2001, The
  Astrophysical Journal, 556, L91

\bibitem[{Spitzer(1962)}]{spitzer1962jr}
Spitzer, L. 1962, Jr., Physics of fully ionized gases

\bibitem[{Springel(2005)}]{springel2005cosmological}
Springel, V. 2005, Monthly notices of the royal astronomical society, 364, 1105

\bibitem[{Springel {et~al.}(2005)Springel, Di~Matteo, \&
  Hernquist}]{springel2005modelling}
Springel, V., Di~Matteo, T., \& Hernquist, L. 2005, Monthly Notices of the
  Royal Astronomical Society, 361, 776

\bibitem[{Springel \& Hernquist(2003)}]{springel2003cosmological}
Springel, V. \& Hernquist, L. 2003, Monthly Notices of the Royal Astronomical
  Society, 339, 289

\bibitem[{{Tornatore} {et~al.}(2007){Tornatore}, {Borgani}, {Dolag}, \&
  {Matteucci}}]{tornatore2007simulating}
{Tornatore}, L., {Borgani}, S., {Dolag}, K., \& {Matteucci}, F. 2007, \mnras,
  382, 1050

\bibitem[{Tornatore {et~al.}(2004)Tornatore, Borgani, Matteucci, Recchi, \&
  Tozzi}]{tornatore2004simulating}
Tornatore, L., Borgani, S., Matteucci, F., Recchi, S., \& Tozzi, P. 2004,
  Monthly Notices of the Royal Astronomical Society, 349, L19

\bibitem[{Van~Rossum \& Drake~Jr(1995)}]{van1995python}
Van~Rossum, G. \& Drake~Jr, F.~L. 1995, Python tutorial, Vol. 620 (Centrum voor
  Wiskunde en Informatica Amsterdam, The Netherlands)

\bibitem[{Virtanen {et~al.}(2020)Virtanen, Gommers, Oliphant, Haberland, Reddy,
  Cournapeau, Burovski, Peterson, Weckesser, Bright,
  {et~al.}}]{virtanen2020scipy}
Virtanen, P., Gommers, R., Oliphant, T.~E., {et~al.} 2020, Nature methods, 17,
  261

\bibitem[{Vogelsberger {et~al.}(2013)Vogelsberger, Genel, Sijacki, Torrey,
  Springel, \& Hernquist}]{vogelsberger2013model}
Vogelsberger, M., Genel, S., Sijacki, D., {et~al.} 2013, Monthly Notices of the
  Royal Astronomical Society, 436, 3031

\bibitem[{{Voit} {et~al.}(2015){Voit}, {Donahue}, {Bryan}, \&
  {McDonald}}]{voit2015ti}
{Voit}, G.~M., {Donahue}, M., {Bryan}, G.~L., \& {McDonald}, M. 2015, \nat,
  519, 203

\bibitem[{Wadsley {et~al.}(2008)Wadsley, Veeravalli, \&
  Couchman}]{wadsley2008treatment}
Wadsley, J., Veeravalli, G., \& Couchman, H. 2008, Monthly Notices of the Royal
  Astronomical Society, 387, 427

\bibitem[{Wiersma {et~al.}(2009)Wiersma, Schaye, \& Smith}]{wiersma2009effect}
Wiersma, R.~P., Schaye, J., \& Smith, B.~D. 2009, Monthly Notices of the Royal
  Astronomical Society, 393, 99

\bibitem[{Yuan \& Narayan(2014)}]{yuan2014hot}
Yuan, F. \& Narayan, R. 2014, Annual Review of Astronomy and Astrophysics, 52,
  529

\bibitem[{Yuan \& Han(2020)}]{yuan2020dynamical}
Yuan, Z. \& Han, J. 2020, Monthly Notices of the Royal Astronomical Society,
  497, 5485

\bibitem[{ZuHone {et~al.}(2012)ZuHone, Markevitch, Brunetti, \&
  Giacintucci}]{zuhone2012turbulence}
ZuHone, J., Markevitch, M., Brunetti, G., \& Giacintucci, S. 2012, The
  Astrophysical Journal, 762, 78

\end{thebibliography}
%

\begin{appendix}
\section{Dependency of cool core fractions on the temperature ratio threshold}  \label{sec:appendixA}

In this appendix we show how the cool-core fractions depend on the temperature ratio threshold, set to unity in the main part of the analysis. Although we believe that the choice of $T_{\text{ratio,500}} < 1$, is the most straightforward one to asses the overall cool-core cluster population (including weak, mid and strong cool-core clusters), with a clear physical meaning that the core regions are on average cooler than the cluster average, still it can provide valuable insights to see how the cool-core fractions change depending on this threshold.

To this end we consider two additional thresholds: $T_{\text{ratio,500}} < 0.98$ and $T_{\text{ratio,500}} < 0.95$, which can also be interpreted as the fraction of mid-to-strong and strictly strong cool cores, respectively. Notice that it is not possible to reliably study the population of even stronger cool-cores (e.g. $T_{\text{ratio,500}} < 0.90$) given that we are already separating the data per mass bin, which reduces the amount of objects at the scales of massive clusters for both the simulated and observed samples and limits the possibilities to assess in a statistically robust manner the population of the most rare objects (really strong cool cores at scales of massive clusters).

\begin{figure*}[ht]
    \centering
    \includegraphics[width=0.45\textwidth]{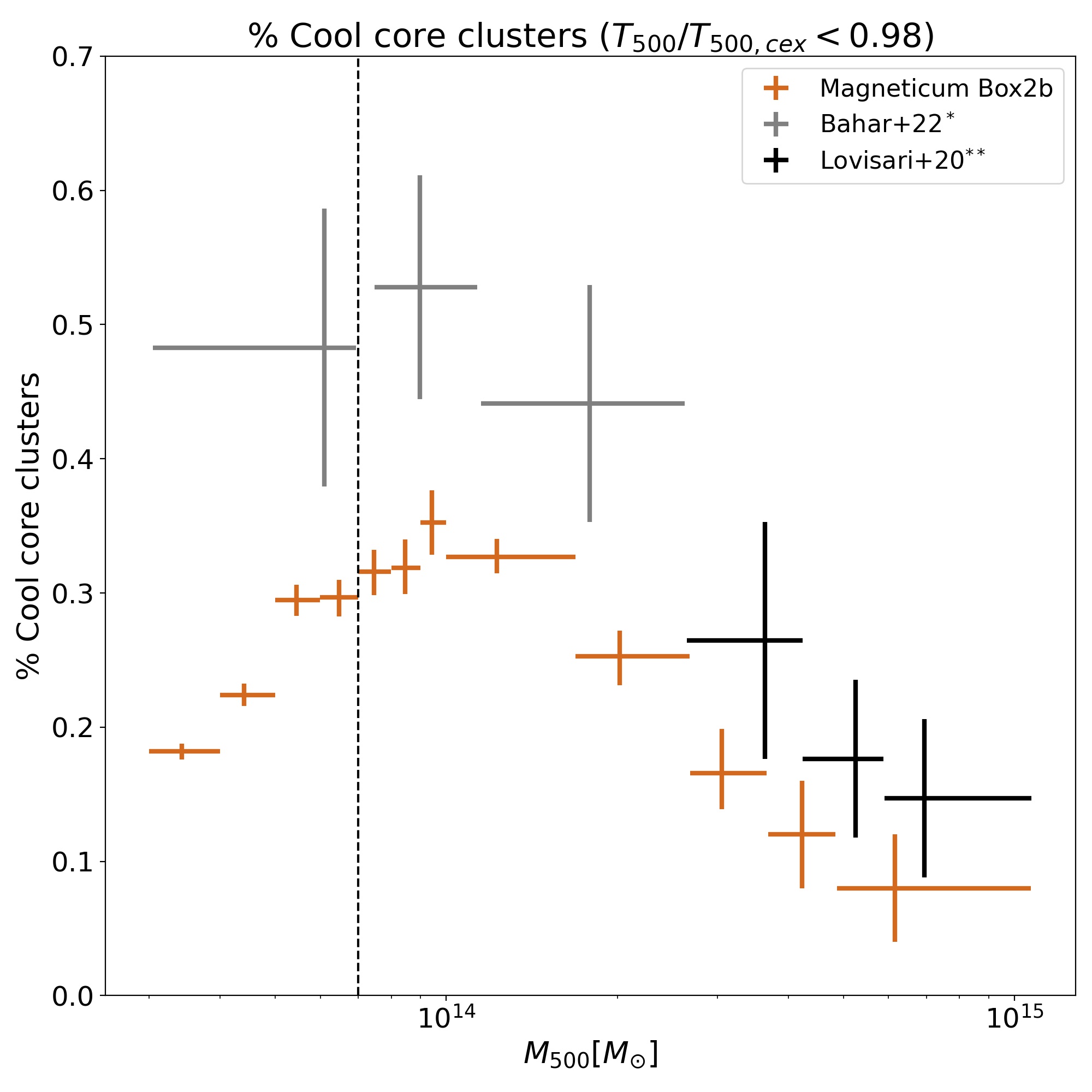}
    \includegraphics[width=0.45\textwidth]{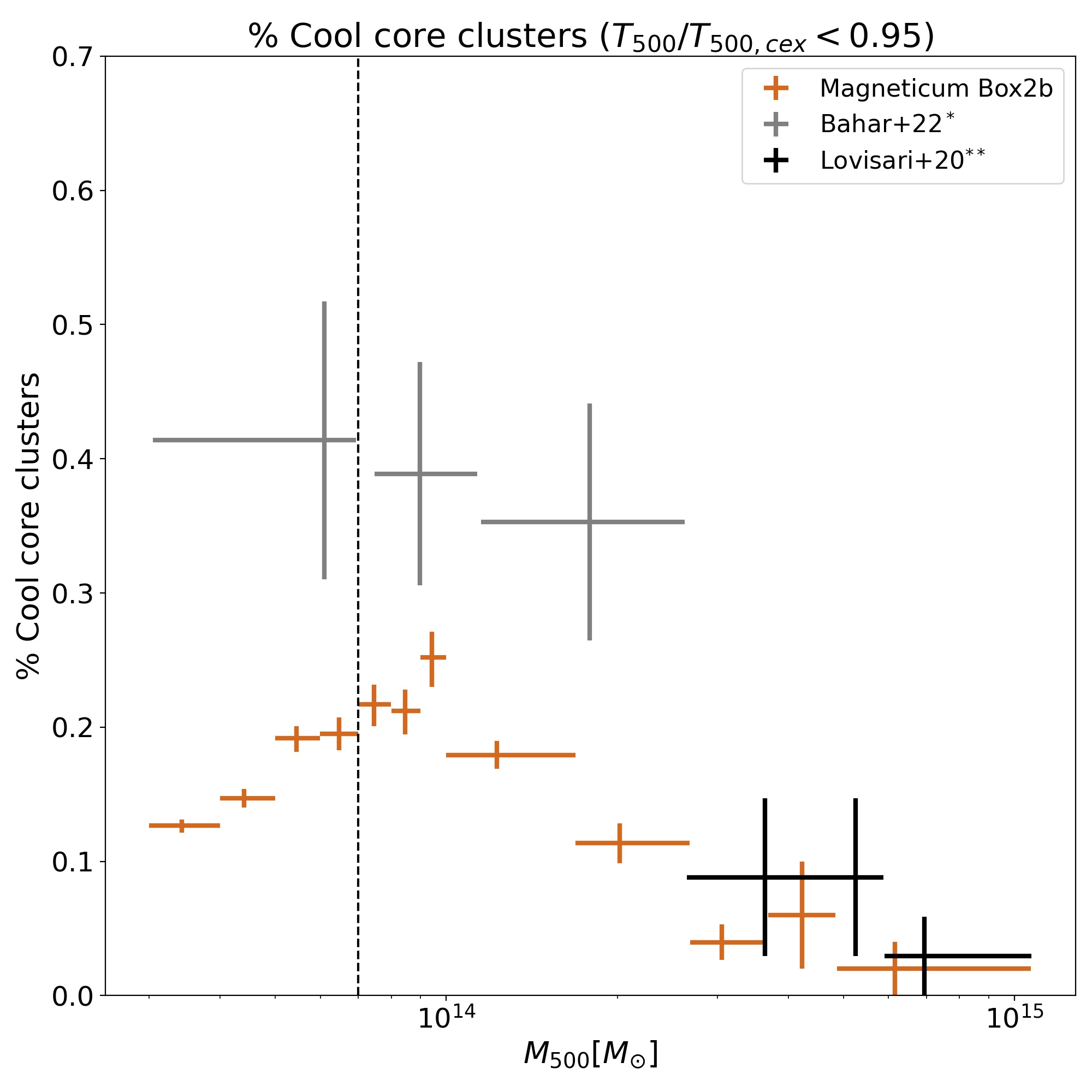}
    \caption{Cool core fractions (same as Fig. \ref{fig:CCFractions}) but using a different temperature ratio threshold. The vertical dashed line corresponds to $M_{500c} = 0.7 \cdot 10^{13} M_\odot$, above which the eFEDS survey is expected to be complete for redshifts below z < 0.3 \citep{comparat2020full}. \textbf{Left panel}: Mid-to-strong cool-core fractions determined by the number of clusters for which the temperature ratio between the total temperature, including the core region, and the core-excised temperature is less than 0.98 ($Tx_{500}/Tx_{500,cex} < 0.98$). \textbf{Right panel}: Strictly strong cool-core fractions determined by the number of clusters for which the temperature ratio between the total temperature, including the core region, and the core-excised temperature is less than 0.95 ($Tx_{500}/Tx_{500,cex} < 0.95$). }
    \label{fig:CCFractionsAlt}
\end{figure*}

Fig. \ref{fig:CCFractionsAlt} shows the results, where we can see that although the characteristic cool-core fractions curve seen in the simulation data is preserved, the cool-core fractions of mid-to-strong and strictly strong cool cores produced by the simulation are significantly lower than that of the observational data at low and mid mass ranges. On the other hand, the simulation still produces cool core fractions compatible with observations at the high mass range bins $M_{500c} > 2.65 \cdot 10^{14}M_\odot$ for all cases (weak, mid, and strong cool-core clusters).

This result extends the conclusions presented in Sec. \ref{sub:CoolCoreFractionComparison}, in that the overheating problems caused by the current AGN feedback which suppresses the overall cool-core cluster population at the low mass range, are also visible at the mid mass range when considering the mid and strong cool-core cluster population. Still, both problems should be addressed by the new AGN total feedback efficiency model presented in Sec. \ref{sub:FeedbackEfficiencies}, which produces an effective total AGN efficiency decreasing towards the mid and low mass range.

\newpage 

\section{Temperature profiles following the ACCEPT centering procedure}  \label{sec:appendixB}

In this appendix we want to explore how the temperature profiles would look like when following the same centering procedure as the ACCEPT sample described in \cite{cavagnolo2008bandpass}. That is, choosing the X-ray peak as the center, unless it is separated by more than 70 kpc from the X-ray centroid, in which case the latter is used as a center.

For this purpose, we have used the average rest-frame band XMM-eFEDS described in Sec. \ref{sec:ComparingObservationsSimulations} to obtain the X-ray peak and centroid based on emissivity weights. Then we have followed the ACCEPT centering procedure with the simulated data, using a threshold of $70 \cdot \sqrt{3/2}$ kpc, which is the 3-dimensional version of 70 kpc when accounting for projection effects, to switch from X-ray peak to X-ray centroid as a center for the radial profiles.

\begin{figure*}[ht]
    \centering
    \includegraphics[width=2\columnwidth]{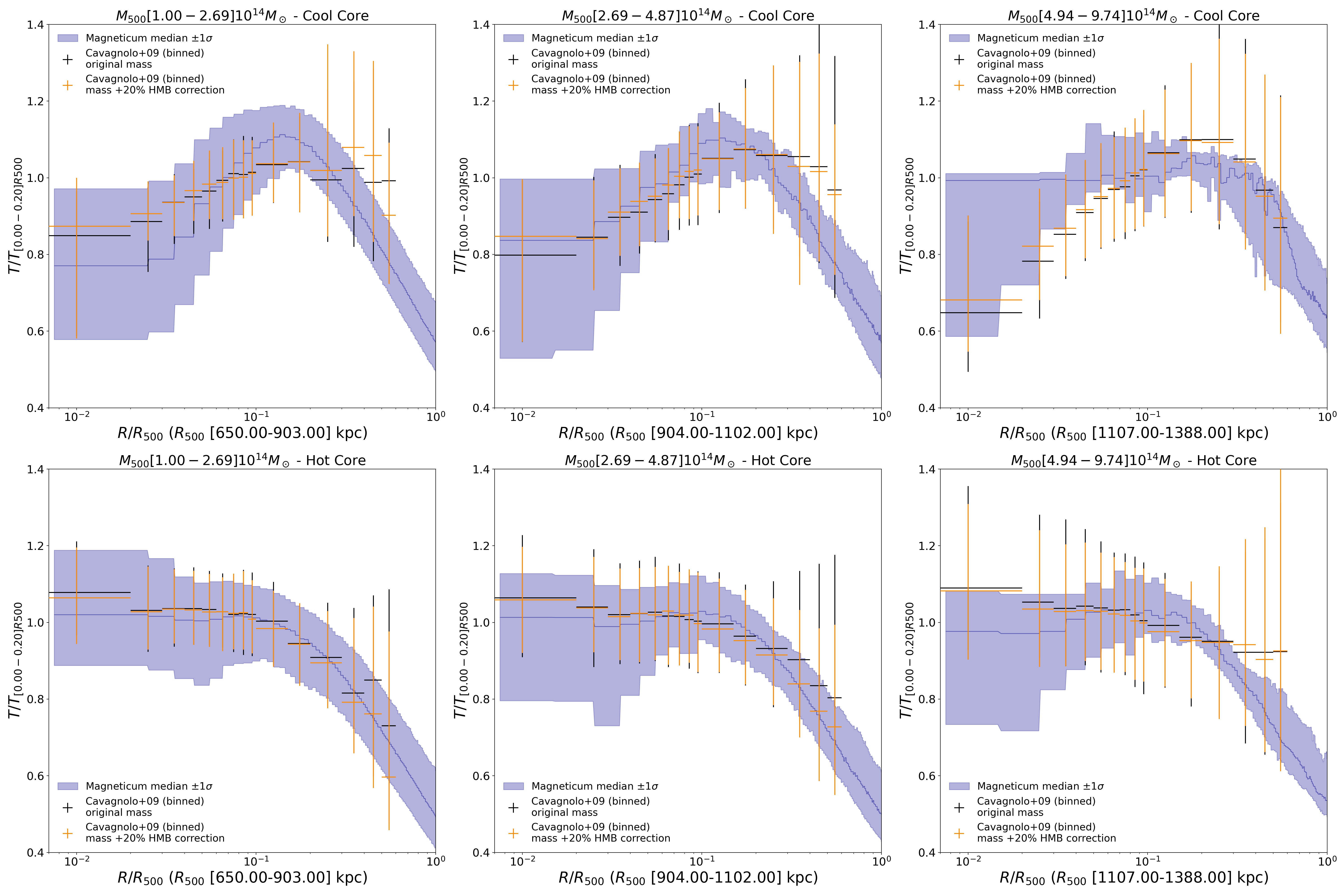}
    \caption{Projected X-ray temperature profiles (same as Fig. \ref{fig:TemperatureProfile}) but using the same centering procedure as the ACCEPT sample described in \citep{cavagnolo2008bandpass}. Columns are sorted by mass range, as indicated on top of each panel (increasing mass range from left to right). The upper row panels correspond to cool-core clusters, and the lower row panels to hot-core clusters.}
    \label{fig:TemperatureProfileAlt}
\end{figure*}

The results are shown in Fig. \ref{fig:TemperatureProfileAlt}, where we see that the hot-core clusters show a stronger isothermal core (with fewer fluctuations) in comparison with the default temperature profiles shown in Fig. \ref{fig:TemperatureProfile}, which use the most gravitationally bound particle, equivalent to the deepest point of the gravitational potential as the center. The explanation for this behavior is relatively simple:

\begin{itemize}
    \item The deepest point of the gravitational potential harbors cool, dense gas that falls by precipitation and also hosts the central AGN, which heats the immediate surroundings; hence the temperature profiles swing from cool to hot as we go through the envelope of cool, dense gas into the immediate surroundings of the central AGN.
    \item On the other hand, the X-ray centroid is distant from the X-ray peak for disturbed clusters (e.g. two dominant galaxies merging), in which case the X-ray centroid is located in between the merging structures, in a rather shallow region of the gravitational potential that does not support strong temperature gradients, harborage of cold dense gas, or stable conditions to host an AGN. All these factors contribute to an isothermal structure at the X-ray centroid of disturbed clusters.
\end{itemize}

\end{appendix}

\end{document}